\documentclass[12pt]{article}
\usepackage[utf8]{inputenc} 
\DeclareUnicodeCharacter{2212}{-}

\usepackage{draft}
\usepackage{cite} 
\usepackage{tabularx}

\usepackage{mathtools}
\usepackage[T1]{fontenc}
\usepackage{esint}

\usepackage{soul}

\usepackage[margin=1in]{geometry}
\usepackage{setspace}
\usepackage{color}
\usepackage{braket}
\usepackage{titling}
\usepackage{graphicx}
\usepackage[bf,margin=20pt,font={small}]{caption}
\usepackage{subcaption}
\usepackage{youngtab}
\usepackage{epigraph}
\usepackage{tikz}
\usetikzlibrary{math}

\usepackage{amsmath,bm}
\usepackage{amssymb,dsfont,amsfonts}
\usepackage{upgreek}
\allowdisplaybreaks
\renewcommand{\tr}{\operatorname{tr}}
\renewcommand{\Tr}{\operatorname{tr}}

\definecolor{royalblue}{rgb}{0.00000,0.44700,0.74100}
\definecolor{royalorange}{rgb}{0.85000,0.32500,0.09800}
\definecolor{royalyellow}{rgb}{0.92900,0.69400,0.12500}
\definecolor{purple}{rgb}{0.5804, 0.0, 0.82745098}
\definecolor{applegreen}{rgb}{0.55, 0.71, 0.0}
\definecolor{bittersweet}{rgb}{1.0, 0.44, 0.37}

\DeclareMathAlphabet{\mathpzc}{OT1}{pzc}{m}{it}

\usepackage{putex}

\usepackage{siunitx}

\usepackage{pgfplots}
\pgfplotsset{compat=1.10}
\usepgfplotslibrary{fillbetween}
\usetikzlibrary{patterns}

\def\XXint#1#2#3{{\setbox0=\hbox{$#1{#2#3}{\int}$ }
		\vcenter{\hbox{$#2#3$ }}\kern-.6\wd0}}

  \usepackage{adjustbox}
\usepackage{multirow}
\usepackage{tikz-cd}
\usepackage[setpagesize=false,pagebackref=false, linktocpage, bookmarksopen=true, colorlinks=true, linkcolor=blue,citecolor=blue,urlcolor=blue]{hyperref}
\usepackage{hyperref}

\numberwithin{equation}{section}

\def\<{\langle}
\def\>{\rangle}

\def\pa{\partial}

\def\ep{\epsilon}

\usetikzlibrary{decorations.pathmorphing}

\tikzset{snake it/.style={decorate, decoration=snake}}

\newcommand{\leftrarrows}{\mathrel{\raise.75ex\hbox{\oalign{%
				$\scriptstyle\leftarrow$\cr
				\vrule width0pt height.5ex$\hfil\scriptstyle\relbar$\cr}}}}
\newcommand{\lrightarrows}{\mathrel{\raise.75ex\hbox{\oalign{%
				$\scriptstyle\relbar$\hfil\cr
				$\scriptstyle\vrule width0pt height.5ex\smash\rightarrow$\cr}}}}
\newcommand{\Rrelbar}{\mathrel{\raise.75ex\hbox{\oalign{%
				$\scriptstyle\relbar$\cr
				\vrule width0pt height.5ex$\scriptstyle\relbar$}}}}

\makeatletter
\def\leftrightarrowsfill@{\arrowfill@\leftrarrows\Rrelbar\lrightarrows}
\newcommand{\xleftrightarrows}[2][]{\ext@arrow 3399\leftrightarrowsfill@{#1}{#2}}
\makeatother

\institution{NYU}{Center for Cosmology and Particle Physics, New York University, New York, NY 10003, USA}

\title{
Beyond $N=\infty$ in Large $N$ Conformal Vector Models at Finite Temperature
}

\authors{Oleksandr Diatlyk, Fedor K.~Popov, 
and Yifan Wang 
}

\abstract{
We investigate finite-temperature observables in three-dimensional large $N$ critical vector models 
taking into account the effects suppressed by $1\over N$. Such subleading contributions are captured by the fluctuations of the Hubbard-Stratonovich auxiliary field which need to be handled with care due to a subtle divergence structure which we clarify. 
The examples we consider include the scalar $O(N)$ model, the Gross-Neveu model, the Nambu-Jona-Lasinio model and the massless Chern-Simons Quantum Electrodynamics. We present explicit results for the free energy density to the subleading order in $1\over N$, which captures the thermal one-point function of the stress-energy tensor to this order. We also include the dependence on a chemical potential. We determine the Wilson coefficient in the thermal effective action that is sensitive to global symmetry for the first time directly in interacting CFTs, which produces a symmetry-resolved asymptotic density of states. 
We further provide a formula from diagrammatics for the one-point functions of general single-trace higher-spin currents. We observe that in most cases considered, these subleading effects lift the apparent degeneracies between observables in different models at infinite $N$, while in special cases the discrepancies only start to appear at the next-to-subleading order.
}

\date{\today}

\begin{document}
\maketitle 

\tableofcontents

\section{Introduction and Summary}

The understanding of strongly coupled quantum matter at finite temperature is an important problem in physics. Its broad applications range from the basic experimental needs, since any realistic quantum critical point would necessarily be at non-zero temperature, to fundamental theoretical questions such as the thermalization of many-body quantum systems, where highly excited states of the system are conjectured to be universally approximated by
the finite-temperature (thermal) state \cite{Srednicki:1994mfb,deutsch1991quantum},
and furthermore to profound connections between quantum gravity and quantum field theory dictated by the holographic principle \cite{Maldacena:1997re, Gubser:1998bc,Witten:1998qj}, where finite-temperature systems are believed to be dual to black holes, and therefore would provide valuable information about the quantum nature of the latter. 

Conformal Field Theory (CFT), which provides a non-perturbative formulation of critical phenomena,
offers a powerful approach to investigate the finite-temperature  quantum system near its quantum critical point. 
While the second-order quantum phase transition occurs at zero temperature and is described by the CFT on the flat spacetime $\mR^{d-1,1}$, 
to turn on finite temperature $T$ amounts to compactifying the Euclidean time $\tau \sim \tau+\B$ with  $\B=\frac{1}{T}$ and imposing suitable periodicity conditions along $\tau$. At thermal equilibrium, this boils down to studying the  Euclidean CFT (from Wick rotation) on ${\rm S}^1_\B \times \mR^{d-1}$ (known as the thermal background), and the basic observables are correlation functions of local operators on this background, the simplest of which being the one-point function $\la T_{\m\n}\ra_\B$ of the stress-energy tensor which measures the free energy of the thermal state.
On the one hand, such thermal observables behave qualitatively different from the conventional flat space CFT correlation functions 
due to the explicit breaking of conformal symmetry by the thermal background, and for the same reason they appear much harder to determine in interacting models. On the other hand, these thermal observables present a universal coarse-grained description of the flat space operator data, through the asymptotic density of states in the CFT Hilbert space on ${\rm S}^{d-1}$ and  operator-product-expansion (OPE) coefficients averaged over high energy states \cite{Iliesiu:2018fao} (see also the recent work \cite{Benjamin:2023qsc}). In light of the AdS/CFT correspondence \cite{Maldacena:1997re, Gubser:1998bc,Witten:1998qj}, this translates to the expectation that the black hole solution in gravity is a universal coarse-grained description
that encodes a large number of underlying microstates. 
Nonetheless deriving such universal formulas directly from the flat space OPE data is very challenging in interacting CFTs of spacetime dimension $d\geq 3$. A main focus of this work is to provide explicit solutions to thermal observables using field theory techniques in interacting CFTs in $d=3$.\footnote{It would be interesting to study our results in relation to the flat space OPE data. We leave that to future work.}

There is an alternative interpretation of the CFT on ${\rm S}^1_\B \times \mR^{d-1}$  entirely as a zero-temperature quantum system (or a classical statistical system in $d$ dimensions) and consequently a reinterpretation of the results we present in this paper. Instead of taking ${\rm S}^1_\B$ to be the compactified Euclidean time, we can choose it to be a compactified spatial circle (and keep the new Euclidean time along one of the $\mR^{d-1}$ directions).
This way, it describes the quantum system confined to a finite interval of length $\B$ with certain periodic boundary conditions, also known as the
Kaluza-Klein compactification of the CFT.
Here an important and universal observable is the Casimir force, which underpins
the quantum nature of the system. For instance, if we confine free electromagnetic field in cavity, the walls of the cavity would feel the pressure due to the quantum fluctuations of the vacuum despite being at zero temperature. Such an effect was firstly predicted by Hendrik Casimir in 1948 and later confirmed experimentally. The Casimir effect exists for any quantum field theory, but in the absence of a mass-gap, the Casimir pressure is expected to depend on the geometric moduli of the cavity algebraically. Consequently the Casimir force will be long-ranged and more easily detected in a gapless phase, in contrast to a gapped phase where the Casimir pressure decays exponentially as we increase the cavity size.  Therefore,  the Casimir force (pressure) serves as a salient observable when the system undergoes a quantum phase transition (or a second-order phase transition in the statistical model),  which is often described by a CFT. It is again measured by the one-point function $\la T_{\m\n}\ra_\B$ but with the time and space directions swapped. Similar reinterpretations hold for more general observables on ${\rm S}^1_\beta \times \mR^{d-1}$.

In this work we study $d=3$ CFTs on the  background ${\rm S}^1_\beta \times \mR^2$. The dimensionful parameter $\beta$ breaks the conformal symmetry explicitly to the Euclidean isometry on $\mR^2$ and translation symmetry along ${\rm S}^1_\B$. As an immediate consequence, the one-point functions of local operators are no longer constrained to be zero. Instead the   residual symmetry and dimensional analysis require the one-point function of a primary operator $\cO^{\m_1\dots \m_\ell}$ of scaling dimension $\Delta$ and spin $\ell$ to take the following form with an overall constant $b_\cO$ \cite{Iliesiu:2018fao},
\ie 
\la \cO^{\m_1\dots \m_\ell}(x) \ra_\B=\frac{b_\cO}{\B^\Delta}( e^{\m_1}\dots e^{\m_\ell} -{\rm traces})\,,\quad e^\m=(1,0,\dots,0)\,,
\label{thermal1PF}
\fe
and the descendant operators all have vanishing one-point functions. The dimensionless one-point function coefficients $b_\cO$ are the most basic building blocks in the finite temperature CFT and they determine  the most general correlation functions 
together with the flat OPE data.

For instance, the stress-energy tensor has the following one-point function at finite temperature (we keep $d$ general for the moment),
\ie 
    \la T_{00} \ra_\B  = (d-1)\frac{f}{\beta^d}\,, \quad  \la T_{ij} \ra_\B = -\frac{f \delta_{ij}}{\beta^d}\,,\quad \la T_{0i}\ra_\B=0\,.
    \label{T1PF}
\fe 
where the constant $f=\frac{b_T}{d}$ determines the free energy density of the CFT thermal state via
\ie 
F(\B)= \dfrac{f}{\B^d}\,,
\label{freeEcoeff}
\fe
which is negative by the positivity of energy \cite{Iliesiu:2018fao}.\footnote{Note that the Lorentzian energy density (which is positive) is related to the temporal component of the Euclidean stress tensor (in \eqref{T1PF}) by Wick rotation $T^{ \rm Lorentzian}_{00}= i^2 T^{\rm Euclidean}_{00}$.} When the ${\rm S}^1_\B$ is regarded as a spatial circle, $f$ is referred as the critical Casimir amplitude which determines the Casimir pressure from changing the circle size $\B$. 
This universal quantity $f$
 have been measured for a variety of systems in experiments or from the Monte-Carlo simulations (see \cite{Dantchev:2022hvy} for an extensive review). Yet only limited results are available from the theoretical side for $d\geq 3$ 
\cite{Chubukov:1993aau,Petkou:1998wd,Christiansen:1999uv,Kaul:2008xw,Katz:2014rla,Iliesiu:2018zlz,Iliesiu:2018fao,Petkou:2018ynm}.
 One main purpose of this paper is to 
 compute the free energy coefficient (equivalently critical Casimir amplitude) $f$ explicitly in interacting $d=3$ CFTs that are solvable in some regime, and similarly for one-point functions \eqref{thermal1PF} of more general local operators.

Importantly these one-point functions also give access to flat space CFT  
data that are hard to obtain otherwise. Regarding the background ${\rm S}^1_\B\times \mR^{d-1}$ as a limit of ${\rm S}^1_\B\times {\rm S}^{d-1}_R$ as ${R\over \B}\to \infty$, we can see the free energy coefficient controls the asymptotic density of high energy CFT states on ${\rm S}^{d-1}$, equivalently heavy local operators by the state-operator correspondence. Explicitly, the partition function ${\rm S}^1_\B\times {\rm S}^{d-1}_R$ that counts states in the Hilbert space $\cH_{{\rm S}^{d-1}}$ graded by dilatation operator $\Delta$ is determined by the free energy coefficient $f$ in this limit,
\ie 
Z_{{\rm S}^1_\B \times {\rm S}^{d-1}_R}\equiv {\rm Tr\,}_{\cH_{{\rm S}^{d-1}}} e^{-{\B\over R} \Delta} 
=
\int d\Delta\, \rho(\Delta)e^{-{\B\over R} \Delta} \xrightarrow[]{{R\over \B}\to \infty} Z_{{\rm S}^1_\B \times \mR^{d-1}}= e^{
- {V_{d-1}\over \B^{d-1}}  f }\,,
\label{Zsphere}
\fe
where $\rho(\Delta)$ is the density of states and $V_{d-1}\equiv S_{d-1} R^{d-1}={2\pi^{d\over 2}\over \Gamma \left(d\over2\right)} R^{d-1}$ is the area of ${\rm S}^{d-1}_R$.
Performing an inverse Laplace transform, this implies the following asymptotic density of heavy operators \cite{Shaghoulian:2015kta},
\ie 
 \log \rho(\Delta) \xrightarrow[]{\Delta\gg 1} {d\over (d-1)^{d-1\over d}}(-fS_{d-1})^{1\over d}\Delta^{d-1\over d}\,.
 \label{asympDENSITY}
\fe
Similarly, the one-point function of a general local operator $\cO^{\m_1\dots\m_J}$ on ${\rm S}^1_\B\times {\rm S}_R^{d-1}$ is determined by the thermal one-point function $\la \cO\ra_\B$ for $\cO\equiv e^{\m_1}\dots e^{\m_J}\cO^{\m_1\dots\m_J}$ in the same limit,
\ie 
\la \cO(x) \ra_{{\rm S}^1_\B\times {\rm S}_R^{d-1}}
\equiv {1\over Z_{{\rm S}^1_\B\times {\rm S}_R^{d-1}}} \sum_{\phi\in \cH_{{\rm S}^{d-1}}} \la \phi|\cO(x)|\phi\ra e^{-{\B\over R}\Delta_\phi}
\xrightarrow[]{{R\over \B}\to \infty}
\la \cO(x) \ra_\B = {b_\cO\over \B^{\Delta_\cO}}\,,
\fe
where $\phi$ labels an orthonormal basis in $\cH_{{\rm S}^{d-1}}$.
The inverse Laplace transform then produces an asymptotic formula for the averaged OPE coefficients 
\cite{Gobeil:2018fzy},
\ie 
\left.\overline{\la \phi|\cO|\phi\ra} \right|_{\Delta}\equiv {\sum_{\phi\in \cH_{{\rm S}^{d-1}}}\la \phi|\cO|\phi\ra \delta(\Delta_\phi-\Delta)
\over \rho(\Delta)}
\xrightarrow[]{~\Delta \gg 1~}
b_\cO \left (   \Delta \over (1-d)f  S_{d-1}  \right)^{\Delta_\cO\over d}\,.
\label{asympOPE}
\fe
Note that all of the above are completely determined by the one-point function coefficient $b_\cO$ (including the free energy coefficient $f$ as a special case).

In general, CFTs are strongly coupled  and thus inaccessible via small perturbations near some integrable theory. A class of $d=3$ models with a large number $N$ of scalars and fermions governed by $O(N)$ (or $U(N)$) invariant interactions, which we will refer to collectively as 
large $N$ vector models,  circumvent this obstacle by admitting an expansion in $\frac{1}{N}$ and thus providing an ideal playground to study interacting CFTs (see \cite{Moshe:2003xn} for a review). Surprisingly, this simple-looking vector model has proven to be an excellent description of real-world critical systems. In the case of scalar vector models, the UV Lagrangian for the   scalar fields $\phi^i$ with $i=1,2,\dots,N$ takes the following form
\ie 
S=\int d^3 x \left(
{1\over 2} \pa_\m \phi^i \pa^\m \phi^i + \frac12 m_0^2 \phi^i\phi^i+{\lambda_0\over 4 } (\phi^i\phi^i)^2
\right) \,.
\label{phi4}
\fe
The special cases
with $N=1,2,3$ correspond to the Ising model, the XY model and the Heisenberg model respectively, each of which has experimental realizations both as zero-temperature quantum phase transitions and classical (thermal) phases transitions in statistical systems and can also be simulated on a lattice.\footnote{See \cite{Henriksson:2022rnm} for a comprehensive review on the $O(N)$ CFTs including their conformal data and physical applications.} Furthermore, the limit $N\to 0$ describes the statistics of polymers \cite{deGennes:1972zz}. In the case of the Ising and the XY model, the free energy coefficient $f$ in  \eqref{freeEcoeff} (equivalently the critical Casimir amplitude)  
has been computed by the Monte-Carlo techniques \cite{vasilyev2007monte,vasilyev2009universal} and using the $\epsilon$ expansion and functional renormalization group \cite{Jakubczyk:2012iza}. The results are summarized below,
\begin{gather}
    f_{N=1}^{\rm MC} = −0.1527\,, \quad    f_{N=2}^{\rm MC} = −0.3066\,, \quad f_{N=1}^{\rm RG}= -0.181\,,  \quad f_{N=2}^{\rm RG} = -0.353\,.
\end{gather}
In the large $N$ limit, the leading contribution to the coefficient $f$ was determined in \cite{Sachdev:1993pr},
\begin{gather}
   f_N \sim  -\frac{2\zeta(3)}{5\pi} N =  −0.153051 N\,,
\end{gather}
up to subleading corrections in ${1\over N}$.
Notably this is in rather good agreement with the Monte-Carlo simulations and the RG computations for the Ising model ($N=1$) and the XY model ($N=2$) despite the small values of $N$. It is then natural to ask about the nature of the ${1\over N}$ corrections. 

Famously, the ${1\over N}$ expansion in the large $N$ scalar vector model (similarly for fermionic models) coincides with a semi-classical expansion  where ${1\over N}$ plays the role of the Planck constant $\hbar$ \cite{Moshe:2003xn}. This semi-classical expansion is facilitated by introducing the Hubbard-Stratonovich auxiliary field $\sigma$ and rewriting the action \eqref{phi4} in a form that is quadratic in the elementary  fields $\phi^i$. Path-integrating over $\phi^i$ then produces a non-local effective Lagrangian $\cF(\sigma)$ for the $\sigma$ field which is amenable to saddle-point analysis due to the small $\hbar\sim {1\over N}$. In the scaling region, the saddle-point $\sigma=\sigma_*$ in the $N\to \infty$ limit and the value of the effective Lagrangian $\cF(\sigma_*)$ determines the $O(N)$ CFT completely to the leading order in $N$. The first correction comes from the one-loop determinant around the saddle $\sigma=\sigma_*$, which is suppressed by ${1\over N}$.

For the large $N$  critical $O(N)$ scalar model, the  ${1\over N}$ correction to the free energy coefficient $f$ in \eqref{freeEcoeff}  was first computed in \cite{Chubukov:1993aau}. The calculation requires a subtle numerical procedure which 
 we will clarify here. We also provide several approaches to access the same observable. Together they lead to the following expression for the free energy coefficient for the $O(N)$ model 
 including the first ${1\over N}$ correction,
\ie 
f_{O(N)}=&
-\frac{2\zeta(3)}{5\pi} N
+ {0.06399553} + \cO(N^{-1})\,,
\fe
which improves significantly on the precision of the result from \cite{Chubukov:1993aau}. It is immediate to note that the correction is comparable to the leading term at small $N$. This is related to the fact that we solve these models using the saddle-point approximation, which generally produces asymptotic series in the expansion parameter.

We further analyze the ${1\over N}$ correction to the free energies of fermionic large $N$ vector models using our numerical procedure. These models all arise from the critical points of $N$ Dirac fermions with different types of four-fermion interactions.
They include the Gross-Neveu (GN) model with the maximal global symmetry and its closely related variations with reduced symmetry such as the chiral Ising Gross-Neveu (cGN) model  and the Nambu-Jona-Lasinio (NJL) model. 
The results are summarized below,
\ie 
f_{{\rm GN}_N}=f_{{\rm cGN}_N}=&
-\frac{3\zeta(3)}{4\pi} N
- {0.01340099} + \cO(N^{-1})\,,\\
f_{{\rm NJL}_N}=&
-\frac{3\zeta(3)}{4\pi} N
-0.02680198 + \cO(N^{-1})\,.
\label{fEfermion}
\fe 
We see that all these models have the identical free energy density at leading order in the large $N$ expansion, which coincides with that of the free fermions \cite{Petkou:1998wd,Christiansen:1999uv}. 
While the NJL model is distinguished from the GN and the cGN models at the subleading order in ${1\over N}$. The free energy density of the cGN model only starts to differ from the GN model at the next-to-subleading order in $1\over N$.

The fermionic large $N$ models present additional subtleties that are absent in the bosonic models. For instance, the large $N$ saddle-point equation in the scaling region has multiple solutions in addition to the one producing \eqref{fEfermion}. The analog of the Hubbard-Stratonovich field here is a pseudoscalar $\varphi$ and the solution $\varphi=0$ gives \eqref{fEfermion}.
The additional solutions $\varphi\neq 0$ break the parity symmetry (and come in a complex conjugate pair) 
lead to the following free energy density
\ie 
f_{{\rm GN}^{'}_N}=&
-\kappa   N
+ {0.14222693} + \cO(N^{-1})\,,
\\
f_{{\rm NJL}^{'}_N}=&
-\kappa N
-0.44297273 + \cO(N^{-1})\,,
\fe 
with 
\ie 
\kappa\equiv \frac{2}{3} {\rm Cl}_2(\pi/3)-\frac{\zeta (3)}{3 \pi }=0.54908554\,,
\fe
where ${\rm Cl}_2$ is the second Clausen function (see \eqref{clausen}). Since these solutions have a lower free energy compared to \eqref{fEfermion} one may naively expect them to describe the actual CFT at finite temperature. However a careful inspection of these fermionic large $N$ models reveals that these saddle-points are not on the steepest descent contour that is obtained from deforming the original integration contour for $\varphi$  that defines the unitary QFT \cite{Zinn-Justin:1991ksq,Moshe:2003xn}. Therefore these saddle-point solutions are spurious for the study of the unitary models and do not describe the corresponding unitary fermionic CFT. Instead we propose that they potentially describe non-unitary cousins (analogs of the Lee-Yang CFT) after a rotation of the defining integration contour.

We also study an interesting class of 3d CFTs that share the same $U(N)$ global symmetry. They are described by the 3d  Chern-Simons Quantum Electrodynamics (CSQED) with $N$ charge one Dirac fermions and Chern-Simons level $k$. In the 't Hooft limit, namely $N,|k|\to \infty$ with the 't Hooft coupling $\lambda={4\pi N\over k}$ fixed, they give rise to a one-parameter family of CFTs labelled by $\lambda$. We find that while the free energy coefficient in this case is independent of $\lambda$ in the leading large $N$ order, the subleading piece is a nontrivial function $g(\lambda)$ as in,
\ie 
f_{{\rm CSQED}_{N,k} }=& -\frac{3 \zeta (3)}{4 \pi } N + g(\lambda)
+ \cO(N^{-1})\,,
\fe
which has the following limiting behaviors,
\ie 
\lim_{\lambda=\infty} g(\lambda)=-0.21211735\,,\quad \lim_{\lambda=0} g(\lambda)=0\,,
\fe
corresponding to QED with vanishing Chern-Simons level $k=0$ and the free limit (with infinity Chern-Simons level) respectively. The former agrees with the previous result in \cite{Kaul:2008xw}.

In addition to the free energy coefficient $f$ in \eqref{freeEcoeff}, we also develop diagrammatic methods in the large $N$ vector models to compute the one-point functions of other primary operators as in \eqref{thermal1PF}. Because of the unbroken global symmetry, only singlet operators with respect to the symmetry acquire nontrivial one-point functions. In the large $N$ vector models, a family of such operators are known as the single-trace higher-spin currents $J^s_{\m_1\dots\m_s} $ which takes the following schematic form in the scalar vector model \cite{Lang:1992zw,Maldacena:2011jn,Maldacena:2012sf,Alday:2015ota,Skvortsov:2015pea,Giombi:2016hkj},
\ie 
J^s_{\m_1\dots\m_s} \sim \phi^i \pa^{\m_1}\dots \pa^{\m_s} \phi^i -{\rm traces}\,,
\fe
with even spin $s$. These operators are conserved currents in the $N=\infty$ limit but develop anomalous dimensions at the order ${1\over N}$. Their thermal one-point function coefficients $b_s$ in the $N=\infty$ limit of the scalar $O(N)$ model has been previously computed in \cite{Iliesiu:2018fao} using the inversion formula. Here we provide a direct diagrammatic derivation of $b_s$ for both bosonic and fermionic large $N$ vector models in the leading large $N$ limit. We also describe explicitly the procedure to obtain their ${1\over N}$ corrections where the inversion formula in \cite{Iliesiu:2018fao} does not obviously apply.

It was shown recently in
\cite{Kang:2022orq} that by incorporating global symmetry twists one can obtain a refined version of the asymptotic density of states \eqref{asympDENSITY} labelled by the representations of the symmetry group. 
More specifically, let us consider a CFT with continuous global symmetry $G$ on ${\rm S}^1_\B\times {\rm S}_R^{d-1}$ as around \eqref{Zsphere}. We focus on
a $U(1)$ subgroup of the full symmetry generated by a charge $Q$ and turn on a $U(1)$ holonomy $g=e^{i\mu}$ along the ${\rm S}^1$ factor. The resulting symmetry twisted partition function is
 \ie 
 Z_{{\rm S}^1_\B \times {\rm S}^{d-1}_R}(g)\equiv & {\rm Tr\,}_{\cH_{{\rm S}^{d-1}}} e^{-{\B\over R} \Delta + i \mu Q} 
=
\sum_{{\bf r}\in {\rm Irrep}(G)}  {\chi_{\bf r}(g)\over \dim {\bf r}}\int d\Delta\, \rho(\Delta,{\bf r})e^{-{\B\over R} \Delta}  \,,
\label{Ztwist}
\fe 
where the sum is over irreducible representations $\bf r$ of $G$, $\chi_{\bf r}$ is the corresponding group character and $\rho(\Delta,{\bf r})$ is the symmetry-resolved density of states of the CFT. In the high temperature limit, symmetry twist leads to addition contributions in the thermal free energy,
\ie 
 Z_{{\rm S}^1_\B \times {\rm S}^{d-1}_R}(g) \xrightarrow[]{{R\over \B}\to \infty} Z_{{\rm S}^1_\B \times \mR^{d-1}} (g)= e^{
- {V_{d-1}\over \B^{d-1}}  \left( f+{b \over 4} \Tr(\mu^2) +\dots \right) }\,,
\label{freetwisted}
\fe 
and the leading effect at small $\m$ valued in the Lie algebra $\mf{g}$ is captured by the Wilson coefficient $b$ above as introduced in \cite{Kang:2022orq} where $\Tr(\cdot)$ denotes the Killing form on $\mf{g}$.\footnote{Note that our normalization of $b$ differs from that of \cite{Kang:2022orq} by the volume a unit sphere (i.e. $b_{\rm there}=S_{d-1} b_{\rm here}$).} In particular $b$ is argued to be positive based on the relation to the domain wall tension for the $g$-twisted sector \cite{Kang:2022orq}. As was derived in \cite{Kang:2022orq}, the twisted free energy is related to the one without symmetry twist by
\ie 
{Z_{{\rm S}^1_\B \times \mR^{d-1}} (g)
\over 
Z_{{\rm S}^1_\B \times \mR^{d-1}} (1)
}= \left({4 \pi \B^{d-1}\over b V_{d-1}}\right)^{{\dim G}\over 2}
\sum_{{\bf r} \in {\rm Irrep}(G)} \dim {\bf r}\, \chi_{\bf r}(g)e^{-{1\over b V_{d-1}}c_2({\bf r})  \B^{d-1} +\dots }
\label{Zratio}
\fe 
where $c_2({\bf r})$ is the second Casimir for $\bf r$ and we have omitted terms suppressed by higher powers of $\B$.
Using \eqref{Ztwist} and \eqref{freetwisted} together with \eqref{Zratio}, we can determine the asymptotic behavior of the refined density of states $\rho(\Delta,{\bf r})$ normalized by \eqref{asympDENSITY},
\ie 
\log\left[ {\rho(\Delta,{\bf r})\over  
  \rho(\Delta) (\dim {\bf r})^2}\right] \xrightarrow[]{\Delta\gg 1}  
-{c_2({\bf r})\over b} \left( (1-d)f S_{d-1}\over \Delta \right)^{d-1\over d} +\dots \,,
\label{srdos}
\fe 
where we have omitted terms that are representation independent in this limit. In \cite{Kang:2022orq}, the $b$ coefficient was determined for free theories and holographic CFTs from the gravity dual. 
Here by studying the large $N$ vector models at finite temperature with nonzero chemical potential, we provide the first results for the $b$ coefficient directly from interacting CFTs, as summarized below for the $O(N)$ global symmetry of the $O(N)$ CFT, the $O(2N)$ symmetry of the the GN CFT and the $O(N)$ symmetry of 
the NJL CFT (with $N$ Dirac fermions),
\ie 
b_{O(N)}=  &\,  \frac{2\arccos\left(\frac{3}{2}\right)^2+2\left(5+2 \operatorname{arccosh} \left(\frac{3}{2}\right)\right)
   \operatorname{arcsinh}\frac12}{\sqrt{5} \pi }  -{0.42424\over N} \,,
\\
b_{{\rm GN}_N}=&\,{2\log 2\over \pi } -{0.1338\over N} \,,
\\
b_{{\rm NJL}_N}=&\,{2\log 2\over \pi } -{0.2676\over N}\,,
\fe 
where we have included the leading ${1\over N}$ correction. We have also determined the coefficient $b$ for the non-unitary cousins of the GN CFT (see \eqref{GNpb}), as well as for the CSQED to leading order in $N$ (see \eqref{QEDb}).

Our normalization is such that for a free complex scalar or a Dirac fermion with $U(1)$ symmetry, the corresponding $b$ coefficient is\footnote{
Note that for the free scalar the divergent zero mode contribution is removed (e.g. by an orbifold).}
\ie 
b_{\rm scalar}={3\over \pi}\,,\quad b_{\rm fermion}={2\log 2\over \pi }\,.
\fe

We now discuss a number of potential applications and future directions of our work.
As the Monte-Carlo simulations are being pushed to study vector models at higher $N$, it would be interesting to see how our results on the ${1\over N}$ corrections compare to these numerical investigations. It would also be interesting to see if these large $N$ systems can be realized in experiments (e.g. by stacking and twisting existing finite $N$ setups), which would measure the ${1\over N}$ corrections we have found. Another interesting direction involves treating the parameter $N$ as a coupling constant and finding the non-perturbative corrections to the large $N$ results using the techniques developed by Lipatov \cite{Lipatov:1976ny,Suslov:1999cg} and renormalons \cite{Zichichi:1979upj}. 
As special instances of the AdS/CFT correspondence \cite{Maldacena:1997re,Gubser:1998bc,Witten:1998qj},
the $d=3$ large $N$ vector models are expected to be dual to certain versions of Vasiliev's
higher-spin gravity on ${\rm AdS}_4$ \cite{Vasiliev:1990en,Vasiliev:1992av,Vasiliev:1995dn,Klebanov:2002ja,Giombi:2009ek} (see also 
\cite{Aharony:2020omh,Aharony:2022feg} for recent works). The critical large $N$ vector model at finite temperature should then correspond to a black brane solution of Vasiliev's higher-spin gravity. Such a solution is known in the linearized limit, but not yet in the full non-linear theory due to the high complexity of the bulk interactions in the higher-spin gauge theory \cite{Didenko:2021vdb,Didenko:2021vui}. Furthermore, it is not known how to reproduce even the leading $N$ free energy from a bulk action for the higher-spin gravity.\footnote{Note that the collective field method of \cite{
Das:2003vw,deMelloKoch:2010wdf,Jevicki:2011ss,Jevicki:2014mfa,deMelloKoch:2014mos,deMelloKoch:2018ivk} and the recent development in \cite{Aharony:2020omh,Aharony:2022feg} rely on gauge-fixing to the ${\rm AdS}_4$ background in the bulk which cannot describe the (deconfined) thermal state.} 
Nonetheless in light of the successful matching between one-loop effects (i.e. order $\cO(N^0)$ contributions to the free energy) for the CFT on $S^3$ and the higher-spin gravity on ${\rm AdS}_4$ in \cite{Giombi:2013fka}, it would be interesting to compare the one-loop contributions from higher-spin gauge fields on a putative thermal geometry and the finite temperature ${1\over N}$ corrections we find on the CFT side, which 
will provide a nontrivial test on the bulk solution and offer some insights on the structure of the bulk solution. 
Finally, it would be interesting to compare our predictions for the asymptotic density of states obtained from the thermal effective action (which includes refinement by global symmetries) to explicit operator counting.

The rest of the paper is organized in the following way. In  Section~\ref{sec:ONall},  we consider the critical scalar $O(N)$ model at finite temperature in detail. We compute the two-point function of the $\sigma$ field at finite temperature, and then numerically obtain the first correction to the free energy coefficient in the large $N$ expansion and comment on subtleties in the numerical procedure. We also include the dependence of the free energy on a particular $U(1)$ chemical potential. In addition, we show explicitly how the same result follows from the computation of the one-point function of the stress-energy tensor by explicitly summing Feynman diagrams.  We then generalize the diagrammatic  analysis to higher-spin currents of even spin $s>2$. In  Section~\ref{sec3}, we apply the developed techniques to study fermionic vector models including the Gross-Neveu model and the Nambu-Lasino-Jonas model. In Section~\ref{sec:QED}, we investigate thermal one-point functions in the $d=3$ Quantum Electrodynamics with a large number of flavors and Chern-Simons level $k$ in the 't Hooft limit. In the Appendices, we provide technical details for intermediate steps in the main text.

\section{Scalar \texorpdfstring{$O(N)$}{ON} Vector Model}
\label{sec:ONall}

\subsection{Review of the Critical \texorpdfstring{$O(N)$}{TEXT} Model and Large $N$ Expansion} 
Here we review some basics facts about the scalar $O(N)$ vector model. We start in general spacetime dimensions and later specialize to $d=3$.
The action of the model is defined as
\ie 
S=\int d^d x \left(
{1\over 2} \pa_\m \phi^i \pa^\m \phi^i + \frac12 m_0^2 \phi^i\phi^i+{\lambda_0\over 4 } (\phi^i\phi^i)^2
\right) \,,
\fe
where $m_0$ and $\lambda_0$ are the bare mass and coupling and $\phi^i$ belongs to the vector representation of the $O(N)$ global symmetry group. Famously, the model admits a large $N$ limit where physical observables such as the correlation functions of local operators can be extracted using a saddle-point approximation (see \cite{Moshe:2003xn} for an extensive review). To see that, we apply the Hubbard–Stratonovich (HS) trick using the auxiliary $\sigma$ field, also known as the HS field, whose original integration contour is along the imaginary axis \cite{Moshe:2003xn}. Then up to a constant shift, we have
\ie 
S=\int d^d x \left(
{1\over 2} \pa_\m \phi^i \pa^\m \phi^i + {1\over 2} \sigma \phi^i\phi^i-{  \sigma^2\over 4\lambda_0}    +\frac{1}{2} r_0 \sigma
\right) \,,
\fe
with $r_0 = \frac{m_0^2}{\lambda_0}$. Note that $\sigma$ field can be thought as a mass of the field $\phi^i$. Integrating out the fields $\phi^i$, we arrive at the following effective Lagrangian  for $\sigma$,
\begin{gather}
   \cF(\sigma) =   \frac12 N \log \det \left(-\Box + \sigma \right) - \frac{\sigma^2}{4\lambda_0}    +\frac{1}{2} r_0\sigma  \,.
    \label{eq:leadN}
\end{gather}
By demanding $\lambda_0 = \frac{\lambda^{t}_0}{N}, r_0 = N r_0^{t}$ with $\lambda_0^t,r_0^t$ held fixed in the large $N$ limit, we see that $\frac{1}{N}$ plays the role of an effective Planck constant which gives rise to a semi-classical expansion. For that purpose we first need to solve the equation  of motion for $\sigma$,
\begin{gather}
   G_\phi(x,x)  = \frac{\sigma(x)}{\lambda^{t}_0} - r_0^{t}\,,
\end{gather}
where the LHS coincide with the coincident limit of the propagator for $\phi^i$,
\ie 
  \left(-\Box + \sigma\right)G_\phi(x,y) =\delta(x-y)\,.
  \label{Gphi}
\fe
Assuming the homogeneous ansatz $\sigma(x) = \sigma$, the saddle-point equation for $\sigma$ is given by,
\begin{gather} \label{eq:gapeq}
    \int \frac{d^d k}{(2\pi)^d} \frac{1}{k^2 + \sigma} = \frac{\sigma}{\lambda^{t}_0} - r_0^{t}\,,
\end{gather}
which is also known as the gap equation since $\sigma$ determines the mass gap for the scalar fields.
The LHS of the above equation  is divergent for $d\geq 2$ and needs to be regularized. It is straightforward to check that the divergences could be absorbed in the redefinition of the bare coupling constants $r^{t}_0$. To study the CFT, we bring the system to a critical point by further fine-tuning this parameter. Note that for $d<4$ near the free  Gaussian point, the composite operators $\phi^i\phi^i$ and $(\phi^i\phi^i)^2$  are both relevant and thus we should fine-tune them at the same time to reach the free fixed point. Perturbed away from the Gaussian point by the $(\phi^i\phi^i)^2$ operator, the theory  can flow to an interacting critical point, where $(\phi^i\phi^i)^2$ becomes irrelevant. In this case, we  only need to fine-tune the mass $r_0$. 
This critical point describes the second-order phase transition between the ordered and the disordered phases of the $O(N)$ model.
The actual value of the parameter $r_0$ where the system becomes critical is scheme-dependent and we will label the regularization scheme by $\cR$. The phase transition occurs at $\sigma=0$ (for $\sigma < 0$ there are tachyonic instabilities) since $\sigma$ controls the correlation length. It is thus more convenient to parameterize the bare coupling $r_0^{t}$ as
\begin{gather}
r_0^{t} = r^{t} - \int_\cR \frac{d^d k}{(2\pi)^d} \frac{1}{k^2}, \label{eq:req}
\end{gather}
where the RHS is computed with the chosen regularization scheme in the UV. Note that this equation has a solution only for $d>2$ since at $d=2$ the integral acquires an IR divergence. This is a manifestation of the Coleman-Mermin-Wagner theorem stating that there are no Goldstone modes in $d\leq 2$. Substituting the value of $r_0^{t}$ \eqref{eq:req} in the gap equation \eqref{eq:gapeq}, we obtain
\begin{gather}
     \int_\cR \frac{d^d k}{(2\pi)^d} \left( \frac{1}{k^2 + \sigma} - \frac{1}{k^2}\right) = \frac{\sigma}{\lambda^{t}_0} -r^{t}\,. 
\end{gather}
Now by tuning $r^t$ we can bring the system to the critical
phase transition. Indeed, with $\sigma > 0$, we obtain the following equation
\begin{gather} \label{eq:lamdaeq}
     \frac{r^{t}}{\sigma}-\frac{1}{\lambda^{t}_0} = \int_\cR \frac{d^d k}{(2\pi)^d} \frac{1}{k^2(k^2 + \sigma)}\,.
\end{gather}
Expanding the RHS of the above equation for small $\sigma$ at $2<d<4$ gives \cite{Moshe:2003xn}
\begin{gather}
  r^{t} - \frac{\sigma}{\lambda_0^{t}} \sim K_d\sigma^{\frac{d}{2}-1} + I_\cR \sigma +\ldots,
  \label{smallsigmaexp}
\end{gather}
where $K_d$ is a scheme-independent constant
\ie 
K_d\equiv -{\Gamma(1-{d\over 2})\over  (4\pi)^{d\over 2}}\,,
\label{Kdexp}
\fe
and $I_\cR$ is a constant that depends on the regularization scheme $\cR$. The terms that are subleading when $\sigma \to 0$ are omitted in \eqref{smallsigmaexp}. Near the critical point, we have the scaling behavior $r^{t} \sim  K_d \sigma^{\frac{d}{2}-1}$. If we further set $\lambda_0^{t} = -\frac{1}{I_\cR}$ then the corrections due to the finite-size effects would become suppressed. This value should be carefully chosen when adopting a numerical lattice regularization scheme (see Appendix~\ref{sec:latticereg} for further comments). For the analytical treatment, we find the dimensional regularization to be the most convenient.  For $2<d<4$, this gives,
\ie  
 \int\limits_{\rm  dimreg} \frac{d^d k}{(2\pi)^d} \frac{1}{k^2(k^2 + \sigma)}  = K_d \sigma^{\frac{d}{2}-2}\,, \quad \frac{1}{\lambda_0^{t}} = 0\,, \quad r^{t} =K_d\sigma^{\frac{d}{2}-1}\,,
 \label{crittune}
\fe 
so to reach the critical point we set $r^t =  0$ (equivalently $r_0^{t}=0$) and to cancel subleading corrections we pick $\lambda_0^t=\infty$ in this scheme. Consequently, the corresponding CFT on flat space is governed by the following action 
\ie 
S=\int d^d x \left(
{1\over 2} \pa_\m \phi^i \pa^\m \phi^i + {1\over 2} \sigma \phi^i\phi^i 
\right)\,,
\label{CFTaction}
\fe
and similarly the effective Lagrangian for $\sigma$ at the critical point is
\ie 
\cF(\sigma) =  \frac12 N \log \det \left(-\Box + \sigma \right)  \,.
\label{CFTsigmaaction}
\fe

To determine observables at the subleading order in the ${1\over N}$ expansion, 
we also need the propagator for $\sigma$, which follows from the large $N$ Lagrangian \eqref{eq:leadN},
\begin{gather} \label{eq:sigmaprop}
    G^{-1}_\sigma(x,y) = -{\dfrac{N}{2}} \frac{1}{\lambda^{t}_0} \delta^{(d)}(x-y) -  {\dfrac{N}{2}}G^2_\phi(x,y).
\end{gather}
The negative signs reflect the fact that the original integration contour for $\sigma$ runs parallel to the imaginary axis \cite{Moshe:2003xn}. 
At the critical point in the dimensional regularization scheme (see \eqref{crittune}), 
we have from \eqref{Gphi},
\begin{gather}
    G^{-1}_\sigma(p) =
    \frac{  \pi   \Gamma \left(\frac{d}{2}\right)}{(4\pi)^{d\over 2}\sin \left(\frac{\pi  d}{2}\right)\Gamma (d-1)}
   N p^{d-4} \,,\quad G_\sigma(x,0) =\frac{\sin \left(\frac{\pi  d}{2}\right) \Gamma (d-1)}{\pi ^5  \Gamma \left(\frac{d}{2}-2\right) \Gamma \left(\frac{d}{2}\right)} \frac{1}{N x^4}\,,
\end{gather}
so that the scaling dimension of the composite operator $\sigma \propto (\phi^i)^2 $ is $\Delta_\sigma=2$ and independent of the spacetime dimension (thus different from the mean-field value for general $d$, a hallmark of interacting CFT).

Here we are mostly interested in the $d=3$ CFT on the background ${\rm S}^1_\B \times \mR^2$ which is flat but has nontrivial topology. This is described by the same action \eqref{CFTaction} with periodic boundary conditions for the fields along ${\rm S}^1_\B$. The gap equation (saddle-point equation) for $\sigma$ now takes the following form,\footnote{In the following we will simply write $\vec k$ as $k$ for the two dimensional momentum to simplify the notation.}
\ie 
0=  {1\over \B}  \sum_{n\in \mZ}  \int { d^2 \vec k  \over |\vec k|^2+\omega_n^2+\sigma} \label{eq:thermalgap}
\fe
where $\omega_n\equiv {2\pi n\over \B}$ are the bosonic Matsubara frequencies and the UV divergence is regulated in the same way as in flat space.
As we review in the next subsection, the saddle-point solution $\sigma=\sigma_*$ is no longer zero for finite $\B$ and corresponds to a finite mass-gap generated by quantum effects. Equivalently, it determines the one-point function of $\sigma\sim (\phi^i)^2$ in the large $N$ limit,
\ie 
\la (\phi^i)^2_{\rm ren}\ra_\B 
=\sqrt{\frac{\pi^5  \Gamma \left(\frac{d}{2}-2\right) \Gamma \left(\frac{d}{2}\right)}{\sin \left(\frac{\pi  d}{2}\right) \Gamma (d-1)}}{ N^{1\over 2}\sigma_*\over \B^2}\,,
\fe
where $(\phi^i)^2_{\rm ren}$ is the renormalized primary operator with the normalized two-point function on flat space
\ie 
\la (\phi^i)^2_{\rm ren}(x)(\phi^i)^2_{\rm ren}(y) \ra={1\over (x-y)^4}\,.
\fe

\subsection{Large $N$ Free Energy and the Subleading Correction} \label{sec:ONfreeE}

Here we derive the free energy density of the scalar $O(N)$ CFT on ${\rm S}^1_\B\times \mR^2$,
\ie 
F(\B)\equiv - {1\over \B V_2} \left. \log Z_{{\rm S}^1_\B\times \mR^2} \right
|_{\rm non-extensive}\,,
\label{Fdensity}
\fe
where $V_2$ regulates the infinite spatial volume and we focus on the non-extensive (in ${\rm S}^1_\B$) part of $\log Z_{{\rm S}^1_\B\times \mR^2}$ which is free from counterterm ambiguities. The free energy density has the following large $N$ expansion,
\ie 
F(\B)=F_0(\B)+F_{-1}(\B)+\cO(N^{-1})\,,
\label{Fexp}
\fe
where $F_0$ denotes the leading contribution of order $\cO(N)$ and $F_{-1}$ is the first subleading contribution at $\cO(1)$.

We start with the effective Lagrangian for $\sigma$,
\ie 
\cF(\sigma)= {N\over 2 } \tr \log (-\Box+\sigma) 
={N\over 2 \B} 
\sum_{n\in \mZ} \int {d^2  p \over (2\pi)^2} \log \left (
 p^2
+
\omega_n^2+ \sigma\right ), 
\label{Fsigmastart}
\fe
where $\omega_n\equiv {2\pi n\over \B}$. In the large $N$ limit, the leading free energy density is determined by the saddle-point $\sigma=\sigma^*$,
\ie 
F_0(\B)= \cF_{\rm ren}(\sigma_*) \,,
\fe
of the renormalized Lagrangian $\cF_{\rm ren}(\sigma)$ for $\cF(\sigma)$,
\ie 
\cF_{\rm ren}(\sigma)={N\over 2}
\left (
{1\over \B}\sum_{n\in \mZ} \int_\cR {d^2  p \over (2\pi)^2} \log \left (
 p^2
+
\omega_n^2+ \sigma\right )
+r_0^t \sigma - 
 \int_\cR {d^2 p d \omega\over (2\pi)^3} \log \left(p^2+\omega^2 \right) 
\right)\,,
\label{Fren}
\fe
where we work in a general regularization scheme $\cR$, keep only terms that are non-vanishing for constant $\sigma$ (sufficient for the leading large $N$ analysis) and restore $r_0^t$ (which vanishes in the dimensional regularization). The last term in the above expression is the cosmological constant counterterm.

A key point is that the sum-integral in 
\eqref{Fren} has the following large energy momentum expansion,
\ie 
\cF_{\rm ren}(\sigma)={N\over 2}\left( \sigma   \left( 
{1\over \B}\sum_{n\in \mZ} \int_\cR {d^2 p \over (2\pi)^2} {1\over p^2+\omega_n^2} + r_0^t
\right)+\dots
\right)\,,
\fe
where the omitted terms are absolutely convergent thus independent of the regularization scheme $\cR$. Furthermore, the first term in the bracket is also absolutely convergent which follows from \eqref{eq:gapeq} at $\sigma=0$ (zero temperature). Consequently $\cF_{\rm ren}(\sigma)$ is scheme-independent.

To facilitate the analytic calculations, we perform dimensional regularization on the two-dimensional momentum integral and zeta function regularization on the Matsubara sum. Implementing this procedure to \eqref{Fsigmastart}, we first obtain
\ie 
\cF(\sigma)
\xrightarrow{\rm dimreg} -{N\over 8\pi \B}\sum_{n\in \mZ} (\omega_n^2+\sigma) \left(\log ( \omega_n^2+\sigma)-1\right) \,.
\label{Fsigmadimreg}
\fe 
The following identity (which holds in zeta function regularization) will be useful \cite{Laine:2016hma},
\ie 
\sum_{n\in \mZ} \log \left (\sigma  +\left({2\pi n \over \B }+\A \right)^2 \right) =\B \sqrt{\sigma} + \log (1-e^{-\B \sqrt{\sigma} +i \B \A })
+\log (1-e^{-\B \sqrt{\sigma}-i\B\A})\,.
\label{sumlogid}
\fe
Integrating in $\sigma$, we then obtain the renormalized Lagrangian\footnote{Note that in a general regularization scheme we will need to subtract off the extensive piece of the free energy to obtain the scheme-independent part of the free energy density  $F$ (see \eqref{Fdensity}). Such an extensive contribution to $F$ in a CFT on ${\rm S}^1_\B\times \mR^2$ can only come from the cosmological constant ($\B$ independent). For our choice of regularization scheme here, such a constant is absent, since the last two terms in \eqref{Fren} vanish in this scheme.}
\ie 
\cF_{\rm ren}(\sigma)
=&-{N\over 4\pi \B^{3}} 
\left(
{1 \over 3} \left(\B^{2}\sigma\right)^{\frac32}+2 \B\sqrt{\sigma} {\rm Li}_2(e^{-\B \sqrt{\sigma}})
+2 {\rm Li}_3(e^{-\B \sqrt{\sigma}})
\right)\, .
\label{Frenleading}
\fe
The integration constant is fixed by requiring,
\ie 
\cF_{\rm ren}(0)=-\frac{N }{2 \pi  \beta ^3}\zeta (3)\,, 
\label{freeFren0th}
\fe
which follows  directly from \eqref{Fsigmadimreg} by zeta function regularization.
 
Extremizing $\cF_{\rm ren}(\sigma)$ with respect to $\sigma$, we find the thermal gap equation (the regularized version of \eqref{eq:thermalgap})
\ie 
0=\B \sqrt{\sigma} + 2\log (1-e^{-\B \sqrt{\sigma}   })\,,
\label{reggapeqn}
\fe
with the following solution 
\ie 
\sigma_*=\frac{\Delta^2}{\beta^2}\,,\quad \Delta=\log \left(\frac{1}{2} \left(\sqrt{5}+3\right)\right)\,, \ \ \ 
\cF_{\rm ren}(\sigma_*)=
-{2\zeta(3)\over 5\pi}{N\over \B^3}\,.
\label{ONsaddle}
\fe
Consequently, the leading free energy density for the $O(N)$ CFT is 
\ie 
F_0(\B)=-{2\zeta(3)\over 5\pi}{N\over \B^3}\,.
\label{leadingONF}
\fe

Let us briefly comment on a subtle point in this calculation. The CFT is Lorentz invariant at zero temperature (i.e. $\B=\infty$). It is a priori not obvious whether the short-distance regulator implemented above (a combination of zeta function and dimensional regularizations) respects the Lorentz symmetry at scales much smaller than $\B$. In fact, as we will see momentarily, this regulator is not compatible with the Lorentz symmetry at the subleading order in the ${1\over N}$ expansion. Nonetheless, as explained after \eqref{Fren}, in the leading large $N$ limit, the scheme independence is \textit{enhanced}. 

As a consistency check, we also evaluate \eqref{Fren} with a manifestly Lorentz-invariant regulator \cite{Chubukov:1993aau} by performing sum-integral with the following Lorentz-invariant hard cutoff
\ie 
p^2+\omega_n^2<\Lambda^2\,,
\label{Linvreg}
\fe
and obtain the same results numerically as in \eqref{ONsaddle}.\footnote{We emphasize that $r_0^t\neq 0$ at the critical point in this regularization scheme.}

We are mostly interested in the first subleading correction in ${1\over N}$ denoted by $F_{-1}(\B)$ in \eqref{Fexp}. According to the semi-classical expansion in the large $N$ vector model, this is computed by the log-determinant of the second variation of the effective Lagrangian \eqref{eq:leadN} with respect to $\sigma$ (equivalently from the inverse propagator $G_\sigma^{-1}$), subject to regularization and renormalization that we explain below,
\begin{gather}
   F_{-1}(\beta) \leftarrow \frac{1}{2 \beta}\sum_{n\in \mZ} \int \frac{d^2 p}{(2\pi)^2}  \log |G^{-1}_{\sigma}(\omega_n,p)|\,.
    \label{eq:1ncorr}
\end{gather}
Explicitly, the inverse propagator on ${\rm S}^1_\B\times \mR^2$ in momentum space (from \eqref{eq:sigmaprop} with $\lambda_0^t=\infty$) is proportional to the self-energy $\Pi_\B$ (where a factor of $N$ is extracted for convenience),
\ie 
   G_\sigma^{-1}(\omega_n,p)=-N\Pi_\B (\omega_n,p)\,,
   \fe 
and $\Pi_\B$  takes the following form
   \ie 
   \Pi_\B (\Omega,p) = & \frac{1}{2\beta}  \sum_{n\in \mZ} \int \frac{d^2 q}{(2\pi)^2} \frac{1}{(p-q)^2 + (\Omega - \omega_n)^2 + \sigma_*}  \frac{1}{q^2 +  \omega_n^2 + \sigma_*}\,,
   \label{Pidef}
\fe 
with $\sigma_*$ as in \eqref{ONsaddle}. We take the integral over the spatial momentum using the Feynman parametrization and evaluate the sum over the discrete frequencies  exactly. The detailed computation is presented in  Appendix~\ref{app:ONmodel}, and the resulting expression is
\ie 
\label{propsigma}
 \Pi_{\beta}(\Omega,p) = -{1\over 16\pi}\int \limits_{0}^{1} \dfrac{dx}{\sqrt{\sigma_*+\left(\Omega^{2}+p^2\right)(x-x^{2})}} \dfrac{\sinh{\left(\beta \sqrt{\sigma_*+\left(\Omega^{2}+p^2\right)(x-x^{2})} \right) }}{\cos{(\beta \Omega x)-\cosh{\left(\beta \sqrt{\sigma_*+\left(\Omega^{2}+p^2\right)(x-x^{2})} \right)}}}\,.
\fe 
To evaluate \eqref{eq:1ncorr} using \eqref{propsigma} requires a further regularization. Indeed, we see that the $\sigma$ self-energy at large momentum behaves as (see Appendix~\ref{app:largepON} for details),
\ie 
   {}&\Pi_\B(\Omega,p) =\dfrac{1}{16 P}+ \frac{1}{P^2} \left (- \frac{  \sqrt{\sigma_*}}{4\pi} + \int \frac{d^2 q}{(2\pi)^2} \frac{n(\epsilon_q)}{\epsilon_q} \right)  +\dfrac{2\Omega^{2}-p^{2}}{P^{6}}\dfrac{1-6\gamma}{6\pi}\sigma_*^{\frac32} +\dots \,,
  \label{sigmapexp}
\fe 
where
\ie  
P^2 \equiv  \Omega^2+ p^2\,, \quad \epsilon_q \equiv  \sqrt{ q^2 + \sigma_*}\,, \quad n(\epsilon) \equiv \frac{1}{e^{\beta \epsilon} - 1}\,. 
   \fe  
The first term in this expansion \eqref{sigmapexp} corresponds to the flat spacetime propagator and should be cancelled to obtained the renormalized free energy. We thus arrive at the following expression for the renormalized free energy at the subleading order in ${1\over N}$,
\ie 
    F_{-1}(\beta) =\,& \frac{1}{2 \beta}\sum_n \int \frac{d^2 p}{(2\pi)^2} \log |G^{-1}_{\sigma}(\omega_n,p)| - \frac{1}{2}\int \frac{d^3 P}{(2\pi)^3} \log|G^{-1}_\sigma(P) |   \\
    =\,&  \frac{1}{2 \beta}\sum_n \int \frac{d^2 p}{(2\pi)^2} \log\left[16 \sqrt{p^2 + \omega_n^2} \Pi_{\beta}(\omega_n,p)\right] + \frac{1}{4\pi \beta^3} \zeta(3)\,,
    \label{subleadingF}
\fe
where the last term in the second line follows from \eqref{freeFren0th} (see Appendix~\ref{app:ONmodel} for further details).

The second term in \eqref{sigmapexp} is proportional to the gap equation \eqref{reggapeqn} and  vanishes for the special point $\sigma=\sigma_*$ from \eqref{ONsaddle} relevant for describing the finite temperature CFT. If this term were nonzero,  we would encounter a quadratic divergence in the subleading free energy \eqref{subleadingF} that is linear in the temperature, which would be inconsistent with the structure of UV divergences in local quantum field theory.

The next term in the large momentum expansion \eqref{sigmapexp} no longer vanishes at the critical point $\sigma=\sigma_*$, and consequently
\begin{gather}
 \log\left[ 16 \sqrt{p^2 + \omega_n^2} \Pi_{\beta}(\omega_n,p)\right] = 16\frac{1-6\gamma}{6\pi}\Delta^3 \frac{2\omega_n^2 - p^2}{(\omega_n^2+p^2)^{5/2}}  + \mathcal{O}\left(\frac{1}{P^4}\right)\,.
\end{gather}
The further subleading terms in the above expansion converge absolutely and we don't need to worry about them. On the other hand, the first term on the RHS is dangerous since it contributes an apparent logarithmic divergence to \eqref{subleadingF} and could lead to regularization ambiguities. To see this explicitly, we consider the following sum-integral,
\begin{gather}
I=\dfrac{1}{\beta}\sum_n \int_{\mathcal{M}^n} \frac{d^2 p}{(2\pi)^2}\left( \frac{2\omega_n^2 - p^2}{(\omega_n^2 + p^2)^\frac{5}{2}} \right)
 =\dfrac{1}{\beta}\sum_n {1\over (2\pi)^2}\int_{\partial\mathcal{M}^n_{}} \frac{p_i dS^i}{(p^2+\omega_n^2)^\frac32}\,,
 \label{Itoy}
\end{gather}
where $\cM^n$ for fixed $n$ is a region in the full two-dimensional momentum space and plays the role of a regulator for the momentum integral. In the second equality, we have used the fact that the first integrand is a total derivative. 
If we choose $\cM^n$ to be $\mR^2$ independent of $n$,
naively this sum-integral is regulated to zero. If we instead implement the regularization procedure used earlier (dimensional regularization for the momentum integral and zeta function regularization for the Matsubara sum), we also find the answer is zero. However, as we have already emphasized below \eqref{leadingONF}, in general one has to be careful with choosing the correct regulator that is compatible with the symmetry that is preserved under renormalization. Here we assume Lorentz symmetry (at zero temperature) in the renormalization procedure and consequently should use a Lorentz-invariant UV regulator, such as \eqref{Linvreg}, which corresponds to choosing $\cM^n=\{p|p^2+\omega_n^2<\Lambda^2\}$ in \eqref{Itoy}.\footnote{We thank Subir Sachdev for discussions on this point.} This gives 
\begin{gather}
 I = \dfrac{1}{3\pi^2}+ \mathcal{O}\left(\frac{1}{\beta \Lambda }\right)\,,
\end{gather}
which determines the contribution from this term to the free energy \eqref{subleadingF}. 
In the next section, we will evaluate
\eqref{subleadingF} numerically taking into account this regulator subtlety.

\subsection{Numerical Calculation}\label{sec:numerical}

Now we are in position to compute numerically the subleading correction to the free energy for large $N$ critical vector model, by evaluating the sum-integral in \eqref{subleadingF}. Because of the oscillatory behavior of $\Pi_\B$ (see \eqref{propsigma}), this sum-integral needs to be handled with care. Below we explain the strategy for this numerical evaluation which we implement in \texttt{mathematica}.

We start by rescaling all momentum and energy in \eqref{subleadingF} by $\B$,
\ie 
p \to {p \over \B}\,,\quad \omega_n \to {\omega_n \over \B}\,, 
\fe
such that the $\beta$ dependence is completely factored out and given by ${1\over \B^3}$. The goal is to determine the dimensionless coefficient in $F_{-1}(\B)$.

We first need an efficient way to evaluate the $\sigma$ self-energy $\Pi_\beta(\omega_n,p)$ to high precision. The expression \eqref{propsigma}  for $\Pi_\beta(\omega_n,p)$ is a highly oscillating integral at large momentum, thus converges very slowly. Therefore, to optimize the numerical evaluation of $\Pi_\beta(\omega_n,p)$, we introduce two spherical shells of radius $\Lambda_{0}$ and $\Lambda$, with $\Lambda_{0}<\Lambda$. Where $\Lambda$ is the UV cut-off and $\Lambda_0$ is chosen in such a way that the difference between the actual value of the self-energy \eqref{propsigma} and its large momentum expansion \eqref{PibetalargeP} would be negligible.

For $\omega_n,p$ within the spherical shell $p^2+\omega_n^2<\Lambda_{0}^2$, we evaluate directly the sum-integral of $\log{\left[16\sqrt{\omega_{n}^{2}+p^2}\Pi_\beta(\omega_{n},p)\right]}$ in \eqref{subleadingF} for $\left|\omega_{n}\right| < \Lambda_0$ and  $p< \sqrt{\Lambda^{2}_{0}-\omega_{n}^2}$ to high precision. We find that the optimal range for $\Lambda_{0}$ is $250 \leq \Lambda_{0} \leq 450$, where the difference between the direct computation of \eqref{propsigma} and its large $P=\sqrt{\omega_n^2 +p^2}$ expansion \eqref{PibetalargeP} is of the order $ \sim 10^{-14}$.

Between the two spherical shells $\Lambda^2>\omega_n^2+p^2>\Lambda^2_{0}$, we can use the large $P$ expansion of the $\sigma$ self-energy to compute the sum-integral to the desired precision, by keeping all terms in \eqref{PibetalargeP}. Finally, the relativistic UV cutoff $\Lambda$ is taken to infinity numerically. Note that this resolves the regularization ambiguity which we have discussed near the end of the last section.

Implementing the procedure outlined above in \texttt{Mathematica} and taking  different  values of $250 \leq  \Lambda_{0} \leq  450$ and $10^{4} \leq\Lambda \leq 10^{8}$,  we  find that  the sum-integral in \eqref{subleadingF} evaluates to 
 \begin{gather}
     \frac{1}{2 \beta}\sum_n \int \frac{d^2 p}{(2\pi)^2} \log\left[16 \sqrt{p^2 + \omega_n^2} \Pi_{\beta}(\omega_n,p)\right] =- \frac{0.03166112}{\beta^3}\,,
 \end{gather}
and consequently the subleading correction to the free energy of the $O(N)$ vector model reads,
 \ie 
      F_{O(N),-1}(\beta) =  \frac{0.06399553}{\beta^3}\,.
      \label{FONsubfinal}
      \fe 
\subsection{Turning on Chemical Potential}\label{sec:ONchem}
\begin{figure}
    \centering
    \includegraphics{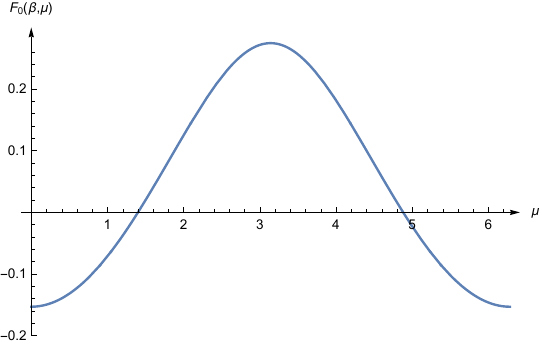}
    \caption{The leading large $N$ free energy $F_0(\B,\m)$ of the $O(N)$ CFT as a function of the imaginary chemical potential $\mu$ (here $\B=1$).}
    \label{fig:ChemBoxR0}
\end{figure}

\begin{figure}[!htb]
    \centering
    \includegraphics{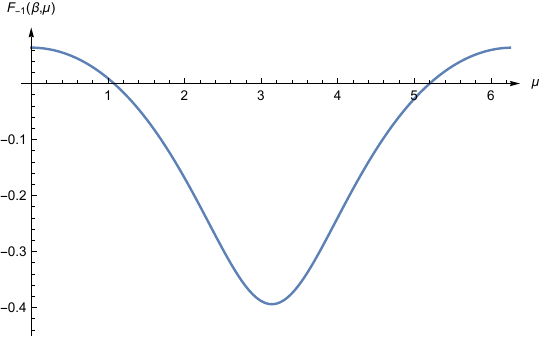}
    \caption{The subleading correction $F_{-1}(\B,\m)$ to the free energy  of the $O(N)$ CFT as a function of the imaginary chemical potential $\mu$  (here $\B=1$).}
    \label{fig:chempotfig}
\end{figure}

One natural extension of our analysis of the finite temperature free energy of the $O(N)$ CFT in the previous section is to include a background for its global symmetry. Here for simplicity, we consider $N$ even and take the $U(1)$ subgroup of $O(N)$ with commutant $SU(N/2)$ (such that the complex scalars $\phi_j+i\phi_{j+N/2}$ for $j=1,\dots,N/2$ have charge 1 under this $U(1)$).
We
turn on an imaginary chemical potential parameterized by $\mu\in [0,2\pi)$  for this $U(1)$ subgroup, via a background gauge field with nonzero temporal component $A_0 = {\mu i\over \B}$. Consequently the energy spectrum for the scalar fields are shifted to $\beta \tilde{\omega}_{n}=2\pi n+ \mu$.
The effective Lagrangian in this case is 
\ie 
\cF(\tilde{\sigma},\mu)=\dfrac{N}{2  \B 
     }\sum_{n}\int \dfrac{d^{2}p}{(2\pi)^{2}}\log{\left(  \tilde{\omega}^{2}_{n}+p^{2}+\tilde{\sigma}\right)}   \,,
\fe 
and its renormalized version reads
\ie 
\cF_{\rm ren}(\tilde{\sigma},\mu)  =&-\dfrac{N}{4\pi \B^3}\left(\dfrac{\B^3\tilde{\sigma}^{3/2}}{3}+\operatorname{Li}_{3}(e^{-\B\sqrt{ \tilde{\sigma}} + i\mu})+\operatorname{Li}_{3}(e^{-\B\sqrt{\tilde{\sigma}} - i\mu})
\right.
\\
&\left.+\B\sqrt{\tilde{\sigma}} \operatorname{Li}_{2}(e^{-\B\sqrt{\tilde{\sigma}}+i\mu}) + \B\sqrt{\tilde{\sigma}} \operatorname{Li}_{2}(e^{-\B\sqrt{\tilde{\sigma}}-i\mu})\right)\,.
\fe 
When $\m=0$, this reduces to the case considered previously (see \eqref{Frenleading}).
The gap equation is given by,
\begin{gather}
\B\sqrt{\tilde{\sigma}}+\log\left(1 - e^{-\B\sqrt{\tilde{\sigma}} + i \mu}\right) +\log\left(1-e^{-\B\sqrt{\tilde{\sigma}} - i \mu}\right) =0\,.
    \label{chpotoN}
\end{gather}
We denote its solution by $\tilde\sigma=\tilde\sigma_*(\m)$ whose explicit form is given below,
\begin{gather}
      \tilde\sigma_*(\mu) ={1\over \B^2}\operatorname{arccosh}^2\left[\frac{1}{2}+\cos\mu\right]\,.
      \label{ONsaddlesolwcp}
\end{gather}
The free energy in the leading large $N$ limit follows from 
\ie 
F_0(\B,\m)=\cF_{\rm ren}(\tilde\sigma_*(\m),\m)\,,
\fe
which is plotted in Figure~\ref{fig:ChemBoxR0}.\footnote{As a consequence of the charge conjugation symmetry in the $O(N)$ CFT, the special points $\m=0$ and $\m=\pi$ are extrema of the free energy as evident from Figure~\ref{fig:ChemBoxR0} and Figure~\ref{fig:chempotfig}.}

Moving onto the subleading order in the ${1\over N}$ expansion, we will determine the free energy $F_{-1}(\B,\m)$ by performing the sum-integral as in \eqref{Fren} which now involves the $\sigma$ free-energy $\Pi_\beta^{\mu}$ that depends on $\mu$ (see \eqref{sec:eq:chempolbos} for its explicit integral representation). 
Implementing the numerical procedure explained in Section~\ref{sec:numerical}, we compute the $F_{-1}(\B,\m)$ as a function of the chemical potential  and the result is presented in Figure~\ref{fig:chempotfig}.

As discussed in the introduction (around \eqref{freetwisted}), in the expansion of the thermal free energy $F(\beta,\mu)$ at small chemical potential\footnote{Here we have used the relative normalization $\Tr ={1\over 2} \tr_{\rm fund}$ between  the Killing form and the trace in the fundamental presentation for $O(N)$.}   
\begin{gather}
    F(\beta,\mu) = F(\beta) + \frac14  {N\over 2}\mu^2 b +\dots \,,
\end{gather}
the positive coefficient $b$ governs the asymptotics density of (high energy) states refined by the $O(N)$ global symmetry of the CFT. From our explicit result for $F(\B,\m)$, we find that up to the first subleading order in the ${1\over N}$ expansion,
\ie 
   {N\over 2} b = N b_0 + b_{-1} + \mathcal{O}(N^{-1})\,, 
   \fe 
   with\footnote{The error for $b_{-1}$ is a standard error that is estimated using function \texttt{NonLinearModel} in \texttt{Mathematica}.}
   \ie 
    b_0 = \frac{\arccos\left(\frac{3}{2}\right)^2+\left(5+2 \operatorname{arccosh} \left(\frac{3}{2}\right)\right)
   \operatorname{arcsinh}\frac12}{\sqrt{5} \pi }\,,\quad b_{-1} = -0.21212 \pm 0.00005\,.
\fe

\subsection{One-point Function of the Stress-energy Tensor} \label{sec:ONT1pf}

As explained in the introduction, the thermal free energy of the CFT contains the same information as the one-point function of the stress-energy tensor at finite temperature (see around \eqref{T1PF}).  
So far we have analyzed the free energy of the $O(N)$ CFT using the semi-classical expansion in the large $N$ limit. It would be useful to understand how the same physical quantity can be computed using the standard Feynman diagrams for this Lagrangian field theory. In particular, this second diagramatic approach will have immediate generalizations 
to determining the thermal one-point functions \eqref{thermal1PF} of more general operators in the CFT, which we will discuss in the next section.

\begin{figure}[!htb]
    \centering
    \begin{tikzpicture}[scale=0.5]
     \draw[very thick] (0,-4) circle (0.5);
     \node[below] at (0,-4.5) {$T_{00}$};
     \fill (0,-4.5) circle [radius=3pt];
     \draw(2.5,-4) circle (0.5);
     \node[below] at (2.5,-4.5) {$T_{00}$};
     \fill (2.5,-4.5) circle [radius=2pt];
     \node[below] at (3.5,-3.5) {$+$};
      \draw(4.5,-4) circle (0.5);
      \draw(4.5,-3) circle (0.5);
     \node[below] at (4.5,-4.5) {$T_{00}$};
     \fill (4.5,-4.5) circle [radius=2pt];
     \node[below] at (5.5,-3.5) {$+$};
    \draw(6.5,-4) circle (0.5);
     \draw(6.5,-3) circle (0.5);
      \draw(6.5,-2) circle (0.5);
     \node[below] at (6.5,-4.5) {$T_{00}$};
     \fill (6.5,-4.5) circle [radius=2pt];
     \node[below] at (7.5,-3.5) {$+$};
     \node[below] at (8.5,-4) {$\ldots$};
      \node[below] at (9.5,-3.5) {$+$};
     \draw (10.5,0) circle (0.5);
     \draw (10.5,-1) circle (0.5);
     \draw (9.5,-1) circle (0.5);
     \draw (8.5,-1) circle (0.5);
     \draw (10.5,-2) circle (0.5);
    \draw (11.5,-2) circle (0.5);
     \draw (10.5,-3) circle (0.5);
     \draw (10.5,-4) circle (0.5);
     \draw (10.5,-2) circle (0.5);
     \draw (10.5,-3) circle (0.5);
     \node[below] at (10.5,-4.5) {$T_{00}$};
     \fill (10.5,-4.5) circle [radius=2pt];
     \node[below] at (11.5,-3.5) {$+$};
     \node[below] at (12.5,-4) {$\ldots$};
     \node[below] at (1.5,-3.5) {$=$};
        \end{tikzpicture}
    \caption{Examples of Feynman diagrams (on the right) containing $\phi^i$ loops that  contribute to the thermal one-point function  $\la T_{00}\ra_{0,\B}$ at the leading order in ${1\over N}$. They resum to the diagram on the left with the \textit{large $N$ exact} propagator for $\phi^i$ represented by a thick line. }
    \label{fig:1setsdiag}
\end{figure}
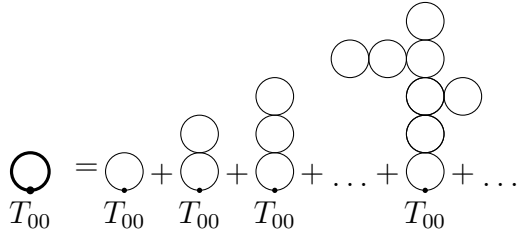

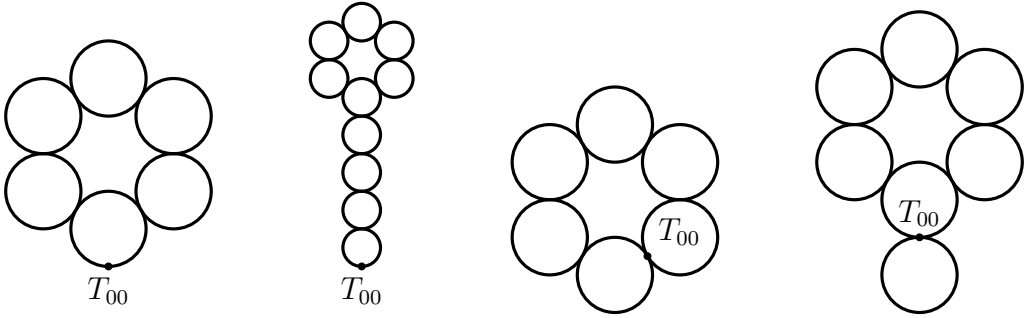
\begin{figure}[!htb]
    \centering
    \begin{tikzpicture}[scale=0.5]
     \draw[very thick] (0,0) circle (1);
     \draw[very thick] (2*sin 60,2-2*cos 60) circle (1);
     \draw[very thick] (2*sin 120,2-2*cos 120) circle (1);
     \draw[very thick] (2*sin 180,2-2*cos 180) circle (1);
     \draw[very thick] (2*sin 240,2-2*cos 240) circle (1);
     \draw[very thick] (2*sin 300,2-2*cos 300) circle (1);
     \node[below] at (0,-1) {$T_{00}$};
     \fill (0,-1) circle [radius=3pt];
    \end{tikzpicture}
    \hspace{1cm}
    \begin{tikzpicture}[scale=0.5]
     \draw[very thick] (0,0) circle (0.5);
     \draw[very thick] (sin 60,1-cos 60) circle (0.5);
     \draw[very thick] (sin 120,1-cos 120) circle (0.5);
     \draw[very thick] (sin 180,1-cos 180) circle (0.5);
     \draw[very thick] (sin 240,1-cos 240) circle (0.5);
     \draw[very thick] (sin 300,1-cos 300) circle (0.5);
     \draw[very thick] (0,-1) circle (0.5);
          \draw[very thick] (0,-2) circle (0.5);
         \draw[very thick] (0,-3) circle (0.5);
              \draw[very thick] (0,-4) circle (0.5);
     \node[below] at (0,-4.5) {$T_{00}$};
     \fill (0,-4.5) circle [radius=3pt];
    \end{tikzpicture}
    \hspace{1cm}
    \begin{tikzpicture}[scale=0.5]
     \draw[very thick] (0,0) circle (1);
     \draw[very thick] (2*sin 60,2-2*cos 60) circle (1);
     \draw[very thick] (2*sin 120,2-2*cos 120) circle (1);
     \draw[very thick] (2*sin 180,2-2*cos 180) circle (1);
     \draw[very thick] (2*sin 240,2-2*cos 240) circle (1);
     \draw[very thick] (2*sin 300,2-2*cos 300) circle (1);
     \node[above right] at (sin 60,1-cos 60) {$T_{00}$};
     \fill (sin 60,1-cos 60)  circle [radius=3pt];
    \end{tikzpicture}
    \hspace{1cm}
    \begin{tikzpicture}[scale=0.5]
     \draw[very thick] (0,0) circle (1);
       \draw[very thick] (0,-2) circle (1);
     \draw[very thick] (2*sin 60,2-2*cos 60) circle (1);
     \draw[very thick] (2*sin 120,2-2*cos 120) circle (1);
     \draw[very thick] (2*sin 180,2-2*cos 180) circle (1);
     \draw[very thick] (2*sin 240,2-2*cos 240) circle (1);
     \draw[very thick] (2*sin 300,2-2*cos 300) circle (1);
     \node[above] at (0,-1) {$T_{00}$};
     \fill (0,-1)  circle [radius=3pt];
    \end{tikzpicture}
    \caption{Examples for the four families of Feynman diagrams that can contribute to the thermal one-point function  $\la T_{00}\ra_{-1,\B}$ at the first subleading order in ${1\over N}$. All internal lines are large $N$ exact propagators for $\phi^i$. }
    \label{fig:3setsdiag}
\end{figure}

From the general structure of the thermal one-point function \eqref{T1PF}, it suffices to focus on the temporal component $T_{00}$ of the stress tensor $T_{\m\n}$, which in the $O(N)$ CFT is given by, up to the improvement total derivative terms,\footnote{Note that the improvement terms are conformal descendants which have vanishing thermal one-point functions and consequently do not affect the subsequent evaluation of $\la T_{00}\ra_{\B}$.}
\begin{gather}
    T_{00}= -\left(-\frac{1}{2} \left(\partial_0 \phi\right)^2 +  \frac{1}{2} \left(\partial_\alpha \phi\right)^2 + \frac{\lambda_0}{4} \left(\phi^i \phi^i\right)^2 +{\rm improvement~terms}\right)\,.
    \label{eq:too}
\end{gather}
The explicit relation between the thermal one-point function and the free energy in the $d=3$ CFT is
\ie 
\la T_{00} \ra_\B=2 F(\B)\,, \label{eq:TFrel}
\fe
and similar to the RHS analyzed previously \eqref{Fexp}, the LHS admits an ${1\over N}$ expansion in the $O(N)$ CFT 
\ie 
\la T_{00} \ra_\B=\la T_{00} \ra_{0,\B}+\la T_{00} \ra_{-1,\B}+\cO(N^{-1})
\label{Texp}
\fe 
which can be seen explicitly by reorganizing the Feynman diagrams at large $N$.

The leading large $N$ result $\la T_{00} \ra_{0,\B}$ for the thermal one-point function only receives contributions from the expectation values of the first two terms in \eqref{eq:too} and comes from resummation of the Feynman diagrams with increasing number of $\phi^i$ loops (see Figure~\ref{fig:1setsdiag}), which induces a non-zero value for the sigma field. Therefore we have, from the one-loop diagram with the large $N$ exact $\phi^i$ propagator,
\ie 
\braket{T_{00}}_{0,\B}=\frac{N}{2 \beta  }  \sum_n \int \frac{d^2 p}{(2\pi)^2} \frac{\omega_n^2 - p^2}{\omega_n^2 + p^2 + \sigma_*} = -  \frac{N}{2} \frac{\partial}{\partial \beta} \left[ \sum_n \int \frac{d^2 p}{(2\pi)^2} \log\left(\omega_n^2 + p^2 + \sigma_*\right) \right],
\fe 
which confirms the relation \eqref{eq:TFrel} upon comparing with \eqref{Fsigmastart} with $\sigma=\sigma^*$ as in \eqref{ONsaddle}.

We now compute the $\frac{1}{N}$ correction to the thermal one-point function $\la T_{00}\ra_\B$. The Feynman diagrams that contribute at this order come in four families which are given in Figure~\ref{fig:3setsdiag} with increasing number of $\phi^i$ loops with \textit{exact} propagators, independent of the temperature (or more general spacetime background).  Each family of these diagrams can be resummed in the IR using the $\sigma$ self-energy (inverse propagator) $\Pi_{\beta}(\omega_n,p)$ (which captures the contributions from a chain of $\phi^i$ loops) given explicitly in \eqref{propsigma} and together they determine $\la T_{00} \ra_{-1,\B}$,
\ie 
\la T_{00} \ra_{-1,\B} =
\la T_{00}^{(1)} \ra_{-1,\B} +\la T_{00}^{(2)}  \ra_{-1,\B} +\la T_{00}^{(3)}  \ra_{-1,\B} +\la T_{00}^{(4)} \ra_{-1,\B}\,.
\label{subleadingTsplit}
\fe 
Explicitly the first two contributions could be written as
\begin{gather}
    \braket{T^{(1)}_{00}}_{-1,\B} = -\dfrac{1}{2 \B^2}  \sum_{n,m} \int \frac{d^2 p}{(2\pi)^2} \frac{1}{\Pi_{\beta}(\Omega_n,p)}\left[ \int \frac{d^2 q}{(2\pi)^2} \frac{1}{\left(\Omega_n - \omega_m\right)^2 + (p-q)^2 + \sigma_{*}} \frac{\omega_m^2 - q^2}{\left(\omega_m^2 + q^2 + \sigma_{*}\right)^2} \right] \,,\notag\\
    \braket{T^{(2)}_{00}}_{-1,\B} = \dfrac{1}{4\B^3} \sum_{n,m,l} \int \frac{d^2 p}{(2\pi)^2} \frac{1}{\Pi_{\beta}(\Omega_n,p)}\left[ \int \frac{d^2 q}{(2\pi)^2} \frac{1}{(\Omega_n-\omega_m)^2 + (p-q)^2 + \sigma_{*}} \frac{1}{\left(\omega_m^2 + q^2 + \sigma_{*}\right)^2} \right]  \notag\\
    \times \frac{1}{\Pi_{\beta}(0,0)} \left[ \int \frac{d^2 k}{(2\pi)^2}  \frac{\omega_l^2 - k^2}{\left(\omega_l^2 + k^2 + \sigma_{*}\right)^2} \right]\,, \label{eq:12partSTE}
\end{gather}
The second equation above can be further simplified to
\ie 
    &\braket{T_{00}^{(2)}}_{-1,\B} = \dfrac{1}{2\B} \sum_n \int \frac{d^2 p}{(2\pi)^2} \frac{1}{\Pi_{\beta}(\Omega_n,p)} \\
    \times&{1\over \B}\sum_m \int \frac{d^2 q}{(2\pi)^2} \frac{1}{(\Omega_n-\omega_m)^2 + (p-q)^2 + \sigma_{*}} \frac{-\sigma_{*} + \frac{2}{\pi} \log\left[2\sinh\left(\frac{\beta \sqrt{\sigma_{*}}}{2}\right)\right] \tanh\left[\frac{\beta \sqrt{\sigma_{*}}}{2}\right]}{\left(\omega_m^2 + q^2 + \sigma_{*}\right)^2}\
    \\
    =& \dfrac{1}{2\B^2} \sum_{m,n} \int \frac{d^2 p}{(2\pi)^2} \frac{1}{\Pi_{\beta}(\Omega_n,p)}  
      \int \frac{d^2 q}{(2\pi)^2} \frac{1}{(\Omega_n-\omega_m)^2 + (p-q)^2 + \sigma_{*}} \frac{-\sigma_{*} }{\left(\omega_m^2 + q^2 + \sigma_{*}\right)^2}\,,
    \fe 
    where we have used in the last equality that 
the saddle $\sigma=\sigma_*$ in \eqref{ONsaddle} satisfies
\ie 
 \frac{2}{\pi} \log\left[2\sinh\left(\frac{\beta \sqrt{\sigma_{*}}}{2}\right)\right] \tanh\left[\frac{\beta \sqrt{\sigma_{*}}}{2}\right]=0\,.
\fe
The contributions from the first two terms in \eqref{subleadingTsplit} then combine to,
\ie 
&\braket{T^{(1)}_{00}}_{-1,\B} +\braket{T^{(2)}_{00}}_{-1,\B}= -\dfrac{1}{2\B^2}  \sum_{n,m} \int \frac{d^2 p}{(2\pi)^2} \frac{1}{\Pi_{\beta}(\Omega_n,p)}
\\
&\times \left[\int \frac{d^2 q}{(2\pi)^2} \frac{1}{\left(\Omega_n - \omega_m\right)^2 + (p-q)^2 + \sigma_{*}} \frac{\omega_m^2 - q^2+\sigma_{*}}{\left(\omega_m^2 + q^2 + \sigma_{*}\right)^2} \right]\,.
\fe 
Similarly, the third family of diagrams in Figure~\ref{fig:3setsdiag} resum to
\ie
    \braket{T^{(3)}_{00}}_{-1,\B} = -{1\over 2\B} \sum_n \int \frac{d^2 p }{(2\pi)^2} \frac{1}{\Pi_{\beta}(\Omega_n,p) } \times  \Pi_{\beta}(\Omega_n,p) \,.
\fe 
Finally, the fourth family of diagrams in Figure~\ref{fig:3setsdiag} vanish, 
\ie 
 \braket{T^{(4)}_{00}}_{-1,\B} 
    =&\dfrac{G_{\rm reg}(x,x)}{4 \B^2\Pi_\B(0,0)} \sum_{n,m} \int \frac{d^2 p }{(2\pi)^2} \frac{1}{\Pi_{\beta}(\Omega_n,p) } 
    \\
    \times &\left[\int \frac{d^2 q}{(2\pi)^2} \frac{1}{\left(\Omega_n - \omega_m\right)^2 + (p-q)^2 + \sigma_{*}} \frac{1}{\left(\omega_m^2 + q^2 + \sigma_{*}\right)^2} \right]  =0\,,
\fe 
where $G_{\rm reg}(x,x)$ is the regularized two-point function of $\phi^i$ at coinciding points which vanishes as a consequence of the gap equation  \eqref{eq:thermalgap}. 
Combing the above all together, we arrive at the full subleading ${1\over N}$ correction to the stress tensor one-point function,
\ie 
    \braket{T_{00}}_{-1,\B} = &   -\dfrac{1}{4 \B^2}  \sum_{n,m} \int \frac{d^2 p}{(2\pi)^2} \frac{1}{\Pi_{\beta}(\Omega_n,p)} 
    \\
    \times &\left( \int \frac{d^2 q}{(2\pi)^2} \frac{1}{\left(\Omega_n - \omega_m\right)^2 + (p-q)^2 + \sigma_{*}} \frac{3 \omega_m^2 + 3\sigma_{*} - q^2}{\left(\omega_m^2 + q^2 + \sigma_{*}\right)^2} \right)\,.
    \label{energfinal}
\fe 
Let us compare the above expression  with that of the free energy. We have from Section~\ref{sec:ONfreeE}
\begin{gather}
    F_{-1}(\beta)= \dfrac{1}{2 \B}  \sum_n \int \frac{d^2 p}{(2\pi)^2} \left.\log \Pi_{\beta}(\Omega_n,p) \right|_{\rm ren},
    \label{Fmoneren}
\end{gather}
where the explicitly renormalized expression is given in \eqref{subleadingF}.
From \eqref{T1PF} and \eqref{Fmoneren}, we have the simple following relation\footnote{Note that the subtraction term in \eqref{subleadingF} disappears after the $\B$ derivative.}
\ie 
\la T_{00} \ra_{-1,\B}=- {\pa \over \pa \B} \left(\B F_{-1}(\B) \right)
=-\dfrac{1}{2}\sum_n \int \frac{d^2 p}{(2\pi)^2} \frac{1}{\Pi_{\beta}(\Omega_n,p)} \frac{\partial}{\partial \beta} \Pi_{\beta}(\Omega_n,p)\,,
\fe
where the last term clearly coincides with \eqref{energfinal} after using \eqref{Pidef}.

\subsection{One-point Function of Higher-Spin Currents}\label{sec:1pfcurrentsON}
Using the ideas from the previous section, we can compute the $\frac{1}{N}$ corrections to the thermal one-point function of more general local operators, which, as emphasized in the introduction, are the basic building blocks for the most general correlation functions in the finite temperature CFT. For illustration, we focus on the so-called
higher-spin current operators in the $O(N)$ CFT, which are $O(N)$ invariant primary operators $J^s_{\mu_1\ldots \mu_s}$ transforming in the rank $s$ symmetric traceless representations of the Lorentz group for positive even integer $s$ (also refers to as the spin $s$ representations) \cite{Lang:1992zw,Maldacena:2011jn,Maldacena:2012sf,Alday:2015ota,Skvortsov:2015pea,Giombi:2016hkj}. These operators are constructed out of $O(N)$ singlet bilinears of the scalar field $\phi^i$ together with $s$ derivatives,
\ie 
 J^s_{\mu_1 \ldots \mu_s} = \left (\dfrac{1}{N}\phi_i \partial_{\mu_1} \ldots \partial_{\mu_s} \phi_i - \text{traces} \right) +{\rm descendants}\,,
 \label{Jsdef}
 \fe 
up to total derivatives which are fixed by demanding the LHS to be a primary operator. 
They generalize the stress-energy tensor which appears at $s=2$ (as well as the singlet scalar operator $(\phi^i)^2_{\rm ren}$ at $s=0$).

Using the residual symmetries of the thermal background, the one-point function of the higher-spin current of rank $s$ is constrained to take the following form
\ie 
    \braket{J^s_{\mu_1\ldots \mu_s}}_\B = \frac{b_s}{\beta^{\Delta_s}} \left(e_{\mu_1} e_{\mu_2} \ldots e_{\mu_s} - \text{traces} \right), \quad e_{\mu} = \left(1,0,0\right)\,,
    \label{Js1pf}
\fe 
with the scaling dimension
\ie 
\Delta_s=s+1+\C_s\,,
\label{anomDelta}
\fe
where $\C_s$ is the anomalous dimension which is suppressed by ${1\over N}$ in the large $N$ limit and thus the $O(N)$ CFT is said to have a slightly broken higher-spin symmetry \cite{Maldacena:2012sf}.
It is convenient to introduce an auxiliary complex null polarization vector $\xi^\mu = \xi \left(1,i,0\right)$ and write\footnote{Note that these are bare operators (also \eqref{Jsdef}) which will be subjected to renormalizations (suppressed by ${1\over N}$). See around \eqref{HS1pfexp} and \eqref{renJ1pf}.}
\ie 
J^s_\xi \equiv   J^s_{\mu_1\ldots \mu_s} \xi^{\mu_1}\ldots \xi^{\mu_s}=\dfrac{1}{N} \phi_i \left(\xi^\mu\partial_{\mu}\right)^s \phi_i+{\rm descendants}\,,
\label{Jxidef}
\fe 
whose one-point function is
\ie 
\braket{J_\xi^s}_\B =  \frac{b_s \xi^s}{\beta^{\Delta_s}}\,,
\label{Jxi1pf}
\fe
where we have used \eqref{Js1pf}.
Therefore to determine the one-point function of the higher-spin current $J^s_{\mu_1\ldots \mu_s}$, it suffices to focus on the component  $J_\xi^s$.

In the large $N$ limit, the thermal one-point function $\la J^s_\xi \ra_
\B$ admits a $1\over N$ expansion similar to \eqref{Texp} for the stress tensor. At the leading order, the same family of diagrams as in Figure~\ref{fig:1setsdiag} contribute and give
\begin{gather}
   \la  J^s_\xi \ra_{0,\B} = {1\over \B}\sum_n \int \frac{d^2 p}{(2\pi)^2} \frac{\left(ip_\mu \xi^\mu\right)^s}{\omega_n^2 + p^2 + \sigma_{*}}\,,
    \label{lo1}
\end{gather}
where $\sigma_*$ is given in \eqref{ONsaddle}.
As usual, the UV divergences in the sum-integral above is regulated by subtracting the  flat space expression in the given regularization scheme.  To simplify the computation, we introduce the  generating function
\ie 
\label{genFG}
   G_{\beta,0}(\xi) \equiv  \sum^\infty_{s=0} \frac{\la J^s_\xi\ra_{0,\B}}{s!} =  {1 \over \B} \sum_{n=-\infty}^{\infty} \int \frac{d^2 p}{(2\pi)^2} \frac{e^{i p_\mu \xi^\mu}}{\omega_n^2 + p^2 + \sigma_{*}}\,,
\fe 
where the regularization of the last term above is implicit.
This generating function is just a specialization of the coordinate-space propagator for $\phi^i$, which for general any spacetime separation $r_\mu = \left(r_0, \vec{r}\right)$ is given by
\ie 
G_{\B,0}(r)=\frac{1}{4\pi} \sum_{m=-\infty}^{\infty} \frac{e^{-\sqrt{\sigma_{*}}\sqrt{\left(r_0 + \beta m \right)^2 + \vec{r}^2}}}{\sqrt{\left(r_0 + \beta m \right)^2 + \vec{r}^2}}\,,
\fe
which follows from \eqref{genFG} by Poisson resummation. Physically, each summand above represents the contribution from a worldline instanton of mass $\sqrt{\sigma_*}={\Delta\over \B}$ propagating along a geodesic on the thermal background that connects the two $\phi^i$ insertions.
The $m=0$ term coincides with the flat space-time propagator in the short distance limit. After subtracting that, we obtain the regularized version of $G_{\B,0}(\xi)$, which gives
\begin{gather}
    \sum^\infty_{s=0} \frac{ \la J^s_\xi\ra_{0,\B}}{s!}= \dfrac{1}{4\pi}\sum_{m\neq 0} \frac{e^{-\sqrt{\sigma_{*}} \sqrt{m\beta (m\beta-2\xi)}}}{\sqrt{m \beta(m \beta-2\xi)}}\,.
\end{gather}
The one-point function coefficient $b_s$ in the leading large $N$ limit (which we will denote as $b_s^0$) of the spin $s$ current is obtained immediately by the small $\xi$ expansion of the RHS above,
\ie 
b_s^0  ={1\over 4\pi}\sum_{n=0}^s\frac{2^{n-s+1} (2s-n)!}{ n!   (s-n)! } \beta^n \sigma_*^\frac{n}{2} \text{Li}_{s+1-n}\left(e^{-\beta\sqrt{\sigma_*}}\right)\,,
\label{bsfinal}
\fe
which agrees with the results obtained in \cite{Iliesiu:2018fao} from a different method (up to an overall normalization factor that only depends on $s$).

Alternatively, we can directly regulate the integrals \eqref{lo1} as follows (setting $\B=1$ below and using \eqref{ONsaddle}),
\ie 
\la J^{s}_\xi \ra_{0,\B}=\,&\lim\limits_{\epsilon \to 0}\sum_{n=-\infty}^{\infty}\int \frac{d^2 p}{(2\pi)^2} \frac{\left(ip_\mu \xi^\mu\right)^s}{\omega_n^2 + p^2 + \Delta^2}e^{-\epsilon (\omega^2_{n}+p^{2}+\Delta^2)^2} \,.
\fe 
By standard contour manipulation, the above can be rewritten as 
\ie 
\la J^{s}_\xi \ra_{0,\B}=- {\xi^s\over 2} \lim_{\ep\to 0}\int \dfrac{d^{2}p}{(2\pi)^2}\oint_\C\frac{dz}{2\pi i} \dfrac{(p_{1}-z)^s}{z^2-\epsilon_{p}^2}e^{-\epsilon(z^2-\epsilon_{p}^2)^2}\left(\coth \left(\frac{z}{2} \right)-1\right)\,,
\label{lo2}
\fe 
where $\epsilon_p \equiv \sqrt{p^2+\Delta^2}$ and $\C$ is a contour that circles the simple poles at $z=2\pi  in$ with $n\in \mZ$ in the counter-clockwise direction.\footnote{The $-1$ shift in \eqref{lo2} does not affect the integral by Lorentz invariance (writing $z=ip_0$ where $p_0$ is  the energy component of the three-momentum).}
Deforming the contour to infinity, we obtain by Cauchy theorem that,
\ie
b_s^0= \int \dfrac{d^{2}p}{(2\pi)^{2}}\dfrac{(p_{1}-\epsilon_p)^s}{\epsilon_{p}} \frac{1}{e^{\epsilon_p}-1}\,,
\label{lo4}
\fe
for positive even $s$, which again reproduces the same result \eqref{bsfinal}, after implementing a straightforward change of variables and using the following integral identity
\ie 
\text{Li}_{s}\left(e^{-\Delta}\right)={1\over  \Gamma(s)}\int_\Delta^\infty dz  {(z-\Delta)^{s-1}\over e^z-1}\,.
\fe
\subsubsection*{The first subleading correction to the one-point function of higher-spin currents}

The subleading correction to the one-point function of higher-spin currents can be computed using Feynman diagrams as in the case of the stress-energy tensor in Section~\ref{sec:ONT1pf}.

The dynamical data in the one-point function \eqref{Jxi1pf} of the higher-spin currents, namely the overall constant coefficient $b_s$ and the anomalous dimension $\C_s$ (see \eqref{anomDelta}) have the following ${1\over N}$ expansions,
\ie 
    b_s = b^0_s + \frac{b^1_s}{N} + \ldots, \quad \C_s = \frac{\Delta^1_s}{N}+\ldots\,.
\fe 
The ${1\over N}$ corrections $\Delta^1_s$ to the scaling dimensions of these operators have been computed in \cite{Lang:1992zw,Skvortsov:2015pea,Giombi:2016hkj,Hikida:2016cla},
\ie 
\Delta_s^1=\frac{16 (s-2)}{3 \pi ^2  (2 s-1)}\,.
\label{anomalousON}
\fe 
Consequently, the first correction in the $\frac{1}{N}$ expansion of the one-point function of bare operator \eqref{Jxi1pf} takes the following form,
\begin{gather}
    \la J^s_{\xi,{\rm bare}} \ra_{\B} =  \left(\frac{b^0_s }{\beta^{s+1}} + \frac{1}{N} \frac{1}{\beta^{s+1}} \left[b^1_s + b^0_s \Delta^1_s \log \left(\beta\Lambda\right)\right]\right) \xi^s +\dots \,,
    \label{HS1pfexp}
\end{gather}
where $\Lambda$ is a UV regulator and the UV divergence could be treated by proper renormalization of the operator. In the above equation, we have reintroduced the subscript on the higher-spin current $J^s_{\xi,{\rm bare}}$ to emphasize this is the bare operator.
For the stress-energy tensor, its scaling dimension does not receive corrections ($\Delta_2^1=0$) and therefore the logarithmic correction is absent, which simplifies the computation.\footnote{Note that the logarithmic term in \eqref{HS1pfexp} indicates a scheme dependence for the one-point function coefficient $b_s^1$ at the subleading order in ${1\over N}$. This scheme dependence can be removed by normalizing with respect to the two-point function of $J^s_\xi$ in the flat spacetime computed in the same scheme.}

Following the strategy of the previous section, we start by incorporating the first $\frac{1}{N}$ correction to the $\phi_i$ propagator as follows,
\begin{gather}
    G^{-1}_\phi(\omega_n,p) = \omega_n^2 + p^2 + \sigma_* + \frac{1}{N} \Sigma(\omega_n,p)\,, 
\end{gather}
where $\Sigma(\omega_n,p)$ is a self-energy part for the matter field $\phi_i$ where $\sigma_*$ is the large $N$ saddle-point for $\sigma$ (as given in \eqref{ONsaddle} in the dimensional regularization). Note that the self-energy $\Sigma(\omega_n,p)$ also contains the information about $1\over N$ corrections to the one-point function of $\sigma \sim (\phi_i)^2$ (i.e. correction to $\sigma_*$).

\begin{figure}[!htb]
    \centering
    \begin{tikzpicture}[scale=0.5]
    \node[below] at (-3,0.75) {$\braket{J^s_{\xi}}_\B=$};
    \draw[very thick] (0,0) circle (1);
     \node[below] at (0,-1) {$J^{s}_{\xi,{\rm bare}}$};
     \fill (0,-1) circle [radius=3pt];
     \node[below] at (1.5,0.75) {$+$};
     \draw[very thick] (3,0) circle (1);
     \draw[very thick,dotted] (2,0)--(4,0);
     \node[below] at (3,-1) {$J^{s}_{\xi,{\rm bare}}$};
     \fill (3,-1) circle [radius=3pt];
     \node[below] at (4.5,0.75) {$+$};
     \draw[very thick] (6,0) circle (1);
     \node[below] at (6,-1) {$J^{s}_{\xi,{\rm bare}}$};
     \fill (6,-1) circle [radius=3pt];
     \draw[very thick] (8.5,0) circle (1);
     \draw[very thick,dotted] (8.5,-1)--(8.5,1);
     \draw[dotted,very thick] (7,0)--(7.5,0);
      \node[below] at (10,0.75) {$+$};
      \draw[very thick] (11.5,0) circle (1);
     \node[below] at (11.5,-1) {$J^{s}_{\xi,{\rm bare}}$};
     \node at (11.5,1) {$\times$};
     \fill (11.5,-1) circle [radius=3pt];
      \node[below] at (13,0.75) {$+$};
     \draw[very thick] (14.5,0) circle (1);
     \node[below] at (14.5,-1) {$J^{s}_{\xi,{\rm bare}}$};
     \fill (14.5,-1) circle [radius=3pt];
     \draw[very thick] (16.5,0) circle (0.5);
      \node at (17,0) {$\times$};
     \draw[dotted,very thick] (15.5,0)--(16,0);
     \node[below] at (17.75,0.75) {$+$};
     \draw[very thick] (19.25,0) circle (1);
         \node at  (19.25,-1) {$\times$};
     \node[below] at (19.25,-1) {$\delta J^{s}_{\xi}$};
    \end{tikzpicture}
    \caption{The Feynman diagrams that contribute to the one-point function of the higher-spin current $\braket{J^s_\xi}_\B$ at the leading and the first subleading order in ${1\over N}$. The dotted lines correspond to the propagator of $\sigma$ field and crosses to the counter-term vertices that are computed in the flat spacetime. }
    \label{fig:HSrenorm}
\end{figure}
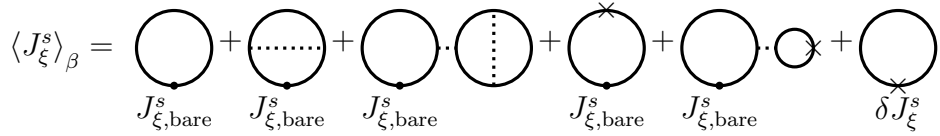

The bare self-energy $\Sigma(\omega_n,p)$ is computed from the second and third diagrams in the Figure~\ref{fig:HSrenorm}, which produce
\ie 
    \Sigma(\Omega_n,p) =& \,{1\over \B} \sum_m \int \frac{d^2 q}{(2\pi)^2} G_\sigma\left(\Omega_n - \omega_m,p-q\right) G_\phi\left(\omega_m,q\right)\\
    &+ {1 \over \B^2} \sum_{m,l} \int \frac{d^2q}{(2\pi)^2}\frac{d^2 k}{(2\pi)^2} \frac{ G_\sigma(0,0) G_\sigma(\omega_m,q) }{\left(k^2+\omega_l^2+\sigma_*\right)^2}\frac{1}{(k+q)^2+(\omega_l+\omega_m)^2+\sigma_*}\,.
    \label{phiselfE}
\fe

The expression \eqref{phiselfE} contains a logarithmic divergence that does not depend on the temperature explicitly and can be explicitly subtracted with the use of counter-terms in the following way in the cut-off scheme with UV regulator $\Lambda$ \cite{Moshe:2003xn},
\begin{gather}
   \Sigma_{\rm ren}(\omega_n,p)   =  \Sigma(\omega_n,p) -  \frac{8}{3\pi^2} \left(\omega_n^2 + p^2 + \sigma_*\right)\log \left(\beta\Lambda\right)\,.
\end{gather}
Then the one-point function to this order in the ${1\over N}$ expansion is then 
\ie 
 \la  J^s_{\xi,{\rm bare}} \ra_{\B} = {1\over \B}\sum_n \int \frac{d^2 p}{(2\pi)^2} \frac{\left(p_\mu \xi^\mu\right)^s}{\omega_n^2 + p^2 + \sigma_{*} + \frac{1}{N} \Sigma_{\rm ren}(\omega_n,p)}+\dots \,.
 \fe 
Note that the above expression contains a UV divergence that renormalizes the operator itself, which could be accounted for by the following counter-term
to the first subleading order in $1\over N$,
\ie 
 \la  J^s_\xi \ra_{\B} = {1\over \B}\sum_n \int \frac{d^2 p}{(2\pi)^2} \frac{\left(p_\mu \xi^\mu\right)^s}{\omega_n^2 + p^2 + \sigma_{*} + \frac{1}{N} \Sigma_{\rm ren}(\omega_n,p)}-  \frac{\Delta^1_s }{N} {\log \left(\beta\Lambda\right)\over \B}\sum_n \int \frac{d^2 p}{(2\pi)^2} \frac{\left(p_\mu \xi^\mu\right)^s}{\omega_n^2 + p^2 + \sigma_{*}} \,,
 \label{renJ1pf}
\fe 
where $\Delta_s^1$ is given in \eqref{anomalousON}. This incorporates contributions from all diagrams in the Figure~\ref{fig:HSrenorm}.

We then obtain the following sum-integral expression for the correction to the one-point function coefficient $b_s$,
\ie 
    b_s^0+\frac{1}{N}b^1_s = {1\over \B} \sum_{n=-\infty}^{\infty} \int \frac{d^2 p}{(2\pi)^2} \frac{(p_\mu \xi^\mu)^s}{\omega_n^2 + p^2 +\sigma_* +\frac{1}{N}\Sigma_{\rm ren}(\omega_n,p)}\,.
    \label{Js1PFbr}
\fe 
Specializing to the case $s=2$ and expanding the integrand, we obtain exactly \eqref{eq:12partSTE} that produces the right answer for the one-point function of the stress-energy tensor.

In \eqref{Js1PFbr}, a Lorentz invariant UV regulator for the sum-integral is implicit (see around \eqref{Linvreg}). More explicitly, assuming that $\Sigma_{\rm ren}(\omega_n,p)$ is analytic and does not contain additional poles in $\omega_n$, we can follow the derivation of \eqref{lo2} to obtain the following contour integral expression at general even spin $s$,
\ie 
    b^0_s+ \frac{1}{N} b^1_s = -\frac{1}{2}\int \frac{d^2 p}{(2\pi)^2} \oint_\C  \frac{dz}{2\pi i} \frac{(p_1-z)^s}{z^2 - \epsilon_p^2 - \frac{1}{N} \Sigma_{\rm ren}(-iz,p)} \coth\left( \frac{z}{2} \right)\,,
\fe 
where $\C$ is the same contour as in \eqref{lo2} and $\epsilon_p=\sqrt{p^2+\Delta^2}$. Pushing the contour to infinity (we have implicitly suppressed the regulator $\ep$ in \eqref{lo2} and the divergences in $\ep$ are cancelled by counter-terms), we obtain
\ie  
    b^1_s =&  \int \frac{d^2 p}{(2\pi)^2}  \frac{(p_1 - \epsilon_p)^{s}}{\left(2 \epsilon_p\right)^3} \, \frac{\Sigma_{\rm ren}(-i\epsilon_p,p) }{\sinh^2 \frac{\epsilon_p}{2}}  \\
   & \times 
   \left(\left(\epsilon_p + \frac{p_1 + (s-1) \epsilon_p}{p_1-\epsilon_p}\, \sinh \epsilon_p\right) + \epsilon_p \, \sinh \epsilon_p\, \partial_\epsilon \log \Sigma_{\rm ren}(-i\epsilon_p,p) \right) \,.
\fe 
The above expression is manifestly finite but still challenging to evaluate numerically due to the complicated expression for $\Sigma(\omega_n,p)$ as in \eqref{phiselfE}.\footnote{One may hope to derive a simpler expression for the one-point function coefficient $b_s^1$ as we have done for the $s=2$ case in \eqref{Fmoneren} (see also \eqref{subleadingF}) directly in terms of the leading self-energy $\Pi_\B(\omega_n,p)$ for the $\sigma$ field whose explicit form is given in \eqref{propsigma}.}

\section{Gross-Neveu Model and Variations}\label{sec3}

We now apply the methods developed in Section~\ref{sec:ONall} to study thermal observables in $d=3$ fermionic CFTs with vector-like large $N$ limits. We will focus on the Gross-Neveu (GN) model as well as its closely related variations.

\subsection{Review of the Gross-Neveu Model and Large \texorpdfstring{$N$}{N} Expansion} 
\label{sec:GNreview}

We start by reviewing the celebrated Gross-Neveu (GN) model \cite{Gross:1974jv} (see also \cite{Moshe:2003xn}), keeping spacetime dimension $d$ general for the moment. This model describes $N$ interacting Dirac fermions $\psi_i$ governed by the following action
\ie 
S_{\rm GN}=-\int d^d x \left(
\bar\psi^i \slashed{\partial} \psi_i -{1\over 2 g_0} (\bar\psi^i \psi_i)^2
\right)\,,
\label{GNaction}
\fe
which has an obvious  $U(N)$ global symmetry that rotates the Dirac fermions, as well as a $\mZ_2$ parity symmetry that acts as,\footnote{This comes from a time-reversal symmetry in the Lorentzian signature after Wick rotation.}
\ie 
\mZ_2: x\to \tilde x \equiv (-x_1,x_2,\dots,x_d)\,,\quad \psi(x) \to \C_1 \psi(\tilde x)\,,\quad \bar\psi(x) \to -\bar\psi(\tilde x)\C_1\,,
\label{parity}
\fe
which forbids the $U(N)$ invariant Dirac mass term $m\bar\psi_i \psi_i$. For $d>2$, the quartic interaction is non-renormalizable and this theory needs a proper UV completion. This is provided for $2<d<4$ \cite{Zinn-Justin:1991ksq} by the Gross-Neveu-Yukawa (GNY) model which has in addition a real pseudoscalar $\varphi$ (which is $\mZ_2$ parity odd) and the following action,
\ie 
\label{actionGNY}
S_{\rm GNY}=\int d^d x \left(
{1\over 2} (\pa \varphi)^2
-\bar\psi^i \slashed{\partial} \psi_i  
+g_1 \varphi \bar\psi^i \psi_i 
+  g_2 \varphi^2+ g_4\varphi ^4
\right)\,.
\fe
Correspondingly, 
in contrast to the $O(N)$ scalar model, the fixed point of the GN model resides in the UV, which coincides with the IR fixed point of the GNY model. This $U(N)$ symmetric and parity invariant critical point is commonly referred to as the GN or the GNY CFT. It describes the second-order order-disorder phase transition with order parameter $\varphi\sim \bar\psi \psi$ and characterized by the spontaneous $\mZ_2$ parity symmetry breaking and dynamical mass generation for the fermions (see \cite{Moshe:2003xn} for a more extensive review).

In $d=3$, which will be the focus here, the GN (or GNY) CFTs of $N$ (two-component) Dirac fermions have an enhanced $O(2N)$ global symmetry that rotates the $2N$ Majorana fermions.\footnote{A small subtlety is that for $N={1\over 2}$, the CFT is properly defined only in the GNY description as the four-fermion coupling vanishes for a single Majorana fermion. Interestingly, this model has emergent $\cN=1$ supersymmetry at the fixed point \cite{Fei:2016sgs} and is also known as the $\cN=1$ super-Ising CFT \cite{Atanasov:2022bpi}.} Relatedly the theory is well-defined for $N \in {\mZ_+\over 2}$ by imposing the Majorana condition on the Dirac fermions.
For different values of $N$, these fermionic CFTs govern the universality classes of quantum phase transitions in a variety of condensed matter systems of interacting fermions, including 
the quantum critical points in $d$-wave superconductors  \cite{PhysRevB.62.6721,Vojta:2000zz,PhysRevLett.86.4672} (for $N=4$) and
the spontaneous breaking of parity (time-reversal) symmetry at the boundary of a topological superconductor (for $N={1\over 2}$) \cite{Grover:2013rc}.  The $O(2N)$ symmetric GN (GNY) CFT also has closely related cousins \cite{Rosenstein:1993zf} defined with a different four-fermion coupling (equivalently a different Yukawa-type coupling in the GNY description with possibly additional scalar fields) and reduced global symmetry, such as the chiral Ising GNY model with $O(N)^2\rtimes \mZ_2^{\rm chiral}$ global symmetry which we discuss in Section~\ref{sec:chiralGNY} and the chiral XY GNY model with $(SO(N)\times U(1))\rtimes \mZ_2^C$ symmetry (also known as the Nambu-Jona-Lasino model) which we will come to in Section~\ref{sec:NJL}. They describe various quantum phase
transitions for spinless and spinful fermions on lattices including graphene \cite{Herbut:2006cs,Herbut:2009vu,Herbut:2009qb,Janssen:2014gea}. 
There has been recent progress in determining the flat space CFT data in these fermionic CFTs using the bootstrap method (see \cite{Erramilli:2022kgp} and references therein).

The large $N$ solution of the GN model at criticality  can be deduced in a similar way as for the critical vector model. We introduce an auxiliary scalar field $\phi$, analogous to the $\sigma$ field for the $O(N)$ vector model and implement a Hubbard-Stratonovich (HS)transformation
\begin{gather}
    S_{\rm GN} = -\int d^d x \left(\bar{\psi}^i \slashed \partial \psi_i + \phi \bar{\psi}^i \psi_i + \frac{g_0 \phi^2}{2} \right)\,.
\end{gather}
We will set $g_0= N  g^{t}_0$ to obtain the proper large $N$ limit with $g^t_0$ held fixed. Integrating out the  fermions, we arrive at the following effective Lagrangian for $\phi$,
\begin{gather}
    \cF_{\rm GN}(\phi)= -N  \left( \tr \log \left(\slashed \partial + \phi \right) + \frac{g^{t}_0 \phi^2}{2} \right)\,.
\end{gather}
As explained in \cite{Zinn-Justin:1991ksq}, the GN and the GNY models coincide in the scaling region where the HS field $\phi$ coincides with the Yukawa pseudoscalar $\varphi$ up to a normalization factor.

In the large $N$ limit we can again argue that the path integral over $\phi$ is dominated by its saddle-point, which satisfies the so-called gap equation
\ie 
    \int_\cR \frac{d^d k}{(2\pi)^d} \tr \frac{1}{i\slashed k + \phi} =  - g_0^t \phi\,,
    \fe 
    or equivalently (for $\phi\neq 0$)
    \ie -\int_\cR \frac{d^d k}{(2\pi)^d} \frac{c_d}{k^2 + \phi^2} = g_0^{t}\,,
    \label{GNgapeqn}
    \fe 
where $c_d\equiv\tr 1 $ counts the components of the Dirac spinor in $d$ spacetime dimensions (i.e. $c_3=2$ in the case of interest). This equation again contains divergences, that we cancel by introducing a renormalized coupling constant $g^t$, 
\ie 
    g^t = g_0^t + c_d\int_\cR \frac{d^d k}{(2\pi)^d} \frac{1}{k^2}  =  c_d\int_\cR \frac{d^d k}{(2\pi)^d} \frac{ \phi^2}{(k^2 + \phi^2) k^2}= c_d K_d \phi^{d-2} + \ldots\,,
\fe 
where $K_d$ is scheme-independent and given by \eqref{Kdexp} and in the last equality above we have neglected further corrections that are scheme-dependent (and suppressed near the critical point). To reach criticality, we fine-tune $g^t$ such that $\phi=0$ so that the fermions become massless.\footnote{For greater $g^t$, the gap equation admits a nonzero solution for $\phi$ which describes the $\mZ_2$ parity symmetry breaking phase \cite{Moshe:2003xn}.} One can check that at this point the HS field $\phi$ is also massless. Indeed, the inverse propagator for $\phi$ is,\footnote{The pseudoscalar $\phi$ here (with propagator $G_\phi$) is not to be confused with the scalar fields $\phi_i$ in the $O(N)$ scalar model studied in Section~\ref{sec:ONall}.} in the dimensional regularization scheme,
\begin{gather}
    G_{\phi}^{-1}(p)
    = -g^{t} - c_d\int_\cR  \frac{d^d k}{(2\pi)^d} \left[ \frac{k\cdot (k+p)}{k^2(k+p)^2} - \frac{1}{k^2}\right] \,.
\end{gather}
The second term on the RHS above vanishes as $p\to 0$  and we have $G_{\phi}^{-1}(0) = -g^{t}$. Therefore  $\phi$ is massive unless $g^t=0$.

\subsection{Free Energy at the Subleading Order } \label{sec:GNFandchem}
The thermal free energy density or equivalently the one-point function of the stress-energy tensor in the GN CFT has the same structure as a series in $\frac{1}{N}$ as in \eqref{Fexp} for the critical $O(N)$ model.
In this section we will compute the leading contribution  $F_{\rm GN,0}  \propto N$ and the first subleading correction $F_{\rm GN,-1} \propto N^0$ in this expansion. The computation of both contributions is similar to the critical $O(N)$ model. We will comment on the differences and the subtleties associated with these fermionic CFTs along the way.

We start with the leading contribution $F_{\rm GN,0}(\B)$. We assume that at finite temperature, the dominant configuration for the HS field $\phi$ is homogeneous and thus the free energy density is
\ie 
\cF_{\rm GN}(\phi)=&-{N\over \B}  \tr \log (\slashed{\pa}+\phi)
=-{N \over \B} \sum_{n\in \mZ} \int {d^2 p\over (2\pi)^2} \log ( p^2+\omega_n^2+\phi^2)\,,
\label{GNF0sumint}
\fe
where the saddle-point value of $\phi$ will be fixed shortly. 
There are two spin-structures for the fermions that are compatible with the $O(2N)$ global symmetry, corresponding to either periodic or anti-periodic boundary conditions along ${\rm S}^1_\B$. Correspondingly the allowed frequencies in the sum of \eqref{GNF0sumint} are,
\ie 
{\rm periodic}:~\omega^+_n={2\pi n\over \B}\,,\quad 
{\rm antiperiodic}:~\omega^-_n={2\pi (n+1/2)\over \B}\,,
\label{freqsumferm}
\fe
and the latter is the standard thermal boundary condition for fermions. In the periodic case, the free energy density \eqref{GNF0sumint} as a function of $\phi$ can be read off  from  \eqref{Frenleading} by comparing \eqref{GNF0sumint} and \eqref{Fsigmastart},
\ie 
\cF_{\rm GN}^{\rm P}(\phi)=-2\cF_{\rm ren}(\phi^2)\,.
\label{FGN0P}
\fe 
One finds from \eqref{ONsaddle} that the saddle-point and the final free energy density are\footnote{Note that there is also a saddle at $\phi=0$ which sub-dominate.}
\ie 
\phi=\sqrt{\sigma_{*}}={1\over \B } \log\left(\frac{3+\sqrt{5}}{2} \right)\,,
\quad F_{\rm GN,0}^{\rm P}(\B)=\cF_{\rm GN}^{\rm P}(\sqrt{\sigma_{*}})= 
 {4 \zeta(3)\over 5\pi }{N\over \B^3}\,.
 \label{GNPsaddle}
\fe 
Note that $F^{\rm P}_{{\rm GN},0}$ is positive, naively implying that the specific heat of such model is negative, but we are considering the periodic boundary condition, which does not define a conventional thermal ensemble.

In the anti-periodic case, following a similar derivation to \eqref{Frenleading}, the free energy density before plugging in the saddle-point for $\phi$ reads (see also \cite{Petkou:1998wd,Christiansen:1999uv}),\footnote{This is also studied for various $d$ in \cite{David:2023uya}.} 
\begin{gather}
\label{FGN0AP}
    \cF_{\rm GN}^{\rm AP}(\phi)
    =\dfrac{N}{2\pi \beta^{3}}\left(\dfrac{1}{3}\left(\beta\phi\right)^{3}+2\beta \phi\,{\rm Li}_{2}\left(-e^{-\beta \phi}\right)+2\,{\rm Li}_{3}\left(-e^{-\beta \phi}\right)\right).
\end{gather}
Formally, $\cF_{\rm GN}^{\rm AP}(\phi)$ has three critical points,
\ie 
\phi=0\,,\quad\phi=\pm \phi_*= \pm {2\pi i \over 3\beta }\,. \label{eq:GNfixpoints}
\fe
The last two solutions $\phi=\pm \phi_*$ are equivalent by the $\mZ_2$ parity symmetry and have lower free energy density than the first solution. 

However we should remember that in the GN model the path integral contour for the HS field $\phi$ is along the real axis (in contrast to the scalar $O(N)$ model) \cite{Moshe:2003xn}. This is obvious from its unitary UV completion, the GNY model \eqref{actionGNY}, since $\phi$ is identified up to a real rescaling with the real pseudoscalar field $\varphi$. Furthermore, one can check that this defining integration contour is a steepest descent contour for constant $\phi$ where $\phi=0$ is the stable saddle-point, whereas the other complex saddle-points $\phi=\pm \phi_*$ are unstable when approached in the real direction.  

Consequently to describe the large $N$ GN CFT at finite temperature, we pick $\phi=0$ in \eqref{eq:GNfixpoints} as our saddle-point solution.
Then the leading contribution to the free energy density 
\ie 
& F_{\rm GN,0}^{\rm AP}(\B) =\cF_{\rm GN}^{\rm AP}(0)= -\frac{3 N \zeta(3)}{4\pi \beta^3} = -\frac{0.28696995 N}{\beta^3}\,,
\label{GNAPleadingF}
\fe 
coincides with that of $N$ free Dirac fermions, as was found in \cite{Petkou:1998wd,Christiansen:1999uv}. Nonetheless, we will see that at the subleading order in $1\over N$, this coincidence is lifted.

The subleading $\frac{1}{N}$ corrections to the free energy density in the GN CFT comes from the fluctuations of $\phi$ around its saddle-point. This is related to the computation of the propagator for $\phi$ in such a background. In this case, as shown in  Appendix~\ref{app:GNselfE}, the $\phi$ propagators  take the following forms for the periodic and anti-periodic spin structures respectively,
\ie 
    \left(G^{\rm P}_\phi\right)^{-1}(\Omega,p,\phi_{+}) = &\, N\left[\frac{1}{\pi\beta}\log{\left(2\sinh{\frac{\beta\phi_{+}}{2}}\right)}+2\left(P^2 + 4\phi^{2}_{+ }\right) \Pi^{+}_{\beta}(\Omega,p)\right]  \\  
    \left(G^{\rm AP}_\phi\right)^{-1}(\Omega,p,\phi_{-}) =  &\,N\left[\frac{1}{\pi\beta}\log{\left(2\cosh{\frac{\beta\phi_{-}}{2}}\right)}+2\left(P^2 + 4\phi^{2}_{- }\right) \Pi^-_{\beta}(\Omega,p)\right]\,,
    \label{defPpm}
\fe 
where $P=(\Omega,p)$ is a short-hand for the three-momentum with $P^2 = \Omega^2 + p^2$, and  $\phi_+=\sqrt{\sigma_*}$ and $\phi_-=0$ are the corresponding saddle-points.
For fermions obeying the periodic boundary condition $\Pi^+_\beta(\Omega,p) = \Pi_\beta(\Omega,p)$ as is given in \eqref{Pidef} (see also \eqref{propsigma}) and for the anti-periodic case $\Pi^-_\beta(\Omega,p)$ is given in \eqref{gnap1}. 

The subsequent $\frac{1}{N}$ correction to the free energy density of the GN CFT with the periodic boundary condition is related to the $O(N)$ answer as follows,
\ie 
\label{FGNsubP}
    F^{\rm P}_{{\rm GN},-1}(\beta) =&\, -\frac{1}{2 \beta}\sum_n \int \frac{d^2 p}{(2\pi)^2} \log\left[G_{\phi}^{\rm P}(\omega_n,p,\sqrt{\sigma_{*}})\right]
   \\
   =&\,  \frac{1}{2 \beta}\sum_n \int \frac{d^2 p}{(2\pi)^2} \log \left(P^2+4\sigma_{*}\right) + F_{O(N),-1}(\beta)\,,
\fe 
where we have already computed the second term in the last line in \eqref{FONsubfinal} and the first term after proper renormalization follows from  \eqref{Frenleading},
\begin{gather}
   \frac{1}{2 \beta}\sum_n \int \frac{d^2 p}{(2\pi)^2} \log \left(P^2+4\sigma_{*}\right) \xrightarrow[{\rm renormalize}]{} 
 {1\over N}\cF_{\rm ren}(4 \sigma_{*})
    =-\dfrac{0.25927703}{\beta^{3}}\,.
\end{gather}
We thus arrive at the final answer of the subleading free energy density for the GN CFT in the periodic spin structure,
\begin{gather}\label{GNPsub}
    F^{\rm P}_{{\rm GN},-1}(\beta) = F_{O(N),-1}(\beta) + {1\over N} \cF_{\rm ren}(4{\sigma_{*}}) =-\dfrac{0.19528150}{\beta^3}\,.
\end{gather} 

For the anti-periodic spin structure,  we have instead
\ie 
    F^{\rm AP}_{{\rm GN},-1}(\beta) = -\frac{1}{2\beta} \sum_n \int \frac{d^2 p}{(2\pi)^2} \log G^{\rm AP}_{\phi}(\omega_n,p 
,0)\,.
\fe 
Proceeding as above, after renormalization, it is given by,
\ie 
F^{\rm AP}_{{\rm GN},-1}(\beta)=\dfrac{1}{2\beta}\sum_{n}\int \dfrac{d^{2}p}{(2\pi)^{2}}\log{\left(\dfrac{8\Pi^{\rm AP}_{\rm GN}(\omega_{n},p,0)}{\sqrt{\omega^{2}_{n}+p^{2}}}\right)}+{1\over 2N}\cF_{\rm ren}(0)  
=\dfrac{-0.01340099}{\beta^3}\,,
    \label{freenrgzeromas}
\fe 
 where the self-energy $\Pi^{\rm AP}_{\rm GN}$ is given in \eqref{gnap1} and in the last step we have applied the numerical procedure as in Section~\ref{sec:numerical} to evaluate the sum-integral above.

\subsection{Chemical Potential Dependence and Phase Transition}
\label{sec:GNchem}

\begin{figure}[!htb]
    \centering
    \includegraphics{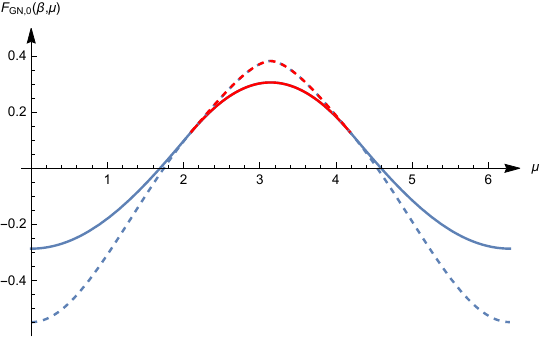}
    \caption{The dependence of the leading   free energy density $F_{\rm GN,0}(\B,\m)$ $F_{\rm GN',0}(\B,\m)$ of the GN (solid) and GN$'$ (dashed) CFT  as a function of the imaginary chemical potential $\m$ (here $\B=1$). The middle portion (colored in red) is dominated by the nontrivial real saddle \eqref{mudepGNmod}.}
    \label{fig:GNmu}
\end{figure}

Analogously to the critical $O(N)$ model analyzed in Section~\ref{sec:ONchem}, here we study the GN model with a non-zero imaginary chemical potential. We consider the $U(1)\subset O(2N)$ global symmetry subgroup for the GN CFT under which the Dirac fermions $\psi_i$ all have charge 1 and 
introduce an external constant electromagnetic potential $A_0={\m i\over \B}$ with $\m\in[0,2\pi)$. It amounts to shifting frequencies of the fermions to $\B \tilde \omega_n\equiv \B \omega_n^-  +  \m$. Note that in particular $\m=\pi$ corresponds to the periodic spin structure for the fermions. The leading free energy density as a function of $\mu$ and the one-point function of $\phi$ field (denoted by $\tilde\phi$ below) is
\ie \label{cFGN}
    {}&  \cF_{\rm GN}(\tilde\phi,\mu)=-{N\over \B}\sum_{n}\int \dfrac{d^{2}p}{(2\pi)^{2}}\log{\left( \tilde{\omega}^{2}_{n}+p^{2}+\tilde\phi^{2}\right)}
      \\
       =&\,\dfrac{N}{2\pi\B^3}\left(\dfrac{\tilde\phi^{3}}{3}+\tilde\phi \operatorname{Li}_{2}(-e^{-\tilde\phi+i\mu}) + \tilde\phi \operatorname{Li}_{2}(-e^{-\tilde\phi-i\mu})+\operatorname{Li}_{3}(-e^{-\tilde\phi + i\mu})+\operatorname{Li}_{3}(-e^{-\tilde\phi - i\mu})\right)\,,
\fe 
which coincides with \eqref{FGN0AP} for $\m=0$ and with \eqref{FGN0P} for $\m=\pi$ as expected.
The gap equation is modified by the nontrivial chemical potential. While the real solution $\phi=0$ remains valid,  the pair of complex solutions in \eqref{eq:GNfixpoints} are deformed to,
\begin{gather}
\tilde \phi_*(\mu) = \operatorname{arccosh}\left[\frac12 -\cos \mu \right]\,,
    \label{mudepGNmod}
\end{gather}
which becomes real for ${2\pi \over 3}\leq \m\leq {4\pi \over 3}$ and is purely imaginary elsewhere for $\m\in [0,2\pi)$. Furthermore for this intermediate range of $\m$ around $\m=\pi$, it is the nonzero saddle \eqref{mudepGNmod} that is stable and dominates the free energy, which is consistent with what we found for the periodic spin structure in \eqref{GNPsaddle}.
This implies a phase transition as we dial $\m$ up from $\m=0$ where the real stable saddles exchange dominance. This transition is first-order and characterized by the nonzero expectation value of $\phi\sim \bar \psi^i \psi_i$, which signals breaking of the spatial parity.\footnote{This is a parity transformation that reflects a spatial direction on $\mR^2$ and is preserved by the chemical potential.} See Figure~\ref{fig:GNmu} for the resulting free energy density $F_{\rm GN,0}(\B,\m)$ in the leading large $N$ limit. 

\begin{figure}[!htb]
    \centering
    \includegraphics{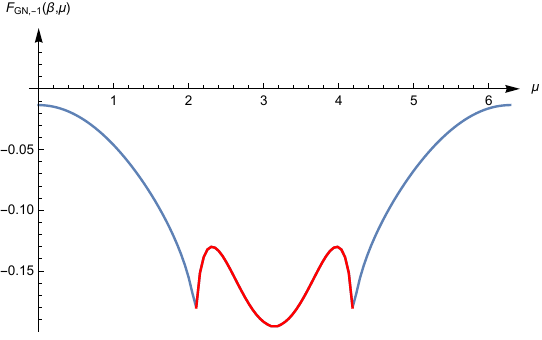}
    \caption{The  subleading contribution to the free energy density $F_{{\rm GN},-1}(\beta,\mu)$  as a function of the imaginary chemical potential $\mu$ of the GN CFT  (here $\B=1$). 
    The middle portion (colored in red) comes from fluctuations around the nontrivial saddle \eqref{mudepGNmod}.}
    \label{fig:chemfm1}
\end{figure}

To compute the ${1\over N}$ correction, we proceed as before using the propagator of the $\phi$ field on such a thermal background with imaginary chemical potential (see \eqref{sec:eq:chempol} for the explicit expression) and the resulting free energy density $F_{{\rm GN},-1}(\beta,\mu)$ is plotted in Figure~\ref{fig:chemfm1}, taking into account the switching of branches at $\m={2\pi \over 3},{4\pi \over 3}$ between which the relevant saddle is given by \eqref{mudepGNmod}.

As before, from the dependence of the free energy on $\mu$, we can extract the Wilson coefficient $b$ in \eqref{freetwisted} which governs the symmetry resolved density of states \eqref{srdos} at high energy. Here for the $O(2N)$ symmetry of the GN CFT, we find  
\ie 
N b=N b_0 + b_{-1} + \mathcal{O}\left(N^{-1}\right), \\
\fe 
with\footnote{As in the case of the critical $O(N)$ model, the error for $b_{-1}$ is a standard error that is computed using \texttt{NonLinearModel} in \texttt{Mathematica}.}
\ie 
 b_0 = \frac{2\log 2}{\pi}\,,\quad    b_{-1} =-0.1338 \pm 0.0002\,.
 \label{GNb}
\fe

\subsection{The Alternative Gross-Neveu Model and Parity Breaking} \label{sec:GNp}

As alluded to before, the large $N$ analysis of the Gross-Neveu model after the HS transformation, suggests an  alternative Gross-Neveu Model defined by a different integration contour for the zero mode of $\phi$ which runs along the imaginary axis. We refer to this new model as ${\rm GN}'$ to distinguish from the usual GN model. As a consequence of the contour rotation, we expect the ${\rm GN}'$ theory to become non-unitary. Nonetheless, at least in the large $N$ limit, the conformal phase at finite temperature exists in the ${\rm GN}'$ model, which shares similarities with the Lee-Yang edge singularity 
(see also \cite{Petkou:2000xx}).

In this case, with the rotated contour, the relevant saddle-point at zero chemical potential is given by $\phi=\pm\phi_*=\pm {2\pi i\over 3\B}$ from \eqref{eq:GNfixpoints} which breaks the $\mZ_2$ parity. 
The leading contribution to the thermal free energy density is
\ie 
 F_{{\rm GN'},0}^{\rm AP}(\B)
=-{N \over 4\pi \B^3} 
\left(
{8\pi \over 3}{\rm Cl}_2(\pi/3)-{4\over 3}\zeta(3)
\right) = -{ 0.54908554 N \over \B^3}  \label{GNpAPleadingF}
\fe 
where the Clausen function is related to usual  polylogarithms by
\ie 
{\rm Cl}_2(z)\equiv \operatorname{Im} \operatorname{Li}_2(e^{i z})\,. 
\label{clausen}
\fe
The leading ${1\over N}$ correction to the free energy can be obtained in a similar way as in the case of the unitary GN model,
\begin{gather}
    F^{\rm AP}_{{\rm GN}',-1}(\beta) =  \frac{1}{2 \beta}\sum_n \int \frac{d^2 p}{(2\pi)^2} \left[\log \left(P^2+4\phi_{*}^2\right) + \log \Pi^-_\beta(P)\right]\,.
\end{gather}
At first glance, for $\phi_*^2<0$ the integrand becomes imaginary for some range of the integral-sum, and therefore seems unphysical. However, we should remember that the  contour of integration for the field $\phi$ needs to be deformed (to be the steepest descent contour) in order to implement a proper saddle-point approximation. To this end, we just need to take the absolute value of the argument of logarithm (that corresponds to a slight tilt of the contour of integration), we then arrive at 
\ie 
\label{GNAPsub}
F^{\rm AP}_{{\rm GN}',-1}(\beta)=&\dfrac{1}{2\beta}\sum_{n}\int \dfrac{d^{2}p}{(2\pi)^{2}}\log{\left(16\sqrt{\omega^{2}_{n}+p^{2}}\Pi^{-}_\beta(\omega_{n},p)\right)}-{1\over 2N}\cF_{\rm ren}(0) +{1\over N}\cF_{\rm ren}(4\phi_{*}^2)
\\
=&\,\dfrac{0.14222693}{\beta^{3}} \,,
\fe 
where we have used from \eqref{Frenleading},
\begin{gather}
\label{ap9}
    \frac{1}{2 \beta}\sum_n \int \frac{d^2 p}{(2\pi)^2} \log \left|P^2+4\phi_{*}^2\right|
  \xrightarrow[{\rm renormalize}]{}
   {1\over N}\cF_{\rm ren}(4\phi_{*}^2)=\dfrac{0.53611329}{\beta^{3}}\,,
\end{gather}
and also \eqref{freeFren0th}.

For the periodic spin structure, a similar calculation gives 
\ie 
 F_{{\rm GN'},0}^{\rm P}(\B)
={\zeta(3)\over \pi} {N\over \B^3}\,,
\fe 
which comes from \eqref{FGN0P} with $\phi=0$. At the first subleading order in ${1\over N}$, taking into account the fluctuations of the HS field $\phi$, we find\footnote{The error in this computation is significantly bigger than in the analogous computation in the unitary GN model due to the spurious divergences (that eventually cancel) when the self-energy of the HS field $\phi$ is computed around $\phi=0$.}
\ie 
    F^P_{\rm GN',-1}(\beta) = -\frac{0.08\pm0.005}{\beta^3}\,.
\fe 

We can also determine the Wilson coefficient  $b$  for the $O(2N)$ global symmetry of the GN$'$ model from the dependence of its free energy on the chemical potential. The result is 
\ie 
   N  b=N b_0 + b_{-1}+ \mathcal{O}\left(N^{-1}\right)
    \fe 
    with 
    \ie
    b_0 = \frac{2}{\sqrt{3}}\,,\quad 
    b_{-1} = -0.562 \pm 0.005\,.
    \label{GNpb}
\fe

 One might wonder how the GN and GN$'$ models could have such different free energies given that these two theories coincide at the perturbative level. These different free energies translate into drastically different behavior for the asymptotic density of states in the putative CFTs. 
 This difference could arise only at the non-perturbative level, which can be studied with the methods developed in \cite{Giombi:2019upv}. We see that if we consider the GN$'$ model and integrate the HS field $\phi$ over the imaginary contour we wouldn't encounter additional instantonic contributions while in the GN model such contributions are present. Consequently, the scaling dimensions of the GN$'$ and the GN models would differ non-perturbatively.

\subsection{One-point Functions of Higher-Spin Currents}\label{sec:GNhs1pf}
As explained before, the free energy density of the CFT determines the thermal one-point function of the stress-energy tensor.  The stress-energy tensor belongs to a tower of spinning operators known as the higher-spin currents, which in the GN model are given by the following, with $s\geq 1$,

\ie 
\tilde J^s_{\mu_1 \ldots \mu_s} = \left (\dfrac{1}{N s}\sum_{\ell=1}^s \bar\psi^i \partial_{\mu_1} \cdots \C_{\m_\ell}\cdots\partial_{\mu_s}  \psi_i - \text{traces} \right) +{\rm descendants}\,,
\label{JsdefGN}
\fe 
similar to those in the $O(N)$ scalar model case (see \eqref{Jsdef}).\footnote{See also recent work \cite{David:2023uya} which studies the thermal one-point functions of higher-spin currents for general $d$.} 
As before, at finite temperature,  the one-point functions are required to take the following form
\begin{gather}
\label{J1pfgenformGN}
    \braket{ \tilde{J}^s_{\mu_1 \ldots \mu_s}}_{\B,\pm} = \frac{\tilde{b}_{\pm,s}}{\beta^{\tilde\Delta_s}} \left(e_{\mu_1} \ldots e_{\mu_s} - \text{traces}\right),
\end{gather}
where $e_\mu$ is a unit vector along the ${\rm S}^1_\beta$ and $\tilde \Delta_s$ is the scaling dimension of the (weakly broken) higher-spin current,
\ie 
\tilde\Delta_s=s+1+\tilde\C_s\,,
\fe
with the anomalous dimension $\tilde\C_s$ that is suppressed by ${1\over N}$ and 
can be found in \cite{Muta:1976js,Manashov:2016uam,Hikida:2016cla,Giombi:2017rhm}.
The $\pm$ subscript in \eqref{J1pfgenformGN}
refers to the choice of spin structure along the ${\rm S}^1_{\B}$.
To compute the coefficients $\tilde{b}_{\pm,s}$ in the leading large $N$ limit, we introduce an auxiliary null three-vector $\xi$ as in Section~\ref{sec:1pfcurrentsON} (see around \eqref{Jxidef}). Following the same procedure there with the same type of Feynman diagrams (the propagating bosons are substituted by the fermions), we find that 
\ie 
\tilde{b}^0_{\pm,s} = \sum_{\omega^{\pm}_{n}} \int \frac{d^2 p}{(2\pi)^2} (ip_{\mu}\xi^{\mu})^{s-1} \tr\left[\dfrac{i\gamma_{\nu}\xi^{\nu}}{i\slashed{p}+\phi_{\pm}}\right]
=2\sum_{n \in \mZ} \int \frac{d^2 p}{(2\pi)^2} \dfrac{(ip_{\mu}\xi^{\mu})^{s}}{(\omega^{\pm}_{n})^{2}
+p^2+\phi_{\pm}^2} \,, \label{hsgn1}
\fe 
where for $\phi_{+}=\sqrt{\sigma_{*}}$ we sum over $\beta\omega^{+}_{n}=2\pi n$, and for $\phi_{-}=\phi_{*}$ we sum over $\beta\omega^{-}_{n}=\pi(2n+1)$ with $n\in \mZ$. The UV divergences in the above expressions are regularized as for the critical $O(N)$ model. 

By comparing \eqref{hsgn1} with \eqref{lo1}, for the periodic boundary condition $\phi_{+}=\sqrt{\sigma_{*}}$ and all one-point functions are related to those in the critical $O(N)$ model simply by\footnote{This is no longer true at the subleading order in ${1\over N}$ as we have seen for the stress-tensor one-point function which follows from \eqref{FGNsubP}.} 
\ie 
\tilde{b}_{+,s}^0 = 2b^0_{s}\,.
\label{bt0P}
\fe For the anti-periodic boundary condition,  we consider the following  generating function similar to \eqref{genFG},
\begin{gather}
   \sum^\infty_{s=0} \frac{  \tilde{b}_{-,s}^{0}\xi^s}{s! \B^{s+1}} =  \dfrac{2}{\beta}\sum_{n \in \mZ} \int \frac{d^2 p}{(2\pi)^2} \frac{e^{i p_\mu \xi^\mu}}{(\omega^{-}_n)^2 + p^2 + \phi_{*}^2} = {1\over 2\pi}\sum_{m\neq 0} (-1)^m \frac{e^{-\phi_{*} \sqrt{m\beta (m\beta-2\xi)}}}{\sqrt{m \beta(m \beta-2\xi)}},
\end{gather}
Expanding the RHS of the above equation we find that $\tilde b_{-,s}^0=0$ for $s$ odd as expected and for $s$ even we have,
\ie 
\tilde{b}^0_{-,s} ={1\over 2\pi}\sum_{n=0}^s\frac{2^{n-s+1} (2s-n)!}{ n!   (s-n)! } \beta^n \phi_*^n \text{Li}_{s+1-n}\left(-e^{-\beta \phi_*}\right)\,,
\label{bt0AP}
\fe
We can also derive an alternative formula for $\tilde{b}^{0}_{-,s}$ starting from \eqref{hsgn1}.
Following the procedure around \eqref{lo2}, we arrive at the following integral expression (setting $\B=1$ below), for even spin $s$,
\begin{gather}
\tilde{b}_{-,s}^{0} =-{2 }{}\int 
\dfrac{d^2p}{(2\pi)^2}\dfrac{(\epsilon_{p}-p_{1})^{s}}{\epsilon_{p}}\frac{1}{e^{\epsilon_{p}}+1}\,,
    \label{hsgn3}
\end{gather}
where $\epsilon_{p}\equiv \sqrt{p^2+\phi_{*}^{2}}$. 
The above integral reproduces the sum expression \eqref{bt0AP} thanks to the following integral identity,
\ie 
\text{Li}_{s}\left(-e^{-\phi_*}\right)=-{1\over  \Gamma(s)}\int_{\phi_*}^\infty dz  {(z-\phi_*)^{s-1}\over e^z+1}\,.
\fe
The subleading ${1\over N}$ corrections $\tilde b^1_{\pm,s}$ to the one-point functions of these higher-spin currents in the fermionic CFT can be in principle computed in a similar way as describe near the end of Section~\ref{sec:1pfcurrentsON} for the $O(N)$ scalar model.

\subsection{The Chiral Ising Gross-Neveu-Yukawa Model}
\label{sec:chiralGNY}
As it was explained in the previous section, another interesting generalization of the usual GN or GNY model is the chiral Ising GN or GNY model with a different symmetry, which has many applications in condensed matter systems including the semimetal-insulator transition in graphene \cite{Herbut:2006cs,Herbut:2009vu,Herbut:2009qb}. The chiral Ising GN model is defined by splitting the $N$ Dirac fermions into two groups as $\psi_i=(\chi^L_a,\chi^R_a)$ where $a=1,\dots,{N\over 2}$
with the following action that has a different quartic interaction compared to \eqref{GNaction},\footnote{Again there is little difference to the description of the model when $N$ is odd for spacetime dimension $d=3$, in which case we simply write the action in terms of the Majorana fermions.}
\ie 
S_{\rm cGN}=-\int d^d x \left(
\bar\chi^{La} \slashed{\partial} \chi_a^L+\bar\chi^{Ra} \slashed{\partial} \chi_a^R -{1\over 2 g_0} (\bar\chi^{La} \chi_a^L-\bar\chi^{Ra} \chi_a^R)^2
\right)\,.
\label{cGNaction}
\fe
Consequently the global symmetry of the model is reduced to $(U(N/2)_L\times U(N/2)_R)\rtimes \mZ_2^{\rm chiral}$ where the so-called ``chiral'' $\mZ_2^{\rm chiral}$ exchanges $\chi_a^L$ and $\chi_a^R$. In $d=3$, this symmetry is enhanced to $O(N)^2 \rtimes\mZ_2^{\rm chiral}$ as mentioned previously.

For $2<d<4$, the UV completion of the model \eqref{cGNaction} is provided by the chiral Ising GNY model, with one additional real pseudoscalar $\varphi$ and the following action,
\begin{gather}
    S_{\rm cGNY} = \int d^d x \left( \frac12\left(\partial  \varphi\right)^2
    -\bar\chi^{La} \slashed{\partial} \chi_a^L-\bar\chi^{Ra} \slashed{\partial} \chi_a^R 
    +g_1\varphi \left(\bar\chi^{La} \chi_a^L-\bar\chi^{Ra} \chi_a^R\right) +   g_2 \varphi^2 + g_4 \varphi^4 \right)\,.
    \label{cGNYaction}
\end{gather}
Note that $\varphi$ is odd under both the ``chiral'' $\mZ_2^{\rm chiral}$ and the parity \eqref{parity} thus invariant under the diagonal combination. The latter is identified with the preserved parity (time-reversal) symmetry at the semimetal-insulator transition in graphene for which $\varphi$ is the order parameter \cite{Herbut:2006cs,Herbut:2009vu,Herbut:2009qb} (see also \cite{Erramilli:2022kgp}). In the scaling region, the chiral Ising GN and GNY models are expected to coincide \cite{Zinn-Justin:1991ksq}.

\begin{figure}[!htb]
    \centering
    \begin{tikzpicture}
        \draw[very thick] (0,0) circle (1);
        \draw[dotted,very thick] (0,-1)--(0,1);
        \draw[dotted,very thick] (3,-1)--(3,1);
        \draw[dotted,very thick] (1,0)--(2,0);
        \draw[very thick] (3,0) circle (1);
        
        \draw[very thick] (6,0) circle (1);
        \draw[dotted,very thick] (7,0)--(7.5,0);
        \draw[dotted,very thick] (6,1)--(8.5,1);
        \draw[dotted,very thick] (6,-1)--(8.5,-1);
        \draw[very thick] (8.5,0) circle (1);
    \end{tikzpicture}
    \caption{The first two diagrams in the $\frac{1}{N}$ expansion whose contributions to the free energy density differ between the GNY and the chiral Ising GNY models (due to the choice of signs at the cubic vertices in the latter). The solid lines denote the propagators for the fermions $\psi_i$ and the dotted lines for the pseudoscalar $\varphi$.}
    \label{fig:cGNYdif}
\end{figure}
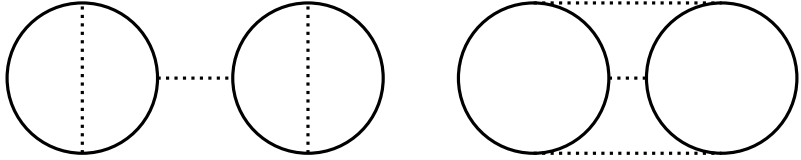

Implementing the HS transformation, the action \eqref{cGNaction} can be recast into the following form
\begin{gather}
    S_{\rm cGN} =- \int d^d x \left(\bar\chi^{La} \slashed{\partial} \chi_a^L+\bar\chi^{Ra} \slashed{\partial} \chi_a^R+ \phi \left(\bar\chi^{La} \chi_a^L-\bar\chi^{Ra} \chi_a^R\right) + \frac12 g_0 \phi^2 \right)\,,
\end{gather}
where the HS field $\phi$ is again identified up to an normalization with the pseudoscalar $\varphi$ in the chiral Ising GNY model \eqref{cGNYaction}.
We then integrate out the fermions and arrive at the following effective Lagrangian for the field $\phi$
\begin{gather}
    \cF_{\rm cGN}(\phi) = - N  \left(\tr \log\left(i \slashed \partial + \phi\right) + \tr \log\left(i \slashed \partial - \phi\right) + \frac12 g_
    0^t \phi^2 \right)\,,
    \label{cGNF}
\end{gather}
where $g_0 = g_0^t N$ as before with $g_0^t$ held fixed in the large $N$ limit. Note the two terms in \eqref{cGNF} where $\phi$ comes with opposite signs. For a general configuration of $\phi$, these two terms are not related to each other and that's why the chiral Ising GNY (or GN) model is different from the usual GNY model. Nonetheless, as we explain below, it is easy to show that the leading and the first subleading corrections to the free energy density coincide.\footnote{In fact the same analysis below easily extends to the free energy on ${\rm S}^1_\B\times {\rm S}^2_R$ and the same conclusion that the free energies for the GN model and its chiral version only differ at the ${1\over N}$ order continue to hold. It would be interesting to study this difference in free energies in relation to the difference in the operator spectrum as was analyzed in \cite{Erramilli:2022kgp}.} This is because for constant $\phi$, the first two terms in \eqref{cGNF} are identical and the saddle-point of $\phi$ obeys the same gap equation as for the GN model,
\begin{gather}
    \int_\mathcal{R} \frac{d^d k}{(2\pi)^d} \tr \frac{1}{i \slashed k + \phi} = -g_0^t \phi\,,
\end{gather}
and by fine-tuning the coupling $g_0^t$ we can bring system to criticality as before. 
The free energy density of the chiral Ising GN model has the same large $N$ expansion as before,
\begin{gather}
    F_{\rm cGN}^\pm(\B) =   F_{\rm cGN,0}^\pm (\B) + F_{\rm cGN,-1}^\pm(\beta)+   F_{\rm cGN,-2}^\pm(\beta) + \ldots\,,
\end{gather}
where $F_{{\rm cGN},\A}$ scales as $N^{1+\A}$. The above discussion indicates that 
 the leading $N$ contributions agree $F_{\rm cGN,0}^{\pm} = F^\pm_{\rm GN,0}$. Furthermore, the first subleading corrections also coincide $F_{\rm cGN,-1}^{\pm} = F_{\rm GN,-1}^{\pm}$ because the large $N$ propagator of the field $\phi$ is an even function of its constant saddle-point value (see \eqref{defPpm}). However it is easy to see that already at the second order, there is a difference between the two versions of the GN model in the free energy density, namely $F_{\rm cGN,-2}^{\pm} \neq  F_{\rm GN,-2}^{\pm}$, which arises from the different contributions of the vacuum diagrams in  Figure~\ref{fig:cGNYdif} at non-zero temperature. 
The same conclusion holds for the thermal one-point function of general higher-spin currents defined in \eqref{JsdefGN}.

\subsection{The Nambu–Jona-Lasinio (NJL) Model}
\label{sec:NJL}

Another important generalization of the GN model reviewed in Section~\ref{sec:GNreview} is the Nambu-Jona-Lasinio (NJL) model. It is commonly defined in terms of four-component Dirac fermions $\Psi_i$ with $i=1,\dots,N_f$ (which naturally arise in the $d=4$ model) with the following action,
\ie
S_{\rm NJL}=-\int d^d x \left(
\bar\Psi^i \slashed{\partial} \Psi_i -{1\over 2 g_0} \left((\bar\Psi^i \Psi_i)^2-(\bar\Psi^i \C_5 \Psi_i)^2\right)
\right)\,,
\label{NJLaction}
\fe
where $\C_5$ is the usual chirality matrix for $d=4$ Dirac spinors.\footnote{We follow the gamma matrix conventions in \cite{Erramilli:2022kgp} (see also \cite{Iliesiu:2015qra}). In particular, the four-fermion interaction of the chiral Ising GN model with $N=2N_f$ discussed in Section~\ref{sec:chiralGNY} takes the form $(\bar \Psi^i \Psi_i)^2$ in terms of these four-component Dirac fermions and the chiral symmetry $\mZ_2^{\rm chiral}$ acts by $\Psi \to \C_5\Psi$.}
The global symmetry of the theory is $(U(N_f)\times U(1))\rtimes \mZ_2^C$ where $U(N_f)$ is the obvious flavor symmetry that rotates the Dirac fermions $\Psi_i$, the extra $U(1)$ ``chiral'' symmetry acts by $\Psi_i \to e^{i \A \C_5} \Psi_i$, and $\mZ_2^C$ is the charge conjugation symmetry. For $d=3$, the global symmetry is enhanced to $(SO(2N_f)\times U(1))\rtimes \mZ_2^C$. 

As is the case for the other fermionic models with four-fermion interactions, the model \eqref{NJLaction} is non-renormalizable for $d>2$ and its UV completion for $2<d<4$ is provided by the Nambu-Jona-Lasinio-Yukawa (NJLY) model which contains one complex scalar field ${\bm\varphi}=\varphi_1+i\varphi_2$ and has the following action \cite{Zinn-Justin:1991ksq},
\ie 
S_{\rm NJLY}=\int d^d x \left(
{1\over 2}|\pa {\bm\varphi}|^2-\bar\Psi^i \slashed{\partial} \Psi_i +g_1 \bar\Psi^i(\varphi_1+i\C_5\varphi_2)\Psi_i
+
g_2 |{\bm\varphi}|^2 +g_4 |{\bm\varphi}|^4
\right)\,.
\label{NJLYaction}
\fe 
The scalar field transforms under the chiral $U(1)$ symmetry as ${\bm\varphi}\to e^{i\A} {\bm\varphi}$. In the scaling region, the NJL and the NJLY models are expected to coincide \cite{Zinn-Justin:1991ksq}, where the theory  undergoes a second-order phase transition with order parameter ${\bm\varphi}$ and spontaneous $U(1)$ chiral symmetry breaking. 

The NJL (NJLY) model is also known as the chiral XY GN (GNY) model in the condensed matter literature, where it describes various quantum phase transitions with complex order parameters (see for example \cite{Roy_2010,Roy_2013,Li_2017,Zerf:2017zqi,Otsuka:2018kcb}). In particular, it governs the superconducting phase transition in graphene at $N_f=2$ \cite{Roy_2010,Roy_2013} and the semimetal-VBS (valence bond solid) transition in graphene and graphene-like material for various values of $N_f$ \cite{Li_2017} as well as the critical surface states of certain topological insulators with emergent supersymmetry 
 at $N_f={1\over 2}$ \cite{lee2007emergence,Roy_2013,zerf2016superconducting}.\footnote{The corresponding CFT has 3d $\cN=2$ supersymmetry and is also known as the $\cN=2$ super-Ising model whose operator spectrum has been analyzed using conformal bootstrap \cite{Bobev:2015vsa}.}
 
To solve the NJL model in the large $N=2N_f$ expansion, we proceed by introducing a complex HS field ${\bm\phi}=\phi_1+i\phi_2$, which is identified (up to an overall normalization) with the complex scalar $\bm \varphi$ in the NJLY model near the critical point. We then integrate out the Dirac fermions and obtain the following effective Lagrangian, with 
\ie 
   \cF_{\rm NJL}({\bm\phi}) = -{N\over 2} \left( \tr \log \left(\slashed{\partial} + \phi_1 + i \gamma_5 \phi_2\right) + g_0^t{\left|{\bm\phi}\right|^2} \right)\,,
\fe 
where the rescaled coupling $g_0^t$ is related to the coupling constant in \eqref{NJLaction} by $g_0=N g_0^t$ and kept fixed in the large $N$ limit. Again 
by fine-tuning $g_0^t$ we can bring the model to its critical point which is described by the chiral XY GN (or GNY) CFT. We exploit the semi-classical approximation in the large $N$ limit as before and derive the gap equation that governs the dominant saddle which we assume to be described by a homogeneous configuration of the complex HS scalar $\bm \phi$.
Taking advantage of the $U(1)$ symmetry, we can always set $\phi_2 = 0$. At zero temperature, the gap equation is then equivalent to the one for the ordinary GN model \eqref{GNgapeqn} with $\phi=\phi_1$ and the same critical value of the coupling $g_0^t$ applies with the CFT vacuum described by ${\bm \phi}=0$. 
The same equivalence holds for the gap equation and its solution at finite temperature for either periodic or anti-periodic spin structures. Consequently, the leading large $N$ contribution to the free energy density of the critical NJL (or NJLY) model with $N_f={N\over 2}$ agrees with that of the critical GN (or GNY) model with $N$ two-component Dirac fermions (see \eqref{GNPsaddle} and \eqref{GNAPleadingF}),
\ie \label{leadingNJLF}
F^{\rm P,AP}_{\rm NJL,0}(\B)=F^{\rm P,AP}_{\rm GN,0}(\B)\,.
\fe
Nonetheless, as we show explicitly below, they start to differ at the first subleading order in ${1\over N}$.

To obtain the $\frac{1}{N}$ correction $F^{\rm P,AP}_{\rm NJL,-1}$ to the free energy density, we compute the induced propagators for the HS fields $\phi_1, \phi_2$, which take the following form 
\ie 
\left(G^+_{\bm\phi}\right)^{-1}(\Omega,p) =   \frac{N}{\pi\beta}\log {\left(2\sinh \frac{\beta \phi_+}{2} \right)}\begin{pmatrix}
   1  & 0
    \\
    0 & 1 
\end{pmatrix} +2N\begin{pmatrix}
   P^2 + 4 \phi_{+}^2  & 0
    \\
    0 & P^2 
\end{pmatrix} \Pi^{+}_{\beta}(\Omega,p)
\,,
\fe 
and 
\ie 
\left(G^-_{\bm\phi}\right)^{-1}(\Omega,p) =  \frac{N}{\pi\beta}\log {\left(2\cosh \frac{\beta \phi_-}{2} \right)}\begin{pmatrix}
   1  & 0
    \\
    0 & 1 
\end{pmatrix} +  2N\begin{pmatrix}
   P^2 + 4 \phi_{-}^2  & 0
    \\
    0 & P^2 
\end{pmatrix} \Pi^{-}_{\beta}(\Omega,p)
\,,
\fe 
with $\phi_\pm$ and $\Pi^{\pm}_{\beta}(\Omega,p)$ the same as around \eqref{defPpm} where the $\pm$ label the periodic and anti-periodic spin structures respectively.  Note that the additional zero of $\left(G^\pm_{\bm\phi}\right)^{-1}_{22}$ at $P^2=0$ corresponds to the Goldstone boson for the chiral $U(1)$ symmetry, which contributes to the ${1\over N}$ correction to the free energy density of the critical NJL model. Comparing with the analysis in Section~\ref{sec:GNFandchem} for the GN model, we find that 
the full $\frac{1}{N}$ correction in the NJL model for the free energy density is given by,
\begin{gather}\label{NJLsubFP}
     F^{\rm P}_{\rm NJL,-1}(\B) =2 F^{\rm P}_{{\rm GN},-1}(\B)-\cF_{\rm ren}(4\sigma_{*})+\cF_{\rm ren}(0)= -\dfrac{0.32259927}{\beta^{3}}\,,
\end{gather}
for the periodic spin structure, and analogously for the 
anti-periodic (thermal) spin structure, the result is 
\ie 
    F^{\rm AP}_{\rm NJL,-1}(\B) =2 F^{\rm AP}_{{\rm GN},-1}(\B)
    =\dfrac{-0.02680198}
    {\beta^{3}}\,.
    \label{NJLsubFAP}
    \fe 
As promised, the above clearly differ from the results in the GN model (see \eqref{GNPsub} and \eqref{GNAPsub}).

Similar to the GN$'$ model obtained from contour rotation compared to the familiar GN model, the NJL model also has an analogous cousin at large $N$ which we refer to as the NJL$'$ model. Once again, to the leading large $N$ limit, its free energy coincides that of the GN$'$ model, 
\ie  
F^{\rm AP}_{\rm NJL',0}(\B)=F^{\rm AP}_{\rm GN',0}(\B)\,,
\fe
whereas this degeneracy is lifted at the subleading order,
  \ie 
     F^{\rm AP}_{\rm NJL',-1}(\B) =2 F^{\rm AP}_{{\rm GN}',-1}(\B)-\cF_{\rm ren}(4\phi_{*}^2)+\cF_{\rm ren}(0)
    =-\dfrac{0.44297273}{\beta^{3}}\,.
\fe

The Wilson coefficient $b$ for the $SO(N)$ symmetry of the NJL model is straightforward to obtain given the above discussion, after turning on a chemical potential. It is easy to see that the leading contribution coincides with the answer for the GN model and the subleading contribution is twice of that for the GN model because of the presence of two real HS fields. Thus we have the following (comparing to \eqref{GNb}),
\ie 
     b =  \frac{2\log 2}{\pi} - \frac{0.2676}{N} \,.
\fe 

Let us now consider more general thermal observables given by one-point function of the higher-spin currents in the critical NJL model. They are defined in a way analogous to \eqref{JsdefGN} but in terms of the four-component Dirac fermions,
\ie 
\hat  J^s_{\mu_1 \ldots \mu_s} = \left (\dfrac{1}{N s}\sum_{\ell=1}^s \bar\Psi^i \partial_{\mu_1} \cdots \C_{\m_\ell}\cdots\partial_{\mu_s}  \Psi_i - \text{traces} \right) +{\rm descendants}\,.
\label{JsdefNJL}
\fe 
The corresponding one-point function coefficients $\hat b_{\pm,s}$ (defined in a similar way as in \eqref{J1pfgenformGN})
can be computed from Feynman diagrams as in Section~\ref{sec:GNhs1pf}. As one may expect from the discussion around \eqref{leadingNJLF} that concerns the free energy density (equivalently the stress-energy tensor one-point function), these more general one-point functions in the critical NJL model also coincide with those in the ordinary GN model (see \eqref{bt0P} and \eqref{bt0AP}) 
to the leading order in the $1\over N$ expansion, namely 
\ie 
\hat{b}^0_{\pm ,s} = \tilde{b}^0_{\pm ,s}\,. 
\fe 
We expect this agreement to fail at the subleading order which we have demonstrated explicitly for $s=2$ (see around \eqref{NJLsubFP} and \eqref{NJLsubFAP}) and for general $s$ 
 by an analysis parallel to that in Section~\ref{sec:1pfcurrentsON}.

\section{Chern-Simons Quantum Electrodynamics}\label{sec:QED}

We now study finite temperature observables in $d=3$ vector-like large $N$ CFTs whose definition involves dynamical gauge fields. For concreteness, we focus on perhaps the simplest class of nontrivial gauge theories, namely the quantum electrodynamics (QED) with $N$ massless Dirac fermions and Chern-Simons coupling $k$ (which we refer to as CSQED), and present explicit results for its free energy density at the subleading order in the $1\over N$ expansion as a function of the 't Hooft coupling $\lambda={4\pi N\over k}$.

\subsection{Review of the CSQED and Large $N$ Thermal Observables}
The CSQED model in $d=3$ describes the interaction of $N$ two-component Dirac fermions with the electromagnetic field $A_\mu$, 
\begin{gather}
\label{CSQEDaction}
    S_{\rm QED}=\int d^3 x \left[\frac{1}{4 e_0^2} F_{\mu\nu}^2 - \bar{\psi}^i \gamma^\mu \left(\partial_\mu + i A_\mu\right) \psi_i - m_0 \bar{\psi}^i \psi_i +  {k\over 4\pi} \epsilon^{\mu\nu\rho} A_\mu \partial_\nu A_\rho\right]\,,
\end{gather}
with Maxwell coupling $e_0$, fermion bare mass $m_0$ and Chern-Simons coupling $k$ that satisfies the quantization condition $k+{N\over 2}\in \mZ$.\footnote{Note that the mass term in \eqref{CSQEDaction} is parity-odd and $SU(N)$ invariant. The CS level $k$ here (also known as the effective CS level)  includes the formal ${N\over 2}$ shift from the gauge invariant regularization of the fermion determinant using the eta invariant $\exp\left({{N\over 2}\pi i\eta(A)}\right)$ \cite{Witten:2015aba}. The theory is parity invariant only when $N$ is even and the effective CS level vanishes $k=0$.
} The theory has a $U(N)/\Gamma\rtimes \mZ_2^C$ global symmetry where the $U(1)$ subgroup of $U(N)$ comes from the topological symmetry (under which the monopole operators are charged), the $SU(N)$ subgroup is the flavor symmetry of the Dirac fermions, $\mZ_2^C$ is the charge conjugation symmetry and the discrete quotient $\Gamma\subset U(1)$ ensures that the symmetry acts faithfully on the local operators \cite{Benini:2017dus,Cordova:2017kue}.\footnote{For $k<{N\over 2}$, $\Gamma=\mZ_{N/2-k}$ and otherwise $\Gamma=\mZ_{N/2+k}$. See \cite{Benini:2017dus} for details.}

There is substantial evidence that at sufficiently large $N$, upon tuning the bare mass $m_0$, the 3d CSQED flows to a nontrivial $U(N)$ invariant CFT at long distance \cite{Pisarski:1984dj,Appelquist:1988sr,Nash:1989xx,Chen:1993cd,Hermele:2004hkd,DiPietro:2015taa}, which we will refer to as the conformal CSQED.\footnote{These abelian gauge theories are also related by conjectured IR dualities to certain nonabelian Chern-Simons-matter theories \cite{Son:2015xqa,Aharony:2015mjs,Karch:2016sxi,Murugan:2016zal,Seiberg:2016gmd,Hsin:2016blu}. Consequently our results for the abelian theories in the large $N$ limit give rise to predictions for the corresponding thermal observables in these nonabelian theories, which are strongly coupled, due to the very nature of these level/rank type dualities.} This is particularly well-established by considering the limit of $N \to \infty$ and studying the ${1\over N}$ expansion (see for example \cite{Borokhov:2002ib,Kaul:2008xw,Giombi:2015haa,Chester:2016ref,Chester:2017vdh,DiPietro:2023zqn}).\footnote{At small enough even $N\leq N_*$ for even critical flavor number $N_*$, the QED model with vanishing Chern-Simons level $k=0$ is believed to exhibit spontaneously ``chiral'' symmetry breaking of $SU(N) \to SU(N/2)\times SU(N/2) \times U(1)$ 
due to certain parity-even $SU(N)$ invariant scalar operator (quartic in the fermions) becoming relevant and inducing a 
nonzero expectation value for the order parameter (quadratic in the 
fermions) \cite{Pisarski:1984dj,Appelquist:1988sr,Dagotto:1989td,DiPietro:2015taa}.} See also \cite{Chester:2016wrc,Li:2018lyb,He:2021sto,Albayrak:2021xtd} for recent works on bootstrapping the corresponding CFTs directly at finite $N$. Such CFTs are known to describe exotic phases of quantum matter in two spatial dimensions including the Dirac spin liquid \cite{Rantner:2000wer,Rantner:2002zz,Herbut:2002yq,Franz:2002qy,Hermele:2005dkq,Hermele:2008afa} for vanishing Chern-Simons level $k=0$ and various values of $N$ depending on the underlying two-dimensional lattice. For nonzero Chern-Simons level $k\neq 0$, they also emerge near the quantum phase transitions between different quantum hall states 
\cite{Lee:2018udi}. Below we explain how to extract finite temperature observables for these quantum critical points in the large $N$ limit, taking into account the subleading effects. 

By integrating out the fermions, we arrive at the following effective Lagrangian for the gauge field,
\begin{gather}
\cF_{\rm CSQED}(A)=  \frac{N}{4 g_0^2} F_{\mu\nu}^2 - N \operatorname{tr} \log\left[\slashed \partial + i \slashed A_\mu +m_0\right]+ \frac{N}{\lambda} i\epsilon^{\mu\nu\rho} A_\mu \partial_\nu A_\rho\,,
\end{gather}
where we have rescaled the couplings
\ie 
e_0^2 = \frac{g_0^2}{N}\,,\quad k = \frac{4\pi N}{\lambda}\,,
\fe
and we take the large $N$ limit with $g_0$ and $\lambda$ kept fixed.
Note that the gauge field $A_\m$ here plays the role of the $\phi$ field in the case of the GN model (similarly the $\sigma$ field for the scalar $O(N)$ model).

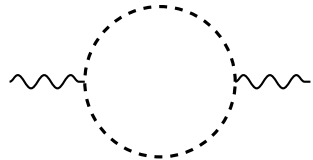
\begin{figure}[!htb]
    \centering
    \begin{tikzpicture}
       \draw [snake it,thick]   (-4,0) -- (-3,0);
        \draw[very thick,dashed] (-2,0) circle (1);
        \draw [snake it,thick]   (-1,0) -- (0,0);
     \end{tikzpicture}
    \caption{The leading diagram in the $\frac{1}{N}$ expansion that contributes to the photon self-energy. The dashed line corresponds to the fermion loop and the curved lines correspond to the photon legs.}
    \label{fig:QEDloop}
\end{figure}

It is easy to see in the large $N$ limit that the theory has a nontrivial fixed point in the IR for any initial coupling $g_0^2$. The quantum correction to the fermion mass in the IR can be cancelled by properly fine-tuning the bare mass parameter $m_0$ (and in the dimensional regularization scheme this corresponds to setting $m_0=0$).\footnote{No fine-tuning is required for $k=0$ and $N$ even due to the extra parity symmetry.} The propagator for the gauge field $A_\mu$ can be computed from standard Feynman diagrams (see Figure~\ref{fig:QEDloop}) and in the Lorentz  gauge is given by (see also \cite{Kaul:2008xw}),
\ie 
    D^{-1}_{\mu\nu}(P) = \frac{1}{e_0^2} \left(P^2 g_{\mu\nu} - P_\mu P_\nu\right) + \frac{k}{4\pi} \epsilon_{\mu\nu\rho}P^\rho +\frac{N}{16 P} \left(P^2 g_{\mu\nu} - P_\mu P_\nu\right)\,,
    \fe 
    and at large distance,
    \ie 
    D^{-1}_{\mu\nu}(P) \xrightarrow[ P\to 0]{} \frac{1}{\lambda} \epsilon_{\mu\nu\rho}P^\rho +\frac{1}{16 P} \left(P^2 g_{\mu\nu} - P_\mu P_\nu\right)  \,,
\fe 
which indeed takes the right form that is required for conformal symmetry. In particular, it implies that the gauge invariant operator $F_{\m\n}$ is a primary operator of scaling dimension $\Delta=2$ in the large $N$ limit. This also means that in the IR we can neglect the Maxwell action since it is irrelevant.

We proceed as before to study this CFT on ${\rm S}^1_\beta \times \mathbb{R}^2$ to determine its thermal observables. The free energy density of the conformal CSQED has the same type of ${1\over N}$ expansion as in \eqref{Fexp} for the critical $O(N)$ model, and we will be interested in the first two terms $F_{\rm CSQED,0}(\B)$ at order $\cO(N)$ and $F_{\rm CSQED,-1}(\B)$ at order $\cO(N^0)$.

 We can again argue that in the large $N$ limit we can pursue a semi-classical approximation where $A_\mu$ is homogeneous. Using gauge transformations we can set the spatial components along $\mR^2$ to $A_1=A_2=0$, but we can not set $A_0=0$ because of the non-trivial topology of the thermal background. 
Nonetheless, we can fix
\ie 
A_0 = u \equiv \frac{1}{\beta}\int\limits^\beta_0 d\tau A_0(\tau,x)\,,
\fe 
and compute the effective Lagrangian as a function of the holonomy $u\in [0,2\pi/\B)$,
\ie 
    \cF_{\rm CSQED}(u) = -\dfrac{N}{\beta}
   \sum_n \int \frac{d^2 p }{(2\pi)^2} \log \left[\left({(2n+1)\pi \over \B}+ u\right)^2+ p^2 \right]\,,
\fe 
where we have imposed anti-periodic boundary conditions\footnote{This is without loss of generality since $u$ changes the effective boundary condition for the fermions.} for the fermions along ${\rm S}^1_{\B}$. Implementing the regularization for the sum-integral as before (see in particular \eqref{cFGN}), we arrive at,
\ie 
 \cF_{{\rm CSQED}}(u)
=\dfrac{N }{2\pi \beta^{3}}\left[{\rm Li}_{3}(-e^{i\beta u})+{\rm Li}_{3}(-e^{-i\beta u})\right]\,.
\fe 
The homogeneous saddle-point for the holonomy $u$ relevant for the thermal CFT and the corresponding leading free energy density then follow,
\ie 
u_* = 0\,,\quad F_{\rm CSQED,0}(\beta) =\cF_{{\rm CSQED}}(u_*)= - \frac{3 N \zeta(3)}{4\pi \beta^3}\,,
\label{FSCQEDlead}
\fe  
which coincides with the free energy density of $N$ free Dirac fermions.
We emphasize that $\cF_{{\rm CSQED}}(u)$ has another saddle-point at $u=\pi$ which is a local maximum and thus unstable.  We see that effective anti-periodic boundary conditions on the fermions are energetically favored.

As in the previous sections we can introduce a chemical potential for the global symmetry and determine the Wilson coefficient $b$ from the symmetry twisted free energy. Here we will do so for the $SU(N)$ global symmetry of  conformal CSQED for even $N$, where the symmetry twist is implemented by $g=(e^{i\m}1_{N/2},e^{-i\m}1_{N/2})$. In this case, the effective Lagrangian becomes
\ie 
 \cF_{{\rm CSQED}}(u,\mu)
=\dfrac{N }{4\pi \beta^{3}}\left[{\rm Li}_{3}(-e^{i(\beta u+\mu)})+{\rm Li}_{3}(-e^{-i(\beta u-\mu)})+ {\rm Li}_{3}(-e^{-i(\beta u+\mu)})+{\rm Li}_{3}(-e^{i(\beta u-\mu)})\right]\,.
\fe 
We find that $u=u_*=0$ is the minimum of $\cF_{{\rm CSQED}}(u,\mu)$ for all $\mu$. Therefore, the thermal free energy of the CSQED, as a function of the chemical potential, is
\ie
    \cF_{{\rm CSQED}}(\mu) = \dfrac{N }{2\pi \beta^{3}}\left[{\rm Li}_{3}(-e^{i  \mu})+{\rm Li}_{3}(-e^{-i \mu})\right]= -\frac{3\zeta(3)}{4\pi \B^3}+ \frac{\log 2}{2\pi \B^3} \mu^2 +\dots \,,
    \label{QEDb}
\fe 
which produces the leading large $N$ answer for the $b$ coefficient $b_{\rm CSQED} = \frac{2 \log 2}{\pi }$.

It is easy to repeat the analysis in the previous section to compute the one-point function of the higher-spin currents defined in \eqref{JsdefGN} in the large $N$ limit of the conformal CSQED. The one-point function coefficient $\tilde b_s^0$ for the spin $s$ current follows from \eqref{bt0AP} with $\phi_*=0$, 
\ie 
    \tilde{b}^0_s      =\frac{1}{2\pi} \frac{2^{1-s} (2s)!}{s!} \operatorname{Li}_{s+1}(-1) = -\frac{ \left(2^s-1\right) \zeta (s+1) \Gamma
   \left(s+\frac{1}{2}\right)}{\pi^\frac32}\,,
\fe 
which again coincides with the free fermions as expected in the leading large $N$ limit of the CSQED.

\subsection{Subleading corrections to the Free Energy}

The large $N$ CSQED is of course very different from free fermions. Indeed, the distinctions already manifest at the first subleading order in the $1\over N$ expansion of thermal observables in the CFT. Below we demonstrate this explicitly for the free energy density (equivalently the one-point function of the stress-energy tensor), which depends nontrivially on the 't Hooft coupling $\lambda$.

To obtain the $\frac{1}{N}$ correction to the free energy density, we compute the quantum propagator for the gauge field $A_\m$. We first work with vanishing Chern-Simons level $k=0$. The photon polarization tensor then reads,
\ie 
    \Pi_{\mu\nu}(\Omega_n,p) = {1\over \B } \sum_n \int \frac{d^2 k}{(2\pi)^2} \frac{\tr\left[\gamma^\mu \slashed k \gamma^\nu (\slashed k + \slashed p) \right]}{k^2 (k+p)^2}\,, \quad \Omega_n \equiv  {2\pi n \over \B}\,,
\fe 
from fermions running in the loop.
One can check that it has the following structure as required by the residual symmetry at finite temperature and the Ward identity,
\ie 
\label{photonpolarizationgenform}
   & \Pi_{00}(\Omega_n, q) = \Pi_E(\Omega_n,q)\,, \quad \Pi_{0i}(\Omega_n,q) = - \frac{\Omega_n q_i}{q^2} \Pi_E(\Omega_n,q)\,,\\
    &\Pi_{ij}(\Omega_n,q) = \left(q^2\delta_{ij} - q_i q_j\right) \Pi_M(\Omega_n,q) + \frac{\Omega_n^2 q_i q_j}{q^4} \Pi_E(\Omega_n,q)\,, 
\fe 
where $\Pi_E(\Omega_n,q)$ and $\Pi_M(\Omega_n,q)$ are scalar functions (see Appendix~\ref{app:massNqed} for explicit expressions). After taking into account the Faddeev-Popov ghosts, 
the ${1\over N}$ correction to the free energy density of the conformal QED is given by,
\ie 
    F_{\rm QED,-1}(\beta) =  \frac{1}{2\beta} \sum_n \int \frac{d^2 p}{(2\pi)^2}  \log\left[\Pi_M(\Omega_n,q) \Pi_E(\Omega_n,q)\right]\,.
    \label{QEDsubsumintegral}
\fe 
Using the results of the Appendix~\ref{app:massNqed} and implementing the numerical evaluation of the sum-integral above as explained in Section~\ref{sec:numerical}, we find that,
\ie 
    F_{\rm QED,-1}(\beta) = - \frac{0.21211735}{\beta^3}\,,
    \label{FQEDsub}
\fe 
which agrees with the results in \cite{Kaul:2008xw}.
\begin{figure}[!htb]
    \centering
    \includegraphics{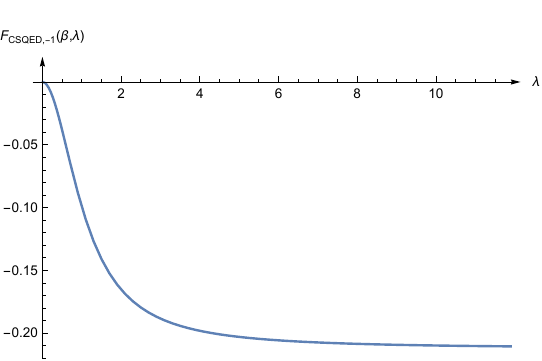}
    \caption{The dependence of the subleading correction $F_{\rm CSQED,-1}(\B,\lambda)$ to the free energy density of the conformal CSQED  on the 't Hooft coupling $\lambda$ (here $\B=1$).}
    \label{fig:thooft}
\end{figure}

We now re-introduce the Chern-Simons coupling $k$ and study the dependence of the free energy density on the 't-Hooft coupling constant $\lambda={4\pi N\over k}$. Recall that the leading contribution to the free energy density $F_{\rm CSQED,0}(\B)$ is $\lambda$ independent. In this case, the photon polarization operator is modified in a simple way to,
\ie 
    \Pi^\lambda_{\mu\nu}(\Omega_n,q) = \Pi_{\mu\nu}(\Omega_n,q) + \frac{1}{\lambda} \epsilon_{\mu\nu\rho} q^\rho\,.
\fe 
Consequently, the free energy density at the subleading order reads,
\ie 
     F_{\rm CSQED,-1}(\beta,\lambda) = \frac{1}{2\beta} \sum_{n} \int \frac{d^2 p}{(2\pi)^2}  \log\left[\Pi_M(\Omega_n,q) \Pi_E(\Omega_n,q) + \frac{1}{\lambda^2}\right]\,,
\fe 
which now depends nontrivially on $\lambda$ and can be computed numerically from the method outlined in Section~\ref{sec:numerical} (see Figure~\ref{fig:thooft}). In particular, it interpolates between the free fermion answer at $\lambda=0$ where there is no subleading correction to the free energy density \eqref{FSCQEDlead} and the pure QED answer \eqref{FQEDsub} at $\lambda=\infty$.

 \section*{Acknowledgements}

We thank Ofer Aharony, Nathan Benjamin, Simone Giombi, Igor R. Klebanov, Hirosi Ooguri, Subir Sachdev, Evgeny Skvortsov and William Witczak-Krempa for insightful discussions and correspondence.
The work of YW was
supported in part by the NSF grant PHY-2210420 and by the Simons Junior Faculty Fellows program.
F.K.P.
is currently a Simons Junior Fellow at New York University and supported
by a grant 855325FP from the Simons Foundation.

\appendix

\section{Lattice Regularization}
\label{sec:latticereg}
 Here we discuss an alternative approach to compute the thermal free energy of the critical $O(N)$ model in $d=3$ using a finite lattice. First, we consider the system on an infinite lattice $\mathbb{Z}^3$ with lattice spacing $a$ \cite{montvay1994quantum}. It amounts to replacing the scalar propagators in the following way
 \ie 
   G^{-1}(p,m^2) = p^2 + m^2 \,\to\, G_a^{-1}(p,m^2) = 6 - 2 \cos p_x- 2 \cos p_y  - 2\cos p_z + m^2\,, 
\fe 
with normalized periodic momenta $p_{x,y,z} \in \left[-\pi,\pi\right]$.
This automatically regulates the UV divergences. The gap equation \eqref{eq:gapeq} becomes,
 \begin{gather}
     \int\limits^\pi_{-\pi} \frac{d^3 k}{(2\pi)^3} G_a(k, \sigma) =  \frac{\sigma}{\lambda_0^t}-r_0^t\,,
 \end{gather}
 expanding the left-hand side of this equation for small $\sigma$ we obtain,
 \ie 
       \int\limits^\pi_{-\pi} \frac{d^3 k}{(2\pi)^3} G_a(k, \sigma) \sim -r_*^t - \frac{1}{4\pi} \sqrt{\sigma} + \frac{\sigma}{\lambda^t_*} + \ldots\,, 
       \fe  
with 
\ie 
       r^t_* =  -\int\limits^\pi_{-\pi} \frac{d^3 k}{(2\pi)^3} G_a(k, 0)= -0.252731\,, \quad \lambda^t_* = -0.012\,,
       \label{rlambda}
\fe 
where $\lambda^t_*$ is determined from numerically fitting the integral. 
To approach the critical point we fine-tune $r_0^t$ to $r^t_*$. In this case $\sigma=0$ is the solution and the system is gapless in the IR. The other coupling $\lambda^t_0$ could be left arbitrary but it is better for the numerics to also tune it to $\lambda^t_0=\lambda_*^t$ to cancel the leading finite-size corrections from the least irrelevant operator. Indeed, the propagator for the HS field $\sigma$ in this limit becomes,
\ie 
   -{2\over N}  G^{-1}_\sigma(p_i) =\frac{1}{\lambda_0^t} + \int \limits^\pi_{-\pi} \frac{d^3 k}{(2\pi)^3} G_a(k,0)G_a(k+p,0) \xrightarrow[p \to 0]{}\frac{1}{8\left|p\right|} + \frac{1}{\lambda_0^t} - \frac{1}{\lambda^t_*} + \mathcal{O}(p)\,, 
\fe 
 and if we set $\lambda^t_0=\lambda^t_*$ the leading correction to the conformal propagator is cancelled.

 Now we study the same model at criticality on a finite lattice $\mathbb{Z}_{n_t} \times \mathbb{Z}_n^2$. For general $n,n_t\gg 1$ this correspond to the lattice approximation for studying the CFT on the torus ${\rm S}^1_{R}\times {\rm S}^1_{R} \times {\rm S}^1_\beta$ with $\frac{R}{\beta} = \frac{n}{n_t}$. If we further take $n\gg n_t$, the lattice setup approximates the thermal background $\mathbb{R}^2 \times {\rm S}^1_\beta$ with inverse temperature $\beta = n_t a$. The scalar propagator on this lattice is 
 \ie 
     G_a^{-1}(p,\sigma) = 6 - 2 \cos p_x- 2 \cos p_y  - 2\cos p_t + \sigma\,, \quad p_{x,y} = \frac{2\pi i_{x,y}}{n}\,,\quad p_t=\frac{2\pi i_t}{n_t} \,.
\fe 
where $i_{x,y,t}\in \mZ$.
 The gap equation is
 \ie 
     \frac{1}{n^2 n_t}\sum_{i_x,i_y=1}^n \sum^{n_t}_{i_t=1} G_a\left(\frac{2 \pi i_x}{n},\frac{2 \pi i_y}{n},\frac{2 \pi i_t}{n_t},\sigma \right) = -r^t_* + \frac{\sigma}{\lambda_0^t}\,,
\fe 
 and we will attempt to set $\lambda_0^t \sim \lambda_*^t$ to reduce the leading finite-size corrections as discussed above.  We have checked numerically that the solution of this gap equation is $\sigma_* = \frac{\Delta^2}{n_t^2}$  which produces the leading free energy density (see \eqref{leadingONF}) with good precision for a lattice of size $n=100$ and $n_t=40$
 \begin{gather}
    {1\over N}F_0 n_t^3 \sim -0.160282036\,.
 \end{gather}
 To obtain the subleading correction to the free energy density, we should study the propagator of the HS field $\sigma$, equivalently the self-energy (see also \eqref{Pidef}),
 \begin{gather}
     \Pi_\beta(q) = \frac{1}{2\lambda_0^t} +  \frac{1}{2n^2 n_t}\sum_{i_x,i_y=1}^n \sum^{n_t}_{i_t=1} G_a\left(p,\sigma_*\right)G_a\left(p+q,\sigma_*\right)\,.
 \end{gather}
 In the limit of the interest $n \gg n_t \gg 1$ and small $q$, we expect that $\Pi_\beta(q) \sim n_t \tilde{\Pi}_\sigma(q n_t)$. Then the subleading correction to the free energy  is 
\ie 
F_{-1}(n,n_t) = \frac{1}{2} \frac{1}{n^2 n_t} \sum^n_{i_x,i_y=1} \sum^{n_t}_{i_t=1}\log\left[ \Pi_\beta\left(\frac{2 \pi i_x}{n},\frac{2 \pi i_y}{n},\frac{2 \pi i_t}{n_t}\right)\right]\,.
\label{Flatticesum}
\fe 
Limited by the precision for the value of $\lambda_*^t$ in \eqref{rlambda} and the size of the lattice we can consider with the available computing power,
 we wouldn't get the desired behavior 
 \ie 
 F_{-1}(n_t) \sim {\rm const} + \frac{f_{-1}}{n_t^3}\,,
 \label{Fsubcriticalform}
 \fe for the free energy density at the critical point (where $f_{-1}$ is the desired CFT observable and the constant needs to be subtracted by numerical fitting).   
 
 To see this more explicitly, we estimate that the self-energy $\Pi_\beta(q)$ behaves for $n_t \gg q n_t \gg 1$ as
\ie 
\label{PIfinitesize}
  \Pi_\beta(q) \sim \frac{1}{\lambda^t_0} - \frac{1}{\lambda_*}+\frac{1}{8 q} + \frac{g_1}{q^2 n_t} + \frac{g_2}{q^4 n_t^3}\,.
\fe 
To achieve a good approximation to the critical behavior, we need to consider a sizable lattice such as $n\sim 500\,, n_t \sim 25$. 
For instance, in this parameter regime, the value of the $\sigma$ field from solving the gap equation is $\sigma_* n_t^2 = 0.9268$ which is close to the critical value \eqref{ONsaddle}. But then to evaluate the sum in \eqref{Flatticesum} accurately requires large processing powers that we don't have at the moment.  For a smaller lattice such as $n\sim 70$ and $n_t \sim 10$ however, we obtain instead $\sigma_* n_t^2 =0.9409$, which reflects a relative error of  $\epsilon_{\sigma_*} \sim 1\%$. Since the free energy density involves a sum over a logarithm of the self-energy which depends on $\sigma$,  such a relative error is quite important.

For smaller lattices, the subleading corrections in \eqref{PIfinitesize} also need to be taken into account, which requires setting $\lambda_0^t=\lambda_*^t$ with high precision. However the fact that we don't know $\lambda_*^t$ (see \eqref{rlambda}) at high precision introduces a numerical uncertainty to the computation of the $\Pi_\beta$ which again propagates to the numerical evaluation of the subleading piece $F_{-1}(n_t)$ of the critical free energy density.
We expect the total error in \eqref{Fsubcriticalform} to be $\Delta_{F_{-1}} \sim 0.01$  for $n_t \sim 10$, which makes it hard to extract the small critical amplitude $f_{-1}$ (see \eqref{FONsubfinal}). To improve the situation, we must know $\lambda_*^t$ at least within a relative error of $\epsilon_{\lambda^t_*} \sim n_t^{-3}$.

\section{Self-energy of $\sigma$ in the $O(N)$ Vector Model}
\label{app:ONmodel}
In this section we analyze the two-point function of the HS field $\sigma$ at finite temperature for the critical $O(N)$ model at large $N$ and provide useful analytic expressions that will help the numerical evaluation of subleading effects for thermal observables in the CFT. 

From diagrammatic analysis (equivalently by saddle-point approximation of the path integral), one finds that the self-energy for the $\sigma$ field (i.e. the inverse propagator) is,
\begin{gather}
\Pi_{\beta}( \Omega, q)=\dfrac{1}{2\beta}\sum_{n=-\infty}^\infty \int \dfrac{d^{2}k}{(2\pi)^{2}}\dfrac{1}{\left(\omega_{n}-\Omega\right)^{2}+\left(k-q\right)^{2}+\sigma_{*}}\dfrac{1}{\omega_n^{2}+k^{2}+\sigma_{*}}\,, 
\label{ONselfenergysumintegral}
\end{gather}
where $\omega_{n}\equiv {2\pi n\over \B}$ denotes the Matsubara frequency and $\sigma_{*}$ is the solution to the gap equation given in \eqref{ONsaddle} (although the expressions below will not depend on its value unless explicitly stated). Using Feynman parametrization, we take the integral over spatial momentum and obtain the following,
\ie 
\Pi_{\beta}( \Omega, q)=\dfrac{\beta}{32\pi^{3}}\int\limits_{0}^{1}dx \sum_{n=-\infty}^{\infty}\dfrac{1}{\left(n-\frac{\beta}{2\pi}\Omega x\right)^{2}+\frac{\beta^{2}}{4\pi^{2}}M^{2}}\,,
\label{ONselfenergy2}
\fe 
where we have introduced
$M^{2}\equiv\sigma_{*}+\left(\Omega^{2}+q^{2}\right)x(1-x) > 0$. Evaluating the sum in \eqref{ONselfenergy2} then gives,
\ie 
&\Pi_{\beta}( \Omega, q)
\\
=&
-\dfrac{1}{16\pi}\int \limits_{0}^{1}\dfrac{dx}{ \sqrt{\sigma_{*}+\left(\Omega^{2}+q^{2}\right)x(1-x)}} \dfrac{\sinh{(\beta  \sqrt{\sigma_{*}+\left(\Omega^{2}+q^{2}\right)x(1-x)}) }}{\cos{(\beta \Omega x)-\cosh{(\beta \sqrt{\sigma_{*}+\left(\Omega^{2}+q^{2}\right)x(1-x)})}}}\,.
\label{ONselfenergysim}
\fe 
If we introduce a chemical potential $\m$ for the $U(1)\subset O(N)$ subgroup with even $N$ as described in Section~\ref{sec:ONchem}, the self-energy becomes,
\ie \label{sec:eq:chempolbos}
&\Pi^\mu_{\beta}\left( \mu,\Omega, q,\sqrt{\tilde{\sigma}_*} \right)
\\
=&
-\dfrac{1}{16\pi}\int \limits_{0}^{1}\dfrac{dx}{ \sqrt{\tilde{\sigma}_*+\left(\Omega^{2}+q^{2}\right)x(1-x)}} \dfrac{\sinh{(\beta  \sqrt{\tilde{\sigma}_*+\left(\Omega^{2}+q^{2}\right)x(1-x)}) }}{\cos{(\beta \Omega x -  \mu)-\cosh{(\beta \sqrt{\tilde{\sigma}_*+\left(\Omega^{2}+q^{2}\right)x(1-x)})}}}\,,
\fe 
where $\tilde{\sigma}_*$ is the solution \eqref{ONsaddlesolwcp} of the modified gap equation
\eqref{chpotoN}.

The self-energy of the $\sigma$ field captures the subleading correction to the free energy density of the critical $O(N)$ model in the following way,
\begin{eqnarray}
\label{Fsumintegralb4reg}
F_{-1}(\beta)=\frac{1}{2 \beta}\sum_{n\in \mZ} \int \frac{d^2 p}{(2\pi)^2}  \log |G^{-1}_{\sigma}(\Omega_n,p)|=\frac{1}{2\beta}\sum_{{n\in \mZ}}\int \dfrac{d^{2}p}{(2\pi)^{2}}\log{\Pi_{\beta}(\Omega_{n},p)}\,,
\end{eqnarray}
up to regularization and renormalization as described in Section~\ref{sec:ONfreeE}.

\subsection{Large Momentum Expansion}\label{app:largepON}
To analyze the UV structure of the self-energy and the sum-integral in \eqref{Fsumintegralb4reg}, it is useful to have the large momentum expansion of \eqref{ONselfenergysumintegral} which we provide here, generalizing the results in \cite{Chubukov:1993aau} to further subleading orders.

For this purpose, it is useful to use the following representation of the self-energy by first implementing the sum in \eqref{ONselfenergysumintegral},
\ie 
    \Pi_{\beta}( \Omega,q)=&-\dfrac{1}{4}\int \dfrac{d^2 k}{(2 \pi)^2}\dfrac{1}{\left(P^{2}+2 q\cdot k\right)^{2}+4\Omega^{2}\omega^{2}_{k}}\left(\left(P^{2}+2q\cdot k\right)\left(\dfrac{1}{\omega_{q+k}}-\dfrac{1}{\omega_{k}}\right)-\dfrac{2\Omega^{2}}{\omega_{q+k}}\right)
    \\
    +&
   \int \dfrac{d^{2} k}{(2\pi)^{2}}\dfrac{n(\beta\omega_{k})}{\omega_{k}}\dfrac{P^{2}+2q\cdot k}{\left(P^{2}+2 q\cdot k\right)^{2}+4\Omega^{2}\omega^{2}_{k}}\,, 
    \label{ONselfenergydonesum}
\fe 
with $P^{2}=\Omega^{2}+q^2$,  $\omega^{2}_{k}=k^2+\sigma_{*}$, $n(\beta\omega)$ denotes the Bose-Einstein distribution
\ie 
n(\beta\omega )\equiv \frac{1}{e^{\beta \omega}-1}\,,
\fe
and we are keeping $\sigma_*$ general.
Note that the first line of 
\eqref{ONselfenergydonesum} coincides with the self-energy at zero temperature (but nonzero mass $\sqrt{\sigma_*}$) and can be evaluated exactly. Thus we arrive at the following expression, 
\ie 
    \Pi_{\beta}( \Omega,q)=\dfrac{1}{8 \pi P}\arctan{\left[\dfrac{P}{2\sqrt{\sigma_{*}}}\right]}+ \frac{1}{P^2}\int \dfrac{d^{2} k}{(2\pi)^{2}}\dfrac{n(\beta\omega_{k})}{\omega_{k}}\dfrac{1+x}{\left(1+x\right)^{2}+y}\,,
    \label{ONselfEsim}
\fe
where we have introduced the following quantities for convenience,
\ie 
x\equiv \dfrac{2 q \cdot k}{P^{2}}\,, \quad  y\equiv \dfrac{4\Omega^{2}\omega^{2}_{k}}{P^{4}}\,.
\fe 
Due to the exponential suppression for large $x,y$ in \eqref{ONselfEsim}, we can simply expand the integrand and integrate over the spatial momentum $k$ to obtain the large momentum expansion of the self-energy,
\ie 
\Pi_{\beta}( \Omega,q)
=&\,\dfrac{1}{16 P}-\dfrac{1}{2\pi \beta P^2}\log{\left[2\sinh{\frac{\Delta}{2}}\right]}+\dfrac{2\Omega^{2}-q^{2}}{\beta^{3}P^{6}}\dfrac{1-6\gamma}{6\pi}\Delta^{3}+\dfrac{q^2}{\pi \beta^{3}P^6}\Delta^{2}\log{\left[2\sinh{\frac{\Delta}{2}}\right]}
\\
+&
\dfrac{1}{\pi \beta^{5}P^{10}}\left(-2 \Omega^{2}q^2\left(12\gamma_{1}+\gamma_{4}\right)+\Omega^{4}\left(8\gamma_{2}-\gamma_{4}\right)+q^{4}\left(3\gamma_{3}-\gamma_{4}\right)\right)+\mathcal{O}\left(\dfrac{1}{P^{8}}\right)\,,
    \label{PibetalargeP}
\fe 
where $\Delta\equiv \beta \sqrt{\sigma_{*}}$ is a dimensionless number. 

The leading ${1\over P}$ term in \eqref{PibetalargeP} is the conformal answer in flat spacetime which is subtracted to obtain the renormalized free energy density and the dangerous ${1\over P^2}$ term goes away once we specialize to $\Delta$ in \eqref{ONsaddle} that solves the gap equation (see around \eqref{subleadingF}).
Furthermore, the various constants in \eqref{PibetalargeP}  are
\ie 
&\gamma=\frac{1}{\Delta^{3}}\int \limits_{\Delta}^{+\infty}d\omega n(\omega)\omega^{2}=2.3241\,,\quad
&\gamma_{1}=\int \limits_{\Delta}^{+\infty}d\omega n(\omega)\omega^{2}(\omega^{2}-\Delta^{2})=22.8244\,,  \\
&\gamma_{2}=\int \limits_{\Delta}^{+\infty}d\omega n(\omega)\omega^{4}=24.7434\,,\quad
&\gamma_{3}=\int \limits_{\Delta}^{+\infty}d\omega n(\omega)(\omega^{2}-\Delta^{2})^{2}=21.3181\,, \\
&\gamma_{4}=\frac{4\Delta^{5}}{5}=0.660574\,,
\label{le8}
\fe 
where the explicit values are computed for $\Delta$ as in \eqref{ONsaddle}. Similarly, in the presence of chemical potential, using the same methods, we can represent \eqref{sec:eq:chempolbos} in the following form 
\ie 
   {}& \Pi^{\mu}_{\beta}(\mu, \Omega,q,\sqrt{\tilde{\sigma}_{*}})=
    \dfrac{1}{8 \pi P}\arctan{\left[\dfrac{P}{2\sqrt{\tilde{\sigma}_{*}}}\right]}
    \\
    &+\int \dfrac{d^{2} k}{(2\pi)^{2}}\dfrac{1}{2\omega_{k}}\bigg(n(\beta\omega_{k}+i\mu)+n(\beta\omega_{k}-i\mu)\bigg)\dfrac{P^2+2k\cdot q}{\left(P^2+2k\cdot q\right)^{2}+4 \Omega^{2}\omega_{k}}\,,
    \label{ONselfEchem}
\fe 
with the large momentum expansion of the form (we set $\beta=1$)
\ie 
     {}&\Pi^{\mu}_{\beta}(\mu, \Omega,q,\sqrt{\tilde{\sigma}_{*}})=\dfrac{1}{16P}-\dfrac{P^4-2 q^2 \tilde \sigma_* }{4\pi P^6} \log{\left(2\left[-\cos{\mu}+\cosh{\sqrt{\tilde{\sigma}_{*}}}\right]\right)} 
    +\dfrac{2\Omega^{2}-q^{2}}{P^{6}}\dfrac{1-6\tilde{\gamma}}{6\pi}\sqrt{\tilde{\sigma}_{*}^{2}}^{\,3}
    \\&
+
\dfrac{1}{\pi P^{10}}\left(-2 \Omega^{2}q^2\left(12\tilde{\gamma}_{1}+\tilde{\gamma}_{4}\right)+\Omega^{4}\left(8\tilde{\gamma}_{2}-\tilde{\gamma}_{4}\right)+q^{4}\left(3\tilde{\gamma}_{3}-\tilde{\gamma}_{4}\right)\right)+\mathcal{O}\left(\dfrac{1}{P^{8}}\right)\,,
\fe 
with the following constant coefficients,
\ie 
&\tilde{\gamma}=\frac{1}{\sqrt{\tilde{\sigma}_{*}}^{\ 3}}\int \limits_{\sqrt{\tilde{\sigma}_{*}}}^{+\infty}d\omega \dfrac{1}{2}\bigg(n(\omega+i\mu)+n(\omega-i\mu)\bigg)\omega^{2}\,,
\\
&\tilde{\gamma}_{1}=\int \limits_{\sqrt{\tilde{\sigma}_{*}}}^{+\infty}d\omega \dfrac{1}{2}\bigg(n(\omega+i\mu)+n(\omega-i\mu)\bigg)\omega^{2}(\omega^{2}-\tilde{\sigma}_{*})\,,  \\
&\tilde{\gamma}_{2}=\int \limits_{\sqrt{\tilde{\sigma}_{*}}}^{+\infty}d\omega \dfrac{1}{2}\bigg(n(\omega+i\mu)+n(\omega-i\mu)\bigg)\omega^{4}\,,
\\
&\tilde{\gamma}_{3}=\int \limits_{\sqrt{\tilde{\sigma}_{*}}}^{+\infty}d\omega \dfrac{1}{2}\bigg(n(\omega+i\mu)+n(\omega-i\mu)\bigg)(\omega^{2}-\tilde{\sigma}_{*})^{2}\,, \\
&\tilde{\gamma}_{4}=\frac{4\sqrt{\tilde{\sigma}_{*}}^{\,5}}{5}\,,
\label{PilargePchempot}
\fe
which coincide with $\eqref{PibetalargeP}$ for $\mu=0$.
\section{Self-energy of $\phi$ in the Gross-Neveu Model}\label{app:GNselfE}

In this section we compute the two-point function for the HS field $\phi\sim \bar \psi^i \psi_i$ in the critical Gross-Neveu model (equivalently the self-energy of $\phi$).  We consider both periodic and anti-periodic boundary conditions for the fermions and with a $U(1)$ chemical potential.

\subsection{Periodic Boundary Condition}
\label{app:GNP}
We start with computation of the self-energy of $\phi$ at finite temperature in the periodic spin structure with generic mass $\phi_{+}$, 
\ie
    \Pi^{\rm P}_{\rm GN}( \Omega, q,\phi_{+})=& \,\frac{1}{\beta}\sum_{\omega_n^+}\int \dfrac{d^{2}k}{(2\pi)^{2}}\Tr{\left(\dfrac{1}{i\slashed k+\phi_{+}}\dfrac{1}{i\slashed k-i\slashed q+\phi_{+}}\right)}
 =
    -\frac{2}{\beta}\sum_{\omega_{n}^+}\int \dfrac{d^{2}k}{(2\pi)^{2}}\dfrac{1}{\left(\omega^{+}_{n}\right)^2+k^2+\phi_{+}^{2}}
    \\
+
&\,
 \dfrac{2}{\beta}\sum_{n\in \mZ}\int \dfrac{d^{2}k}{(2\pi)^{2}}\dfrac{2\phi_{+}^{2}+\Omega^{2}+q^{ 2}-\omega_{n}^+\Omega-q \cdot k}{\left(\omega_{n}^+-\Omega\right)^{2}+\left(k-q\right)^{2}+\phi_{+}^{2}}\dfrac{1}{(\omega_{n}^+)^{2}+k^{2}+\phi_{+}^{2}}\,, 
\label{Bgnc1}
\fe 
where $\slashed{k}\equiv k^\m \C_\m$ includes the time component $k^0=\omega_n^+$ (see \eqref{freqsumferm}). The first term in the second equality above can be further simplified,
\begin{gather}
    -\dfrac{2}{\beta}\sum_{\omega_{n}^+}\int \dfrac{d^{2}k}{(2\pi)^{2}}\dfrac{1}{\left(\omega^{+}_{n}\right)^2+k^2+\phi_{+}^{2}}=\dfrac{1}{\pi \beta}\log\left[2 \sinh \frac{\beta \sqrt{\phi_{+}^{2}}}{2}\right] \,.
    \label{addtermproptosaddlep}
\end{gather}

Using Feynman parametrization and summing over the Matsubara frequency in the second line of \eqref{Bgnc1}, we arrive at the following expression,
\begin{gather}
    \Pi^{\rm P}_{\rm GN}(\Omega,q,\phi_+)=\dfrac{1}{\pi \beta}\log\left[2 \sinh \frac{\beta \sqrt{\phi_{+}^{2}}}{2}\right] +4\left(2\phi_{+}^{2}+\dfrac{P^{2}}{2}\right)\Pi_{\beta}( \Omega, q)\nonumber \\
    -\dfrac{\Omega}{4\pi}\int \limits_{0}^{1}dx \dfrac{\sin{\left(\beta \Omega x\right)}}{\cos{\left(\beta \Omega x\right)}-\cosh{\left(\beta \sqrt{\phi_{+}^{2}+P^{2}x(1-x)}\right)}}\,,
\label{Bgnc4}
\end{gather}
where $P^2\equiv \Omega^2+q^2$ and $\Pi_\beta$ is the self-energy for the $\sigma$ field in the bosonic $O(N)$ CFT (see \eqref{ONselfenergysumintegral}) where we need to replace $\sigma_{*}\rightarrow\phi_{+}^{2}$.
The second line in the above equation vanishes because its integrand is odd under $x\to 1-x$. Therefore, we finally have 
\ie  
\label{Bgnc12}
\Pi^{\rm P}_{\rm GN}(\Omega,q,\phi_{+})=\dfrac{1}{\pi \beta}\log\left[2 \sinh \frac{\beta \sqrt{\phi_{+}^{2}}}{2}\right] +2\left(P^{2}+4\phi_{+}^{2}\right)\Pi_{\beta}( \Omega, q)\,,
\fe 
whose large momentum expansion follows from that of $\Pi_{\beta}$ in \eqref{PibetalargeP}, and has the following form (with $\beta$=1):
\begin{gather}
\Pi^{\rm P}_{\rm GN}(\Omega,q,\phi_{+})=-\dfrac{4\phi_{+}^2}{\pi P^2}\log{\left[2\sinh{\frac{\sqrt{\phi_{+}^{2}}}{2}}\right]}\nonumber\\
+2(P^2+4\phi_{+}^2)\bigg(\dfrac{1}{16 P}+\dfrac{2\Omega^{2}-q^{2}}{P^{6}}\dfrac{1-6\gamma}{6\pi}\phi_{+}^{3}+\dfrac{q^2}{\pi P^6}\phi_{+}^{2}\log{\left[2\sinh{\frac{\sqrt{\phi_{+}^{2}}}{2}}\right]}
+\nonumber\\
+
\dfrac{1}{\pi P^{10}}\left(-2 \Omega^{2}q^2\left(12\gamma_{1}+\gamma_{4}\right)+\Omega^{4}\left(8\gamma_{2}-\gamma_{4}\right)+q^{4}\left(3\gamma_{3}-\gamma_{4}\right)\right)+\mathcal{O}\left(\dfrac{1}{P^{8}}\right)\bigg)\,,
    \label{PiperlargeP}
\end{gather}
where all the constants are defined in \eqref{le8}. The first term in  \eqref{PiperlargeP} is proportional to the gap equation which follows from \eqref{FGN0P} and therefore it vanishes in the CFT. 

\subsection{Anti-periodic Boundary Condition and General Chemical Potential}\label{app:GNAP}
In the anti-periodic (thermal) spin structure, the self-energy of $\phi$ takes a similar form, with the frequency sum from \eqref{freqsumferm} with generic mass $\phi_{-}$,
\ie
\label{modifiedPiAP}
    \Pi^{\rm AP}_{\rm GN}( \Omega, q,\phi_{-})=
&\,\dfrac{1}{\beta}\sum_{\omega_{n}^-}\int \dfrac{d^{2}k}{(2\pi)^{2}}\Tr{\left(\dfrac{1}{i\slashed k+\phi_{-}}\dfrac{1}{i\slashed k-i\slashed q+\phi_{-}}\right)}
= \\
-& \dfrac{2}{\beta}\sum_{\omega_{n}^-}\int \dfrac{d^{2}k}{(2\pi)^{2}}\dfrac{1}{\left(\omega^{-}_{n}\right)^2+k^2+\phi^{2}_{-}}\\
+
&\,\dfrac{2}{\beta}\sum_{\omega_{n}^-}\int \dfrac{d^{2}k}{(2\pi)^{2}}\dfrac{2\phi_{-}^{2}+\Omega^{2}+q^2-\omega^{-}_{n}\Omega-q \cdot k}{\left(\omega^{-}_{n}-\Omega\right)^{2}+\left(k-q\right)^{2}+\phi_{-}^{2}}\dfrac{1}{\left(\omega^{-}_{n}\right)^{2}+k^{2}+\phi_{-}^{2}}\,,
\fe

where the first term in the second equality can be simplified,
\begin{gather}
    -\dfrac{2}{\beta}\sum_{\omega_{n}^-}\int \dfrac{d^{2}k}{(2\pi)^{2}}\dfrac{1}{\left(\omega^{-}_{n}\right)^2+k^2+\phi^{2}_{-}}=\dfrac{1}{\pi \beta}\log\left[2 \cosh \frac{\beta \phi_-}{2}\right] \,.
    \label{adddiv}
\end{gather}
After performing the sum in the second line of \eqref{modifiedPiAP}, we obtain
\ie
\Pi^{\rm AP}_{\rm GN}(\Omega,q,\phi_-)=&\,\dfrac{1}{\pi \beta}\log\left[2 \cosh \frac{\beta \phi_-}{2}\right] + 2 (P^{2}+4\phi_{-}^{2})\Pi^-_\beta(\Omega,q)\,,
\label{gnap1}
\fe 
with 
\begin{gather}
  \Pi^-_\beta(\Omega,q)=\,\dfrac{1}{16 \pi}\int \limits_{0}^{1}dx\dfrac{1}{\sqrt{\phi_{-}^{2}+P^{2}x(1-x)}}\dfrac{\sinh{\left(\beta \sqrt{\phi_{-}^{2}+P^{2}x(1-x)}\right)}}{\cos{\left(\beta \Omega x\right)}+\cosh{\left(\beta \sqrt{\phi_{-}^{2}+P^{2}x(1-x)}\right)}} \,.
\label{gnap11}  
\end{gather}

If we further introduce a chemical potential $\mu$ for the $U(1)\subset O(2N)$ global symmetry as in Section~\ref{sec:GNFandchem}, we have instead
\ie 
    {}& \Pi^{\rm AP}_{\rm GN}(\mu,\Omega,q,\tilde{\phi}_*)=\dfrac{1}{2\pi}\log{\left(2\left[\cos{\mu}+\cosh{\beta\tilde{\phi}_*}\right]\right)}
     \\
 &   +\dfrac{1}{4\pi}\int \limits_{0}^{1}dx\dfrac{2\tilde{\phi}_*^{2}+P^{2}(1-x)}{\sqrt{\tilde{\phi}_*^{2}+P^{2}x(1-x)}}\dfrac{\sinh{\left(\beta \sqrt{\tilde{\phi}_*^{2}+P^{2}x(1-x)}\right)}}{\cos{\left(\beta \Omega x -  \mu\right)}+\cosh{\left(\beta \sqrt{\tilde{\phi}_*^{2}+P^{2}x(1-x)}\right)}}
    \\
    & -\dfrac{\Omega}{4\pi}\int \limits_{0}^{1}dx\dfrac{\sin{\left(\beta \Omega x -  \mu\right)}}{\cos{\left(\beta \Omega x-\mu \right)}+\cosh{\left(\beta \sqrt{\tilde{\phi}_*^{2}+P^{2}x(1-x)}\right)}} \label{sec:eq:chempol}\,,
\fe
where we used that (with $\tilde{\omega}_{n}=\omega^{-}_{n}+\mu$)
\begin{gather}
     -\dfrac{2}{\beta}\sum_{\tilde{\omega}_{n}}\int \dfrac{d^{2}k}{(2\pi)^{2}}\dfrac{1}{\left(\tilde{\omega}_{n}\right)^2+k^2+\tilde{\phi}_*^{2}}=\dfrac{1}{2\pi}\log{\left(2\left[\cos{\mu}+\cosh{\beta\tilde{\phi}_*}\right]\right)} \,.
\end{gather}
\subsection{Large Momentum Expansion of $\Pi_\B^-$} \label{app:GNAPlargeP}
The large momentum (i.e. large $P$) expansion of the self-energy $\Pi^-_\beta(\Omega,q)$ in the anti-periodic spin structure can be obtained by using the same methods developed in Appendix~\ref{app:largepON}. Here we present the resulting formula (with $\beta=1$),
\ie 
    \Pi^{-}_\beta( \Omega,q)
=&\,\dfrac{1}{16 P}-\dfrac{1}{2\pi P^2}\log{\left[2\cosh{\frac{ \phi_{-}}{2}}\right]}+ 
\dfrac{2\Omega^{2}-q^2}{P^{6}}\dfrac{\chi_0}{6\pi}+\dfrac{q^2}{\pi P^6}\phi_{-}^{2}\log{\left[2\cosh{\frac{\phi_{-}}{2}}\right]} \\
+&\,\dfrac{1}{\pi P^{10}}\left(-2 \chi_{1}\Omega^{2}q^{2}+ \chi_{2} \Omega^{4}+ \chi_{3} q^{4}\right)
+\mathcal{O}\left(\dfrac{1}{P^{8}}\right)\,,
    \label{gnap8}
\fe 
where the definitions of the constants involved and their numerical values for $\phi_{-}=\frac{2\pi i}{3}$  are given below,
\ie 
\chi_0=&\, \phi_{-}^3+6\int \limits_{\phi_{-}}^{+\infty}d\omega n_{F}(\omega)\omega^{2}=20.7\,,
\\
\chi_{1}=&-12\int \limits_{ \phi_{-}}^{+\infty}d\omega n_{F}(\omega)\omega^{2}(\omega^{2}- \phi_{-}^{2})+\frac{4 \phi_{-}^{5}}{5}=-401.2977\,,
\\
\chi_{2}=&-8\int \limits_{ \phi_{-}}^{+\infty}d\omega n_{F}(\omega)\omega^{4}-\frac{4 \phi_{-}^{5}}{5}=-146.4644\,,
\\
\chi_{3}=& -3\int \limits_{ \phi_{-}}^{+\infty}d\omega n_{F}(\omega)(\omega^{2}- \phi_{-}^{2})^{2}-\frac{4 \phi_{-}^{5}}{5}=-145.7247\,,
\fe 
and $n_F(\omega)$ is the usual Fermi-Dirac distribution,
\ie 
n_F(\beta\omega)\equiv \frac{1}{e^{\beta \omega}+1}\,.
\fe 
Combining \eqref{gnap1} and \eqref{gnap8} we obtain that at large momentum
\ie 
     \Pi^{\rm AP}_{\rm GN}( \Omega, q,\phi_{-})=&\,- \dfrac{4\phi_{-}^2}{\pi P^2 }\log{\left[2\cosh{\frac{ \phi_{-}}{2}}\right]}
    +
2\left(P^2+4\phi_{-}^2\right)\left[\dfrac{1}{16P}+\dfrac{2\Omega^{2}-q^2}{P^{6}}\dfrac{\chi_0}{6\pi}
\right.\\
+&\left. \dfrac{q^2}{\pi P^6}\phi_{-}^{2}\log{\left[2\cosh{\frac{\phi_{-}}{2}}\right]} +\dfrac{1}{\pi P^{10}}\left(-2 \chi_{1}\Omega^{2}q^{2}+ \chi_{2} \Omega^{4}+ \chi_{3} q^{4}\right)+\mathcal{O}\left(\dfrac{1}{P^{8}}\right)\,\right]\,,
    \label{PolarModdiflargeP}
\fe 
where the first term after the equality sign is proportional the gap equation which follows from \eqref{FGN0AP} and thus vanish for all solutions in \eqref{eq:GNfixpoints}.

In order to find the large momentum expansion of \eqref{sec:eq:chempol} in the presence of chemical potential $\mu$, it is more convenient to use the following expression (with $\beta=1$)
\ie 
    {}&\Pi^{\rm AP}_{\rm GN}(\mu,\Omega,q,\tilde{\phi}_*)=\dfrac{1}{2\pi}\log{\left(2\left[\cos{\mu}+\cosh{\tilde{\phi}_*}\right]\right)}+2\left(P^2+4\tilde{\phi}_*^{2}\right)\bigg[\dfrac{1}{8 \pi P}\arctan{\left(\frac{P}{2\sqrt{\tilde{\phi}_*^{2}}}\right)}\\
   & -\int \dfrac{d^2k}{\left(2\pi\right)^2}\dfrac{1}{2\omega_{k}}\bigg(n_{F}(\omega_{k}+i\mu)+n_{F}(\omega_{k}-i\mu)\bigg)\dfrac{P^{2}+2q\cdot k}{\left(P^{2}+2 q\cdot k\right)^{2}+4\Omega^{2}\omega^{2}_{k}}\bigg]\,,
    \label{APPolchemdifrep}
\fe 
where $\omega^{2}_{k}=k^2+\tilde{\phi}_{*}^{2}$. In the case of $\mu=\pi$, this reproduces \eqref{Bgnc12} and for $\mu=0$ it gives \eqref{gnap1}. Also, one can see that \eqref{APPolchemdifrep} can be written in terms of \eqref{ONselfEchem} as
\begin{gather}
    \Pi^{\rm AP}_{\rm GN}(\mu,\Omega,q,\tilde{\phi}_*)=\dfrac{1}{2\pi}\log{\left(2\left[\cos{\mu}+\cosh{\tilde{\phi}_*}\right]\right)}+2\left(P^2+4\tilde{\phi}_*^{2}\right)\Pi^{\mu}_{\beta}\left(\mu+\pi,\Omega,q,\tilde{\phi}_* \right)\,.
\end{gather}
Using the same methods as above we deduce that its large momentum expansion takes the following form  
\ie 
{}&  \Pi^{\rm AP}_{\rm GN}(\mu,\Omega,q,\tilde{\phi}_*)=-\dfrac{2\tilde{\phi}_*^{2}}{\pi P^2}\log{\left(2\left[\cos{\mu}+\cosh{\tilde{\phi}_*}\right]\right)}+2\left(P^2+4\tilde{\phi}_*^{2}\right)\bigg[\dfrac{1}{16 P} 
+\dfrac{2\Omega^{2}-q^2}{P^6}\dfrac{\tilde{\chi}_0}{6\pi}
\\
&+\dfrac{q^2}{2\pi P^6}\tilde{\phi}_*^2\log{\left(2\left[\cos{\mu}+\cosh{\tilde{\phi}_*}\right]\right)}+ 
\dfrac{1}{\pi P^{10}}\left(-2 \tilde{\chi}_{1}\Omega^{2}q^{2}+ \tilde{\chi}_{2} \Omega^{4}+ \tilde{\chi}_{3} q^{4}\right)+\mathcal{O}\left(\dfrac{1}{P^{8}}\right)\bigg]\,,
\label{largePGNAPchempot}
\fe
where the constants involved are defined below,
\ie 
\tilde{\chi}_0=&\,\tilde{\phi}_*^3+6\int\limits_{\tilde{\phi}_*}^{+\infty} \frac{d\omega}{2}\left[n_{F}\left(w+i\mu\right)+n_{F}\left(w-i\mu\right)\right]\omega^{2}\,,
\\
\tilde{\chi}_{1}=&-12\int\limits_{\tilde{\phi}_*}^{+\infty} \frac{d\omega}{2}\left[n_{F}\left(w+i\mu\right)+n_{F}\left(w-i\mu\right)\right]\omega^{2}(\omega^{2}- \tilde{\phi}_*^{2})+\frac{4 \tilde{\phi}_*^{5}}{5}\,,
\\
\tilde{\chi}_{2}=&-8\int\limits_{\tilde{\phi}_*}^{+\infty} \frac{d\omega}{2}\left[n_{F}\left(w+i\mu\right)+n_{F}\left(w-i\mu\right)\right]\omega^{4}-\frac{4 \tilde{\phi}_*^{5}}{5}\,,
\\
\tilde{\chi}_{3}=& -3\int\limits_{\tilde{\phi}_*}^{+\infty} \frac{d\omega}{2}\left[n_{F}\left(w+i\mu\right)+n_{F}\left(w-i\mu\right)\right](\omega^{2}- \tilde{\phi}_*^{2})^{2}-\frac{4 \tilde{\phi}_*^{5}}{5}\,.
\fe 
\section{Polarization Tensor of the Large $N$ QED}
\label{app:massNqed}

The polarization tensor for the large $N$ conformal QED at finite temperature takes the general form in \eqref{photonpolarizationgenform} and is determined by two scalar functions $\Pi_E(\Omega,p)$ and $\Pi_M(\Omega,p)$ which we refer to as the electric and the magnetic polarizations respectively. Below we provide their explicit expressions and their large momentum expansions which are used in the main text for the evaluation of subleading effects in the conformal (CS)QED at finite temperature.

\subsection{Electric Polarization and Large Momentum Expansion}
We start with the electric polarization in \eqref{photonpolarizationgenform}. After introducing the Feynman parametrization, and explicitly separating the divergent part of the polarization tensor, we arrive at
\ie 
    \Pi_{E}(\Omega,p) =& \, \dfrac{2}{\beta} \sum_{n\in \mZ} \int \frac{d^2 k}{(2\pi)^2} \frac{\omega_n(\Omega+\omega_n) - k \cdot(k+p) }{(\omega_n^2 + k^2) ((\Omega + \omega_n)^2+(k+p)^2)}\\
    =&\,\dfrac{1}{2\pi \beta}\sum_{n\in \mZ}\int \limits_{0}^{1}dx\dfrac{\omega_{n}\left(3\Omega-4\Omega x\right)-2\Omega^{2}x^2-2P^2x(1-x)+\Omega^{2}+p^2(1-x)}{\left(\omega_{n}+\Omega x\right)^2+P^2x(1-x)}
    \\+&
    \dfrac{1}{\pi \beta}\sum_{n\in \mZ}1-\frac{2}{\beta}\sum_{n\in \mZ} \int \frac{d^2 k}{(2\pi)^2}\dfrac{1}{(\omega_n^2 + k^2)}\,,
    \label{El1}
\fe 
with $\omega_n=\omega_n^-$ as in \eqref{freqsumferm}.
After implementing the sum (with zeta function regularization) in the second equality of \eqref{El1}, we obtain the following simplified expression for the electric polarization,
\ie 
\Pi_{E}(\Omega,p)=&\,\dfrac{1}{4\pi}\int \limits_{0}^{1}dx\dfrac{1}{M}\dfrac{\sinh{\left(\beta M\right)}}{\cos{\left(\beta \Omega x\right)}+\cosh{\left(\beta M\right)}}\left(\frac{P^2}{2}-2\left(P^2+\Omega^2\right)x(1-x)\right)\\
+&
\dfrac{1}{\pi}\int \limits_{0}^{1}dx\dfrac{\Omega x \sin{\left(\beta \Omega x\right)}}{\cos{\left(\beta \Omega x\right)}+\cosh{\left(\beta M\right)}}+\dfrac{1}{\pi \beta}\log{2}\,,
    \label{El6}
\fe 
with 
\ie 
M\equiv P\sqrt{x(1-x)}\,.\label{MdefQED}
\fe 
The large momentum expansion of \eqref{El6} can be deduced using the same methods as in Appendices~\ref{app:ONmodel} and \ref{app:GNselfE} and is given below (for convenience we set $\beta=1$),
\ie
&\Pi_{E}(\Omega,p)= 
\dfrac{1}{16}\dfrac{P^2-\Omega^{2}}{P}+\int {d^2k \over \pi^2 |k| }n_{F}(|k|)\dfrac{\left(k^2+k \cdot p\right)\left(P^2+2k \cdot p\right)-k^2\left(\Omega^{2}-p^2-2k \cdot p\right)}{(P^2+2k \cdot p)^2+4\Omega^{2}k^2}\\
&=
\dfrac{1}{16}\dfrac{p^2}{P}+\dfrac{3}{\pi}\zeta(3)\dfrac{p^2}{P^4}+\dfrac{45}{\pi}\zeta(5)\dfrac{p^2\left(p^2-4\Omega^{2}\right)}{P^{8}}+\dfrac{2835}{\pi}\zeta(7)\dfrac{p^2\left(p^4-12p^2\Omega^2+8\Omega^{4}\right)}{P^{12}}+\cO\left(1\over P^8\right)\,. 
    \label{El7}
\fe

\subsection{Magnetic Polarization and Large Momentum Expansion}
We now move onto the magnetic polarization in \eqref{photonpolarizationgenform}. Using rotation symmetry on the plane, we work with $(p_1,p_2)=(0,p)$, then
\ie
     \Pi_{M}(\Omega,p) = \frac{2}{\B p^2} \sum_m \int \frac{d^2 k}{(2\pi)^2} \frac{k_1^2 - \left(\omega_m (\Omega + \omega_m) +  k_2 (k_2 + p_2)\right) }{((\omega_m+\Omega)^2 + (k+p)^2) (\omega_m^2+k^2)}\,,
    \label{Mag1}
\fe 
Employing the same tricks as in the previous subsections, we arrive at the following simplified expression,
\ie 
\Pi_{M}(\Omega,p)=&\,\dfrac{1}{4\pi p^2}\int \limits_{0}^{1}dx\dfrac{1}{M}\dfrac{\sinh{\left(\beta M\right)}}{\cos{\left(\beta \Omega x\right)}+\cosh{\left(\beta M\right)}}\left[2\left(\Omega^{2}+p^2\right)x(1-x)\right]
\\
-&
\dfrac{1}{2\pi p^{2}}\int \limits_{0}^{1}dx\dfrac{\Omega x \sin{\left(\beta \Omega x\right)}}{\cos{\left(\beta \Omega x\right)}+\cosh{\left(\beta M\right)}}\,.
    \label{Mag3}
\fe 
where we have restored rotational invariance and $M$ is defined as in \eqref{MdefQED}.

Similarly the large momentum expansion of the magnetic polarization is given by  (again first with $(p_1,p_2)=(0,p)$),
\ie 
\Pi_{M}(\Omega,p) = &\,
\dfrac{1}{16}\dfrac{P}{p^2}+\dfrac{1}{p^2} \int {d^2 k\over \pi^2 |k|} n_{F}(|k|)\dfrac{\left(k_{2}^{2}-k^{2}_{1}+k_{2}p_{2}\right)\left(P^2+2k_{2}p_{2}\right)-k^2\left(\Omega^{2}-p^2-2k_{2}p_{2}\right)}{(P^2+2k_{2}p_{2})^2+4\Omega^{2}k^2}
\\
=&\,\dfrac{1}{16}\dfrac{P}{p^2}+\dfrac{3}{\pi}\zeta(3)\dfrac{\Omega^2-2p^2}{p^2 P^4}-\dfrac{45}{\pi}\zeta(5)\dfrac{4p^4-27p^2\Omega^2+4\Omega
^4}{p^2P^8}
\\
-&\dfrac{2835}{\pi}\zeta(7)\dfrac{6p^6-101p^4\Omega^2+116p^2\Omega^{4}-8\Omega^6}{p^2 P^{12}} +{1\over p^2}  \cO\left(1\over P^8\right)\,.
    \label{Mag4}
\fe 
Combining \eqref{El7} and \eqref{Mag4}, we obtain the large momentum expansion of the logarithm relevant for the evaluation of the subleading piece of the free energy density in \eqref{QEDsubsumintegral},
\ie 
\log{\left(16^{2}\Pi_{E}\Pi_{M}\right)}=&\,\dfrac{96}{P^{3}}\dfrac{\zeta(3)}{2\pi}\left(1+\dfrac{\Omega^{2}-2p^2}{P^{2}}\right)+\dfrac{1440}{P^{5}}\dfrac{\zeta(5)}{2\pi}\left(\dfrac{p^2-4\Omega^{2}}{P^{2}}-\dfrac{4p^4-27p^2\Omega^2+4\Omega^4}{P^{4}}\right)\\
&-
\dfrac{1}{2P^{6}}\left(\dfrac{96\zeta(3)}{2\pi}\right)^{2}\left(1+\dfrac{\left(\Omega^{2}-2p^2\right)^{2}}{P^4}\right)
+\dfrac{90720}{P^7}\dfrac{\zeta(7)}{2\pi}\left(\dfrac{p^4-12p^2\Omega^{2}+8\Omega^4}{P^{4}}\right.
\\
&\left.-\dfrac{6p^6-101p^4\Omega^{2}+116p^2\Omega^{4}-8\Omega^6}{P^{6}}\right)+\cO\left(1\over P^8\right)\,.
    \label{lerglogexpEM}
\fe 
\bibliographystyle{JHEP}
\bibliography{finiteT}
\end{document}